\documentclass{aa}

\usepackage{graphicx}
\usepackage{txfonts}
\usepackage[colorlinks,citecolor=blue]{hyperref}
\hypersetup{
    colorlinks = true,
    linkcolor = blue,
    anchorcolor = blue,
    citecolor = blue,
    filecolor = blue,
    urlcolor = blue
    }

\usepackage{booktabs}
\usepackage{mathrsfs}
\usepackage{tikz}
\usepackage{booktabs}
\usetikzlibrary{calc}
\usepackage{mathtools}
\usepackage{siunitx}
\usepackage{svg}
\usepackage{ulem}

\newcommand{\teff}{{T_{\text{eff}}}}

\begin{document}

   \title{Quantitative spectroscopy of single and multiple OB-type stars}
   \subtitle{Non-LTE spectrum analysis with machine learning}

   \author{P. Aschenbrenner\inst{1}
          \and
          N. Przybilla\inst{1}
          }

   \institute{Universit\"at Innsbruck, Institut f\"ur Astro- und Teilchenphysik, Technikerstr. 25/8, 6020 Innsbruck, Austria\\
              \email{patrick.aschenbrenner@student.uibk.ac.at; norbert.przybilla@uibk.ac.at}
             }

   \date{Received ; accepted }

  \abstract
   {}
   {The plethora of spectra of OB-type stars in observatory archives and the much larger numbers to come from the WEAVE and 4MOST spectroscopic facilities
   require efficient, but also accurate and precise methods for (semi)automatic quantitative analyses.}
   {Neural networks were used to emulate the spectra of single- and multi-star systems, trained on hybrid non-local thermodynamic equilibrium (non-LTE) models that cover a wide range of atmospheric parameters and chemical compositions. To derive the full set of stellar atmospheric parameters and uncertainties, a Markov chain Monte Carlo algorithm was implemented to fit high-resolution spectra within 3000\AA-10500\AA.}
   {The neural networks and fitting algorithm were bundled into a programme called Spectral Analysis Tool Using Restricted Neural networks ({\sc Saturn}). In its current implementation, {\sc Saturn} facilitates the emulation of synthetic spectra for spectral types O7 to B9, which differ only negligibly from computed models.
   {\sc Saturn} was tested on a number of benchmark stars that have been studied before, including single OB stars and a detached eclipsing binary (DEB) system. Excellent agreement of atmospheric parameters and elemental abundances for up to ten metal species is found with respect to the data in the literature, often with reduced uncertainties of about 1\% in effective temperature and 0.05\,dex in surface gravity (1$\sigma$ values). For DEB components, the uncertainties are larger, in particular for the fainter secondaries when only a single-epoch spectrum is considered, and lower if surface gravities and the effective temperature ratio from previous detailed light curve and orbital dynamics studies are used as additional fit constraints. Uncertainties of elemental abundances are typically $<$0.10\,dex. Some first applications of {\sc Saturn} for analyses of new targets are shown to demonstrate its capabilities, such as fast rotators, including HD\,149757 ($\zeta$~Oph). Consistent results are also found at reduced spectral resolutions relevant for observations with WEAVE and 4MOST.}
   {}


   \maketitle

\section{Introduction}
Quantitative spectroscopy is the most important tool for the study of stellar atmospheres. It allows atmospheric parameters, such as effective temperature $T_\mathrm{eff}$ and surface gravity $\log (g)$, and the chemical composition to be derived from an investigation of the detailed line spectrum and the global spectral energy distribution (SED). Since the first thorough application of the then newly developed theoretical framework of excitation and ionisation in thermodynamic equilibrium by \citet{Payne-Gaposchkin25}, which led to the identification of hydrogen as the main chemical constituent of stars, the field has reached a high degree of maturity. Despite the concept of local thermodynamic equilibrium (LTE) for the description of the local state of matter in stellar atmospheres being questioned early \citep{Gerasimovic29} -- requiring consideration of the so-called non-LTE effects -- LTE has remained a foundation for applications until today because of its simplicity and effectiveness in many cases.

However, a pure LTE description has been identified early as insufficient to characterise the atmospheres of hot massive stars of spectral types O and B (OB-type stars) and their progeny because of their intense radiation fields, with further deviations driven by low collision rates in the low-density plasmas of evolved stars. Non-LTE model atmospheres and spectrum synthesis became more realistic over decades due to methodological progress, the development of fast computers, efficient numerical algorithms for solving the radiative transfer and kinematic equilibrium equations, and the provision of the required atomic data, including opacities; for a historic overview of the landmark developments, readers can refer to the first chapter of \citet{HuMi15}. Currently, several codes exist to compute expanding line-blanketed non-LTE model atmospheres in spherical geometry for OB stars, mainly {\sc Cmfgen} \citep{HiMi98}, {\sc Fastwind} \citep{Pulsetal05,Pulsetal20}, {\sc PoWR} \citep{Graefeneretal02,HaGr03,Sanderetal15},
{\sc wm}-basic \citep{Pauldrachetal01}, and {\sc Metuje} \citep{KrKu17},
which are suited for quantitative analyses throughout the entire OB-star range. They are complemented by the line-blanketed full non-LTE {\sc Tlusty} code \citep{hubeny_88} and a hybrid non-LTE approach consisting of the {\sc Atlas} LTE line-blanketed model atmosphere code \citep{Kurucz93,Kurucz96} coupled to the non-LTE line-formation package {\sc Detail/Surface} \citep[plus further developments]{Giddings81,BuGi85}, abbreviated as {\sc Ads} henceforth, as introduced by \citet{Przybillaetal06}, \citet{NiPr07}, \citet{Wessmayeretal2022}, and \citet{Aschenbrenneretal23}. Both {\sc Tlusty} and {\sc Ads} assume hydrostatic equilibrium and plane-parallel geometry; they are therefore applicable to less-luminous OB stars where the line-forming photospheric layers are not significantly affected by the mass outflow. Overall, good consistency between the results obtained with the different codes is found, despite some different assumptions being made, although variances in some details remain, which are often related to choices in atomic data.

The sample size of analysed OB stars has increased greatly over the past three decades, driven by instrumental developments. While initially only single spectra were recorded using slit or fibre-fed spectrographs, several multiplexing approaches have been followed more recently. Multi-slit spectrographs increased the number of simultaneously recorded spectra to around twenty \citep[e.g.][]{Kudritzkietal12,Kudritzkietal14}, while fibre-fed multi-object spectroscopy provides data for more than a hundred targets simultaneously, such as those used in the VLT-FLAMES (Very Large Telescope Fibre Large Array Multi Element Spectrograph) survey of massive stars and the VLT-FLAMES Tarantula Survey \citep{Evansetal05,Evansetal11}, and independent analyses of these \citep{Urbanejaetal17}, or studies of clusters containing massive stars within the Gaia-ESO survey \citep{Blommeetal22,Moreletal22}. Integral field spectroscopy promises an even higher multiplex \citep[e.g.][]{Castroetal18}, which, however, is restricted by the spatial density of OB stars in reality \citep[e.g.][]{Evansetal19,Gonzalez-Toraetal22}. The limitations of these high-multiplex observations are spectral resolution, wavelength coverage, or both. Important high-resolution quantitative spectroscopy is therefore still based on single-target spectrographs, such as those used within the IACOB project \citep{SimonDiazetal20} or the  XShootU project \citep{Vinketal23}, which considers hundreds of stars. Large spectroscopic surveys of OB stars with new facilities such as the William Herschel Telescope Enhanced Area Velocity Explorer \citep[WEAVE,][]{Jinetal24} and the 4-metre Multi-Object Spectroscopic Telescope \citep[4MOST,][]{deJongetal19} will even provide spectra for tens of thousands of OB stars. These surveys will comprise a variety of objects, slow to fast rotators, single, binary, and multiple objects \citep[most OB stars turn out to have companions,][]{Chinietal12,Sanaetal12}, and chemically peculiar stars. Powerful algorithms are required to address the quantitative analysis.

A data-driven approach, such as that implemented for cool star analysis, trained on the known properties of a set of hundreds of observed reference stars \citep[{\sc The Cannon},][]{Nessetal15}, is difficult to transfer to the analysis of OB stars, as their spectra are sufficiently unique per object; this is in stark contrast to the similarity of the spectra of solar-type stars and red giants. Algorithms that employ classical equivalent width measurements for the analysis of OB star spectra have become rare, with the grid-based deep learning system described by \citet{Floresetal23} being an example. Ubiquitous are grid-based $\chi^2$-minimisation routines that compare synthetic to observed spectra, as, for example, discussed in the context of hot star analyses by the Gaia-ESO survey (\citealt{Blommeetal22}; see also \citealt{Arayaetal25}). Variants that use on-the-fly model calculations instead of precomputed grids exist, such as the genetic algorithms described by \citet{Mokiemetal05} or \citet{Tamajoetal11}.
A different concept is realised in algorithms using principal component analysis (PCA) in combination with a learning database, which, as a main advantage, offers the possibility to drastically reduce dimensionality. One implementation is described by \citet{Blommeetal22}, another one is the code {\sc Maui} \citep[][]{Urbanejaetal08,Urbaneja26}, which, in addition uses a Markov chain Monte Carlo (MCMC) algorithm to constrain uncertainties. Algorithms rarely facilitate the modelling of composite spectra of binaries \citep[e.g.][]{Irrgangetal14}. Instead, the usual approach followed relies on spectral disentanglement with subsequent individual spectrum analysis \citep[e.g.][]{Pavlovskietal23}, but at the cost of time-series spectroscopy.

Usually, these algorithms extract only a restricted set of parameters, such as a subset of atmospheric and wind parameters, and in addition abundances for a few elements (restricted by the availability of trustworthy model atoms). The full information accessible from the multitude of spectral lines, in particular in high-resolution spectra, is typically not obtained. This also affects the reliability of parameter extraction of spectra of faster rotating stars or lower-resolution spectra, where either the broad rotational profiles or the instrumental broadening (or both) smear out the various individual lines into broad, complex blends. However, including all elements present in stellar atmospheres for the spectrum synthesis comes at the cost of increasing computation time, and, in the worst case, an exponentially growing grid of models is needed for classical interpolation methods. By training a neural network to emulate model spectra, the number of models required can be kept manageable (such as in the PCA approach). Such spectral emulators have already been used to analyse spectra for large-scale surveys of late-type stars using LTE models and full spectrum synthesis \citep[{\sc Payne},][]{Tingetal2019}, with refinements of the approach also being discussed \citep[{\sc TransformerPayne},][]{Rozanskietal25}.
Extensions of {\sc Payne} with non-LTE models of late-type stars were done by \cite{Kovalevetal2019} and applied to binary stars in the Large Sky Area Multi-Object fiber Spectroscopic Telescope Medium Resolution Survey \citep[{LAMOST MRS},][]{Liuetal2020} by \cite{Kovalevetal2022}.
In the context of early-type stars of spectral types A, B, and O over 330\,000 objects were analysed in LTE by \citet[{\sc HotPayne}]{Xiangetal2022} and, independently, by \citet[{\sc Slam-Payne}]{Sun25}.
Analyses of high-resolution spectra with LTE models were done with {\sc Zeta-Payne} \citep{Straumitetal2022} for single stars \citep{Gebruersetal2022} as well as a binary system \citep{KovalevStraumit2023}. Ideally, the analysis of early-type stars requires an extension to non-LTE modelling.

Currently missing is a method to extract basically all available information on stellar parameters and elemental abundances in an efficient but still accurate and precise manner from ground-based spectroscopy of OB stars with charge-coupled device detectors, that is, from the atmospheric cut-off in the near-ultraviolet over the optical to the near-infrared. This shall be applicable to slow and fast rotators alike, to single and multiple stars. In the present work, we discuss an approach that solves this issue in a mostly automated fashion.

The paper is structured as follows. An overview of the calculated models and the emulation method with neural networks is given in Sect.~\ref{section:model_atmospheres}. A general method for fitting spectra of single and multi-star systems is introduced in Sect.~\ref{section:spectral_analysis}. The results of the analysis of five single benchmark stars and one binary system, as well as additional science targets, and their discussion are summarised in Sect.~\ref{section:results}. A general summary and outlook is provided in Sect.~\ref{section:summary}. In Appendix~\ref{appendix:A} we provide details on the neural network training process, Appendix~\ref{appendix:B} contains notes on the determination of the microturbulence velocity. The Appendices~\ref{appendix:C} and \ref{appendix:D} concentrate on SEDs and exemplary fits to the observed spectra of some of the sample stars. Finally, Appendix~\ref{appendix:E} focuses on the applicability of the analysis methodology introduced here to data at lower spectral resolution, such as those to be provided by WEAVE and 4MOST.

\begin{figure*}
    \sidecaption
    \includegraphics[width=11.3cm]{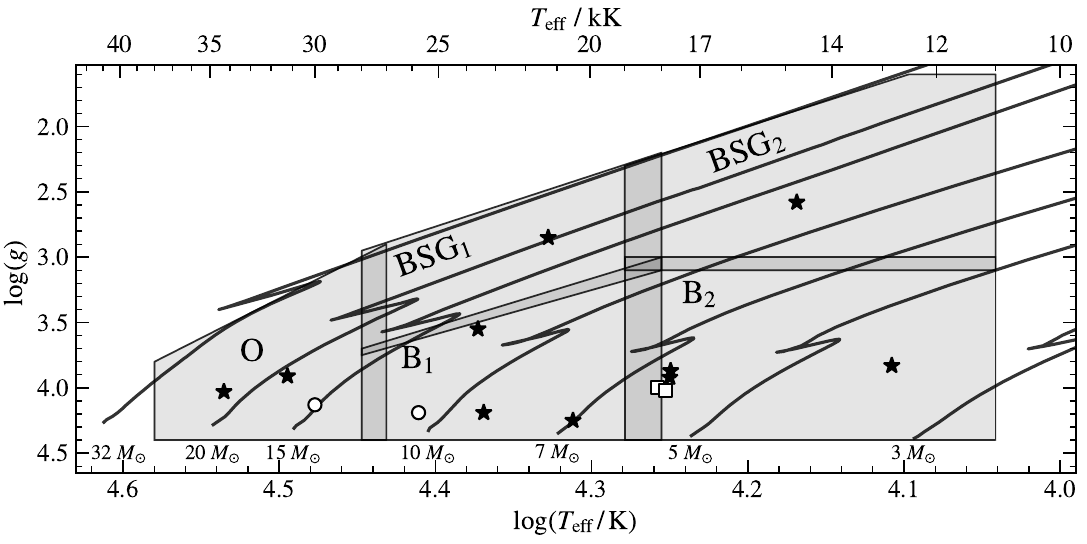}
    \caption{Kiel diagram showing the five different regions of model coverage (O, B$_1$, B$_2$, BSG$_1$, and BSG$_2$).
    Overlayed are \cite{Ekstroemetal12} evolutionary tracks for rotating stars
    ($\varv_\mathrm{ini}/\varv_{\text{crit}}$\,=\,0.4) at solar metallicity (black lines) for different masses, as indicated.
    The five-pointed black stars mark the positions of the single stars analysed in the present work and the matching
    open symbols are the components in the analysed binary systems.}
    \label{fig:Kiel_diagram}
\end{figure*}

\section{Model atmospheres and spectrum synthesis}\label{section:model_atmospheres}
For our model calculations, we subdivided the $T_{\rm eff}$ - $\log (g)$ space covering the OB-type stars that we intend to analyse into five overlapping regions as shown in Fig.~\ref{fig:Kiel_diagram}, labelled O, B$_1$, B$_2$, BSG$_1$, and BSG$_2$. Each region includes different ionisation stages of the elements in the non-LTE calculations that prevail at the given atmospheric parameter ranges, but all models within a region were calculated in a consistent way. The upper $T_\mathrm{eff}$ boundary is defined by the transition towards the appearance of stronger radiatively driven winds, which, similar to the lower $\log (g)$-value envelope, imply a limit for the assumption of hydrostatic equilibrium for the stellar photospheric layers. The upper $\log (g)$ boundary is defined by about the location of the zero-age main sequence (ZAMS), and at the lower $T_\mathrm{eff}$ boundary lines of chemical species that are not covered by our model atoms start to become relevant, for example \ion{Cr}{ii}, so that the spectrum synthesis starts to lose completeness.

\begin{table}
\caption{Model atoms for non-LTE calculations with {\sc Detail}.}
\label{table:modelatoms}
\centering
{\small
\begin{tabular}{llll}
\hline\hline
Ion                  & Terms           & Transitions      & Reference \\
\hline
\ion{H}{i}           & {25}            &  300             & {[}1{]} \\
\ion{He}{i/ii}       & 29+6/20         & 162/190          & {[}2{]}  \\
\ion{C}{i/ii/iii/iv} & 80/68/70/53     & 669/425/373/319  & {[}3{]}  \\
\ion{N}{i/ii/iii/iv} & 89/77/47+3/84+6 & 668/462/410/1123 & {[}4{]}  \\
O\,{\sc i/ii/iii}    & 51/177+2/132+2  & 243/2560/1515    & {[}5{]}  \\
Ne\,{\sc i/ii}       & 153/78          & 952/992          & {[}6{]}  \\
Mg\,{\sc i/ii}       & 88/37           & 471/236          & {[}7{]}  \\
Al\,{\sc ii/iii}     & 54+6/46+1       & 378/272          & {[}8{]}  \\
Si\,{\sc ii/iii/iv}  & 52+3/68+4/33+2  & 357/572/242      & {[}9{]}  \\
P\,{\sc ii/iii/iv}   & 45/25/23        & 358/81/61        & {[}10{]} \\
S\,{\sc ii/iii}      & 78/21           & 302/34           & {[}11{]} \\
Fe\,{\sc ii/iii/iv}  & 265/60+46/65+70 & 2887/2446/2094   & {[}12{]} \\
\hline
\end{tabular}
\tablebib{
  [1] \cite{PrBu04};
  [2] \cite{przybilla05};
  [3]~\cite{Przybillaetal01b}, \citet{NiPr06,NiPr08};
  [4]~\cite{PrBu01}, Przybilla \& Aschenbrenner (in prep.);
  [5]~\cite{Przybillaetal00}, Przybilla \& Butler (in prep.);
  [6]~\cite{MoBu08};
  [7]~\cite{Przybillaetal01a};
  [8] Przybilla (in prep.);
  [9]~Przybilla \& Butler (in prep.);
  [10]~\cite{Aschenbrenneretal25};
  [11]~\citet{Vranckenetal96}, updated;
  [12] \cite{Becker98}, \cite{Moreletal06}.
}
}
\end{table}

\subsection{Models and programmes}
For each region $\sim$5000 model spectra with randomly sampled parameters were calculated.
The free parameters of the models are effective temperature $T_{\rm eff}$, surface gravity $\log (g)$,
microturbulence velocity $\xi$, helium abundance by number $y$, and abundances for ten metals (see Table~\ref{table:modelatoms}). The model parameters were drawn from discrete uniform distributions with step sizes of
\SI{50}{\kelvin} for $T_{\rm eff}$, $0.01$\,dex for $\log (g)$, \SI{1}{\kilo\meter\per\second}
for $\xi$, $0.01$ for $y$, and $0.05$\,dex for the abundances of all other elements. The lower and upper limits
for the parameters dependent on the region and are listed in Table \ref{table:parameter_range}. The helium abundance
for all models is in the interval $[0.07,\,0.15]$ and the limits of the metal abundances, except nitrogen,
are chosen to be $\pm$0.8\,dex of the cosmic abundance standard \citep[CAS,][]{NiPr12,Przybillaetal13,Aschenbrenneretal25}.
The choices for the extent of the parameter ranges in $\xi$, $y$ and for the metal abundances were motivated by typical values that we have found in previous investigations, with some extra security margin, motivated by stellar evolution model predictions (in the case of helium) and the presence of abundance gradients in the Milky Way (metals).
Each model was computed with the hybrid non-LTE approach using {\sc Ads} in an implementation close to that described by \citet{Aschenbrenneretal23}. We briefly mention a few key ingredients: the use of {\sc Atlas12} models allows the effects of turbulent pressure to be considered on the atmospheric structure; use of an accelerated lambda iteration scheme \citep{RyHu91} that allows complex model atoms to be handled; use of an occupation probability formalism \citep{HuMi88,Hubenyetal94} for hydrogen to facilitate a better modelling of the series limits; use of up-to-date Stark-broadening tables for hydrogen and helium \citep{TrBe09,Beauchampsetal97,DiSa90,SchBu89}. For the non-LTE calculations we adopted state-of-the-art model atoms, listed in Table~\ref{table:modelatoms}, which experienced some updates over the years. For each element, the different ions considered are listed with the number of explicit terms (plus superlevels), the number of radiative bound-bound transitions per ion \citep[we note in particular the use of highly accurate and precise oscillator strengths of][]{FFT04,FFTI06}, and references. The model atoms also consider photoionisation cross-sections and large sets of collisional data from ab initio calculations. All model atoms are completed by the
ground energy term of the next-higher ion.

\begin{table}
\caption{\mbox{Parameter ranges coverage by the models in different regions.}}
\label{table:parameter_range}
\setlength{\tabcolsep}{1.5mm}
\small
\centering
\begin{tabular}{lcccc}
\hline\hline
Region & $T_{\rm eff}$ & $\log (g)$ & $\xi$ & N \\
 & K & cgs & km\,s$^{-1}$ & log(N/H)+12 \\
\hline
O    & 27000 - 38000 & 2.90 - 4.40 & 0 - 17 & 7.00 - 8.60 \\
B$_1$   & 18000 - 28000 & 3.00 - 4.40 & 0 - 16 & 7.00 - 8.60 \\
BSG$_1$   & 18000 - 28000 & 2.20 - 3.75 & 0 - 15 & 7.40 - 9.00 \\
B$_2$ & 11000 - 19000 & 3.00 - 4.40 & 0 - 14 & 7.00 - 8.60 \\
BSG$_2$ & 11000 - 19000 & 1.60 - 3.10 & 0 - 14 & 7.40 - 9.00 \\
\hline
\end{tabular}
\end{table}

\begin{figure*}
    \sidecaption
    \includegraphics[width=12cm]{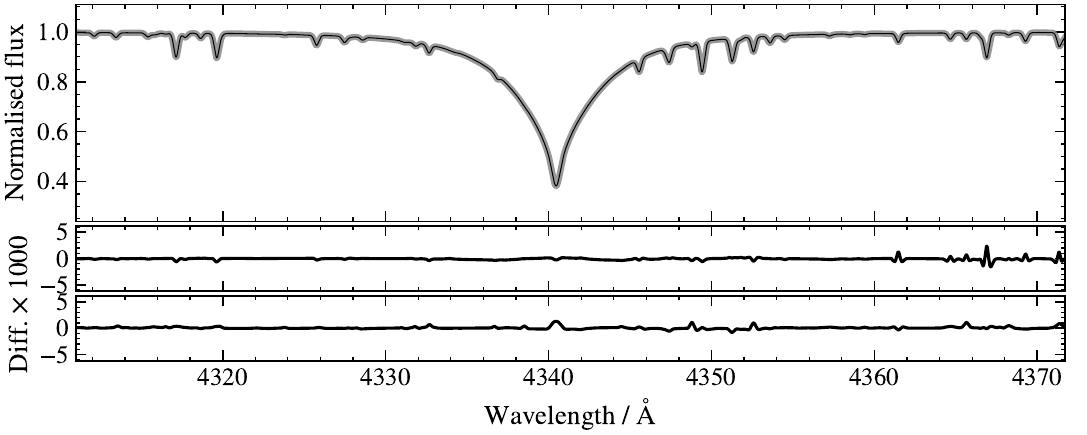}
    \caption{Comparison between a model spectrum calculated with {\sc Ads} and the spectra created with the neural networks. Top panel: the normalised spectrum of an {\sc Ads} model with $T_{\mathrm{eff}}$\,=\,$\SI{18500}{\kelvin}$ and $\log(g)$\,=\,3.10 (grey line) is compared to the spectrum created with the neural network from the B$_1$ region with the same parameters (black line). Middle panel: difference between the two spectra. Bottom panel: difference between the {\sc Ads} model spectrum and the spectrum created with the neural network from the B$_2$ region.}
    \label{fig:overlap_comparison}
\end{figure*}

\subsection{Single-star models}
The normalised flux $f_{\lambda_i}(\Theta)$ at each wavelength point $\lambda_i$ is a function of the
14 different parameters included in the model calculations, where all parameters are
combined into a vector $\Theta$. Based on the universal approximation
theorem \citep{Horniketal89}, we used our calculated models to train five neural networks,
one for each region, to predict the flux for given input parameters. The neural networks
can then be used to generate the models for arbitrary combinations of atmospheric parameters.
The models calculated with {\sc Surface} provide the flux values at a few hundred thousand points $\lambda_i$.
Instead of training a single neural network to predict all spectrum points at once, we split the
entire covered wavelength range into smaller chunks, each one covering $\sim$7500 points and having its own network.
The architecture of these smaller neural networks is a multilayer perceptron and we implemented
it with PyTorch \citep{paszkeetal2019}. The input layer has 14 neurons, one for each parameter, there are
three hidden layers with 32/64/128 neurons, respectively, using the sigmoid activation function
$\sigma(x)=\left( 1+\textrm{e}^{-x} \right)^{-1}$, and the output layer has $\sim$7500 neurons.
The full spectrum is created by concatenating the outputs of the different neural networks.
This approach allows for fitting optimisations since often only small parts of a
spectrum are considered for the analysis.
For the training, we split the calculated models into a training set and a testing set, the details are summarised in Appendix~\ref{appendix:A}.

This procedure was repeated for each of the five parameter regions. For validation, we looked at the overlaps
between the regions and compared a model spectrum computed with {\sc Ads}, which was not used for training,
with the spectra predicted by the neural networks. This is shown in Fig. \ref{fig:overlap_comparison}
for the regions B$_1$ and B$_2$. In general, we found that the absolute difference between model spectra calculated with {\sc Ads} in the testing set and the model spectra generated with neural networks is below \num[output-exponent-marker = E]{4.2e-5} for 50\% for all the spectrum points and below \num[output-exponent-marker = E]{2.8e-4} for 99\% of all the points. The differences reach a maximum value of about \num[output-exponent-marker = E]{2.0e-3} in the cores of a few spectral lines.

\subsection{Multi-star models}
Equation~(1) of \cite{Aschenbrenneretal24} can be employed to construct the normalised composite spectrum $f_{\mathrm{comp}}$ of multi-star systems from the spectra of individual stars
\begin{equation}
f_{\mathrm{comp}} = \frac{\sum_i w_i f_{\mathrm{cont},i} f_i}
                         {\sum_i w_i f_{\mathrm{cont},i}}\,,
\label{eq:composite_spectrum}
\end{equation}
where $f_i$ is the normalised flux of the $i$-th star and $f_{\mathrm{cont},i}$ the corresponding continuum flux, and $w_i$ a weight factor that matches the ratio of effective surface area of each star relative to the primary (for non-interacting stars). Again, we trained a neural network for each of the five regions to emulate the continuum flux of the calculated models. However, the neural networks for continuum emulation do not predict the flux at each wavelength point, only at $\sim$150 pre-defined wavelength points. Then, natural cubic splines are used to interpolate between those points.

\section{Spectral analysis}\label{section:spectral_analysis}
\subsection{Observational data}
In this work, we use high-resolution spectra with a resolving power of $R$\,=\,$\lambda/\Delta\lambda$\,$\approx$\,$\num{40000}$ to~$\num{115000}$ and wide wavelength coverage.
For two stars, as indicated in Table~\ref{table:results}, we used spectra taken with the Fibre Optics Cassegrain Echelle Spectrograph \citep[{FOCES},][]{Pfeifferetal98} on the 2.2\,m telescope at the Calar Alto Observatory in Spain; another three objects were observed with the Fiberfed Extended Range Optical Spectrograph \citep[{FEROS},][]{Kauferetal99} on the 2.2\,m telescope of the Max-Planck-Gesellschaft and the European Southern Observatory (ESO) at La Silla in Chile.
One star was observed with the Fibre-fed Echelle Spectrograph \citep[{FIES},][]{Telting2014} on the 2.5\,m Nordic Optical Telescope at La Palma,
five stars were observed with the Echelle Spectro-Polari\-metric Device for the Observation of Stars \citep[{ESPaDOnS},][]{ManDon03} on the 3.6\,m Canada-France-Hawaii telescope (CFHT) at Mauna Kea in Hawaii,
and the spectrum of one object was taken with the High Accuracy Radial velocity Planet Searcher \citep[{HARPS},][]{Mayoretal2003} at the ESO La Silla 3.6\,m telescope. For test purposes, we also employed intermediate-resolution spectra ($R$\,$\approx$\,5000, 10\,000 and 20\,000) observed with XSHOOTER \citep{Vernetetal11} on the ESO Very Large Telescope.

The data reduction for the spectra of HD\,214680, HD\,35299, HD\,91316, and HD\,164353
is described in previous works, see the references in Table~\ref{table:results}; for the processing of the HD\,160762 raw data see \citet{Irrgang14}.
In the case of HD\,259135, HD\,149757, HD\,87015, and HD\,17081 we downloaded pipeline-reduced spectra from the CFHT
Science Archive at the Canadian Astronomy Data Centre\footnote{\url{https://www.cadc-ccda.hia-iha.nrc-cnrc.gc.ca/en/cfht/}}
and for HD\,93827, HD\,112092, HD\,77464, and for the XSHOOTER data of HD\,35299 we downloaded Phase 3 data from the ESO Science Portal\footnote{\url{https://archive.eso.org/scienceportal/home}}.
The spectra were normalised by fitting a cubic spline function through carefully selected continuum points. The signal-to-noise ratios (S/Ns) measured at $\sim$5000{\AA} are also indicated in Table~\ref{table:results}.

\begin{figure*}
    \sidecaption
    \includegraphics[width=6.1cm]{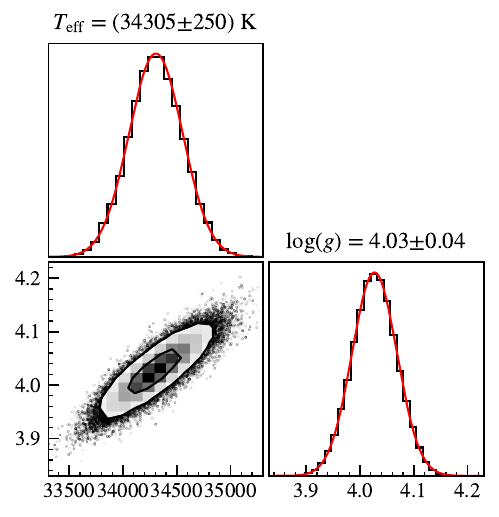}
    \hspace{2mm}
    \includegraphics[width=6.1cm]{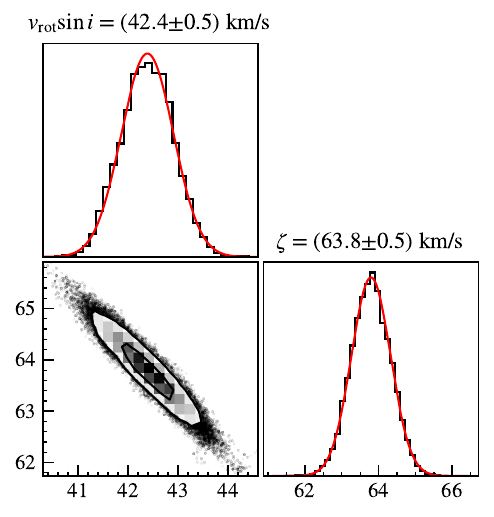}
    \caption{Exemplary corner plots of the MCMC fit results.
    In each example the bottom left shows the 2D histogram with $1\sigma$ and $2\sigma$ contours.
    The two adjacent plots show 1D histograms for the parameters (black) and a
    Gaussian (red) with same mean and standard deviation.
    Left panel set: $T_{\rm eff}$ and $\log (g)$ determination for the benchmark test star HD\,214680.
    Right panel set: $\varv \sin i$ and $\zeta$ determination for HD\,91316.
    }
    \label{fig:corner_plots}
\end{figure*}

\subsection{Model fitting}
To use the neural networks for spectral analysis, we wrote a programme called Spectral
Analysis Tool Using Restricted Neural networks ({\sc Saturn}).
For comparisons between models generated with the neural networks and
the observed spectra, the model spectra are broadened with rotational and radial-tangential macroturbulence
profiles \citep{Gray05}, assuming a linear limb darkening law, and a Gaussian instrumental profile,
which depends on the resolving power of the instrument used for observation.
{\sc Saturn} was implemented in Python (with some sub-routines written in C) and provides a graphical user interface which allows to interactively change
model parameters and to see the effects on the spectrum in real time. Multi-star systems with any number
of stars are supported by implementing composite spectra according to Eq.~\eqref{eq:composite_spectrum}.

The time it takes to generate a model spectrum of a single star using the neural networks
and to apply the broadening to it is $\SI{10}{\milli\second}$ on a fast contemporary central processing unit (vs.~$\sim$2\,h for the single model computation using {\sc Ads}).
With the ability to create model spectra in such a short time, we
use a MCMC fitting algorithm to derive the stellar parameters.
We implemented a random-walk Metropolis algorithm \citep{Metropolisetal53}, where the fit-parameters are
sampled from uniform distributions. The new parameters are accepted with a
probability of $\min \left(1, L(\Theta_i) / L(\Theta_{i-1}) \right)$,
where $L(\Theta_i)$ is the likelihood of the model with new parameters and $L(\Theta_{i-1})$
the likelihood of the model with the previous parameters. We define the likelihood function as
\begin{equation}
    L(\Theta)\propto \mathrm{exp}\left( \sum_{p} \frac{1}{n_{p}} \sum_{\lambda_i}-\frac{\left( f_{\lambda_i,\mathrm{obs}} - f(\Theta)_{\lambda_i,\mathrm{mod}} \right)^2}{2\cdot \left( \sigma^2_{\lambda_i,\mathrm{stat}}+\sigma^2_{\lambda_i,\mathrm{sys}}\right)} \right)\,,
    \label{eq:likelihood}
\end{equation}
where $f_{\lambda_i,\mathrm{obs}}$ is the observed flux at point $\lambda_i$,
$f(\Theta)_{\lambda_i,\mathrm{mod}}$ is the model flux at the same wavelength,
$\sigma_{\lambda_i,\mathrm{stat}}$ the statistical uncertainty of the observed flux and
$\sigma_{\lambda_i,\mathrm{sys}}$ the systematic uncertainty of the model.
We sum over different spectral windows $p$, covering $n_{p}$ wavelength points each.
The factors $1/n_{p}$ correspond to taking the mean log-likelihood of each
spectral window, weighting each window the same. Otherwise, weak metal lines would have almost
no significance compared to the much broader hydrogen and helium lines.
The statistical uncertainty depends on the S/N of the observation
$\sigma_{\lambda_i,\mathrm{stat}} \approx \sqrt{f_{\lambda_i,\mathrm{obs}}}/\mathrm{(S/N)}$ and we assume a
constant systematic uncertainty of $\sigma_{\lambda_i,\mathrm{sys}}=0.001$ for our model spectra.

With {\sc Saturn} any combination of atmospheric parameters, projected rotational
velocity $\varv \sin i$ and macroturbulence velocity $\zeta$ can be fitted simultaneously.
In the case of multi-star systems, the weights and the Doppler shift of the individual stars
can be selected as additional fit-parameters. Although it is possible to fit all
parameters simultaneously, in the present work we follow a more systematic process of iteratively deriving
individual parameters, as described in the following.

\subsection{Atmospheric parameter and abundance determination}
\paragraph{Effective temperature and surface gravity.}
To determine $T_{\rm eff}$ and $\log (g)$, we simultaneously fit the gravity sensitive
wings of hydrogen Balmer lines together with the spectral lines of different ions
for the same element to achieve ionisation equilibrium. The fit-parameters
are $T_{\rm eff}$, $\log (g)$, and the chemical abundances of all elements used
for ionisation equilibrium evaluation. An example of a corner plot of the MCMC fit result for the correlated $T_\mathrm{eff}$ and $\log(g)$ determination in HD\,214680 is shown in Fig.~\ref{fig:corner_plots}. We note the very tight uncertainty distributions.\\[-9mm]
\paragraph{Projected rotational velocity and macroturbulence.}
Both parameters are derived simultaneously by fitting the line profiles of
numerous metal lines. The fit-parameters are $\varv \sin i$, $\zeta$, and the abundances
of the elements involved. We use a slight deviation of Eq.~\eqref{eq:likelihood},
adopting $n_{p}$\,=\,1, corresponding to a classical maximum likelihood fit.
A corner plot of the MCMC fit result of the anti-correlated $\varv \sin i$ and $\zeta$ determination in HD\,91316 is also shown in Fig.~\ref{fig:corner_plots}, again with very tight uncertainty distributions.\\[-9mm]
\paragraph{Microturbulence.}
Usually, the microturbulent velocity $\xi$ is adjusted such that the element
abundance of the individual lines becomes independent of the line strength
(measured via the equivalent width). Instead, $\xi$ can be adjusted so that
the variance of all individual line abundances is minimised, which
is equivalent as we show in Appendix~\ref{appendix:B}. This approach can also be
used to determine $\xi$ in fast-rotating stars and multi-star systems,
where the equivalent widths cannot be determined due to blended lines.
To determine the microturbulence in this work, we fitted multiple
spectral lines of a single element using Eq. \eqref{eq:likelihood},
which yields the right value for $\xi$ based on some assumptions,
as we also show in Appendix~\ref{appendix:B}.\\[-9mm]
\paragraph{Elemental abundances.}
With all other parameters fixed, the abundances of elements are determined by fitting
each individual spectral line of an element and then taking the mean and standard
deviation as the atmospheric abundance and uncertainty.\\[-9mm]
\paragraph{Multi-star systems.}
Analysing spectra of multi-star systems includes an initial step to determine
the radial velocity shift of the individual components. Otherwise, the procedure is exactly
the same as described above, only the number of fit-parameters increases
at each step to account for all stars in the system.

\begin{table*}
\centering
\caption{Properties of the sample stars and their atmospheric parameters.}
\label{table:results}
\small
\setlength{\tabcolsep}{1.2mm}
\begin{tabular}{llllccccccc}
\hline\hline
Object & Sp. Type & Instrument & S/N & $T_\mathrm{eff}$ & $\log (g)$ & $y$
   & \multicolumn{1}{c}{$\xi$} & $\varv \sin i$ & \multicolumn{1}{c}{$\zeta$}
   & Ref. \\
\cline{8-10}
 & & & & K & (cgs) & by number &\multicolumn{3}{c}{km\,s$^{-1}$} & \\
\hline
\multicolumn{4}{l}{benchmark stars:}\\[1mm]
\object{HD 214680} (10 Lac) & O9\,V & FOCES & 500 & 34550$\pm$300 & 4.04$\pm$0.05
           & 0.083$\pm$0.009 & 5$\pm$2 & 14$\pm$1 & 32$\pm$2 & 1 \\
           & & & & 34305$\pm$250 & 4.03$\pm$0.04 & 0.092$\pm$0.010 & 6.1$\pm$0.5
           & 11.3$\pm$0.7 & 34.5$\pm$0.4 \\[1mm]
\object{HD 35299}  & B1.5\,V & FOCES & 350 & 23500$\pm$300 & 4.20$\pm$0.05 & 0.089 & 0$\pm$1 & 8$\pm$1
           & ... & 2\\
           & & & &  23391$\pm$521 & 4.19$\pm$0.06 & 0.092$\pm$0.002 & 0.5$\pm$0.6
           & 1.4$\pm$0.8 & 8.3$\pm$0.2 \\[1mm]
\object{HD 91316} ($\rho$ Leo) & B1\,Iab & ESPaDOnS & 750 & 21700$\pm$200 & 2.87$\pm$0.04
           & 0.092$\pm$0.009 & 15$\pm$2 & 43$\pm$5 & 62$\pm$5 & 3 \\
           & & & &  21270$\pm$201 & 2.85$\pm$0.02 & 0.091$\pm$0.008
           & 15.2$\pm$0.5 & 42.4$\pm$0.5 & 63.8$\pm$0.5 \\[1mm]
\object{HD 160762} ($\iota$ Her)& B3\,IV & FIES & 500 & 17500$\pm$200 & 3.85$\pm$0.05 & 0.089
           & 1$\pm$1 & 6$\pm$1 & ... & 2 \\
           & & & &  17759$\pm$140 & 3.87$\pm$0.03 & 0.086$\pm$0.003
           & 1.1$\pm$0.8 & 0.8$\pm$0.6 & 8.1$\pm$0.2 \\[1mm]
\object{HD 164353} (67 Oph) & B5\,Ib & FEROS & 500 & 14700$\pm$300 & 2.57$\pm$0.05 & 0.091$\pm$0.004
           & 8$\pm$2 & 20$\pm$4 & 32$\pm$5 & 4 \\
           & & &  & 14745$\pm$116 & 2.58$\pm$0.02 & 0.094$\pm$0.007 & 8.3$\pm$1.4
           & 16.0$\pm$1.0 & 33.8$\pm$0.8 \\[1mm]
\object{HD 259135} (V578 Mon)   &  DEB  & ESPaDOnS & 550\\
HD 259135\,A & B1\,V &          &  & 30000$\pm$500 & 4.13$\pm$0.02 & ... & 4$\pm$1 & 117$\pm$4 & ... & 5, 6 \\
& & & & 30117$\pm$457 & 4.12$\pm$0.07 &  &  &  &  \\
& & & & 29829$\pm$243 & 4.13$\pm$0.02 & 0.094$\pm$0.009 & 2.5$\pm$0.2 & 114.8$\pm$0.6 & 16.7$\pm$3.9 \\[1mm]
HD 259135\,B & B2\,V & & & 25750$\pm$435 & 4.19$\pm$0.02 & ... & 2$\pm$1 & 94$\pm$2 & ... & \\
& & & & 25046$\pm$988 & 4.21$\pm$0.15 &  & &  &  \\
& & & & 25587$\pm$213 & 4.19$\pm$0.02 & 0.090$\pm$0.009 & 1.1$\pm$0.8 & 89.6$\pm$2.5 & 57.4$\pm$5.8 \\[1mm]
\multicolumn{4}{l}{new applications:}\\[1mm]
\object{HD 149757} ($\zeta$ Oph) & O9.2\,IVnn & ESPaDOnS & 800 &
 31238$\pm$753 & 3.72$\pm$0.10\tablefootmark{a} & 0.131$\pm$0.011 & 7.7$\pm$3.4 & 382$\pm$4 & ... & \\[1mm]
\object{HD 93827} & B2\,Ib/II & FEROS & 220 & 23590$\pm$950 & 3.44$\pm$0.09\tablefootmark{b}
           & 0.101$\pm$0.009 & 14.2$\pm$3.5 & 241$\pm$1 & 14$\pm$7 \\[1mm]
\object{HD 112092} ($\mu^1$ Cru) & B2\,IV & FEROS & 500 & 20508$\pm$245 & 4.25$\pm$0.05
           & 0.090$\pm$0.006 & 1.8$\pm$1.2 & 31.3$\pm$0.2 & 13.3$\pm$0.6 \\[1mm]
\object{HD 87015} (EO Leo)& B2\,V & ESPaDOnS & 480 & 17782$\pm$365 & 3.92$\pm$0.05
           & 0.101$\pm$0.008 & 2.7$\pm$1.1 & 214$\pm$1 & 45$\pm$4 \\[1mm]
\object{HD 17081} ($\pi$ Cet) & B8\,IV & ESPaDOnS & 600 & 12800$\pm$200 & 3.75$\pm$0.10
           & 0.107$\pm$0.010 & 1.0$\pm$0.5 & 20.2$\pm$0.9 & ... & 7 \\
           & & & & 12819$\pm$76 & 3.83$\pm$0.03 & 0.124$\pm$0.005 & 1.6$\pm$0.1
           & 19.9$\pm$0.1 & 0.4$\pm$0.3 & \\[1mm]

\object{HD 77464} (CV Vel) & DEB & HARPS & 250\\
HD 77464\, A & B2.5\,V &   &  &  18100$\pm$500 & 4.00$\pm$0.01 & ... & ... & 21.5$\pm$2.0 & ... & 8 \\
         & & & & 18085$\pm$455 & 4.00$\pm$0.17 &  \\
         & & & & 18064$\pm$244 & 4.00$\pm$0.01 & 0.094$\pm$0.005 & 1.4$\pm$1.0 & 11.4$\pm$0.4 & 11.3$\pm$0.5 \\[1mm]
HD 77464\, B & B2.5\,V & & & 17900$\pm$500 & 4.02$\pm$0.01 & ... & ... & 21.1$\pm$2.0 & ... & \\
         & & & &  17758$\pm$486 & 4.01$\pm$0.16 &  & \\
         & & & & 17647$\pm$269 & 4.02$\pm$0.01 & 0.102$\pm$0.004 & 2.2$\pm$1.6 & 2.5$\pm$1.2 & 27.7$\pm$0.3\\
\hline\\[-5mm]
 \end{tabular}
 \tablefoot{For the benchmark stars the first row lists the literature values, the second row the values derived in this work. For the detached eclipsing binaries three solutions are given: previous literature results in the first row, the second row provides values derived from the single epoch spectrum adopted here alone, the third row gives our results using the $\log (g)$ values and the $T_\mathrm{eff}$ ratio (if known) from the binary solution based on the light curve and radial-velocity analysis as additional fit constraints. Uncertainties are 1$\sigma$ values.\\
 \tablefoottext{a}{$\log(g_{\rm true})$\,=\,3.91$\pm$0.07, see Sect.~\ref{sect:fundamental};}
 \tablefoottext{b}{$\log(g_{\rm true})$\,=\,3.55$\pm$0.07.}
 }
 \tablebib{(1) \cite{Aschenbrenneretal23}; (2) \cite{NiPr12}; (3) \cite{Wessmayeretal2023};
           (4) \cite{Wessmayeretal2022}; \\ (5) \cite{Garciaetal14}; (6) \cite{Pavlovskietal18}; (7) \cite{Fossatietal2009};
           (8) \cite{Albrechtetal2014}.}
 \end{table*}

\begin{table*}
\caption{Metal abundances $\varepsilon (X)$\,=\,$\log (X/\mathrm{H})+12$ (by number) of the sample stars.}
\label{tab:result_abundance}
\centering
\resizebox{\textwidth}{!}{
\setlength{\tabcolsep}{1.5mm}
\begin{tabular}{llllllllllc}
\hline\hline
Object & C & N & O & Ne & Mg & Al & Si & P\tablefootmark{a} & S & Fe \\ \hline
\\[-3mm]
\multicolumn{4}{l}{benchmark stars:}\\[1mm]
HD\,214680 & 8.36$\pm$0.07 & 8.11$\pm$0.11 & 8.77$\pm$0.06 & 8.24$\pm$0.07 & 7.59$\pm$0.03
           & 6.50$\pm$0.03 & 7.48$\pm$0.10 & 5.25 & ... & ... \\
           & 8.37$\pm$0.08 & 8.15$\pm$0.08 & 8.71$\pm$0.09 & 8.22$\pm$0.08 & 7.62$\pm$0.03
           & 6.52$\pm$0.04 & 7.46$\pm$0.10 & 5.25 & ... & ... \\[1mm]

HD\,35299  & 8.35$\pm$0.09 & 7.82$\pm$0.08 & 8.84$\pm$0.09 & 8.07$\pm$0.10 & 7.53$\pm$0.06
           & ...           & 7.56$\pm$0.05 & 5.18$\pm$0.06 & ...  & 7.53$\pm$0.10 \\
           & 8.35$\pm$0.12 & 7.81$\pm$0.07 & 8.85$\pm$0.09 & 8.10$\pm$0.07 & 7.52$\pm$0.05
           & 6.28$\pm$0.04 & 7.52$\pm$0.08 & 5.19$\pm$0.04 & 7.17$\pm$0.08 & 7.49$\pm$0.10 \\[1mm]

HD\,91316  & 7.94$\pm$0.08 & 8.35$\pm$0.08 & 8.44$\pm$0.08 & 8.06$\pm$0.06 & 7.50$\pm$0.29
           & 6.27$\pm$0.03 & 7.60$\pm$0.09 & ...           & 7.20$\pm$0.05 & 7.27$\pm$0.09 \\
           & 8.02$\pm$0.10 & 8.42$\pm$0.10 & 8.44$\pm$0.08 & 8.04$\pm$0.07 & 7.53$\pm$0.23
           & 6.30$\pm$0.02 & 7.60$\pm$0.13 & 5.11$\pm$0.07 & 7.19$\pm$0.09 & 7.26$\pm$0.09 \\[1mm]

HD\,160762 & 8.40$\pm$0.07 & 7.89$\pm$0.12 & 8.80$\pm$0.09 & 8.05$\pm$0.07 & 7.56$\pm$0.06
           & ...           & 7.51$\pm$0.05 & 5.40$\pm$0.13 & ...  & 7.51$\pm$0.08 \\
           & 8.52$\pm$0.04 & 7.85$\pm$0.06 & 8.81$\pm$0.04 & 8.12$\pm$0.04 & 7.53$\pm$0.02
           & 6.32$\pm$0.07 & 7.53$\pm$0.03 & 5.42$\pm$0.13 & 7.22$\pm$0.11 & 7.52$\pm$0.08 \\[1mm]

HD\,164353 & 8.31$\pm$0.04 & 8.37$\pm$0.06 & 8.81$\pm$0.06 & 8.05$\pm$0.06 & 7.51$\pm$0.05
           & 6.32$\pm$0.04 & 7.65$\pm$0.10 & 5.33$\pm$0.06 & 7.11$\pm$0.11 & 7.63$\pm$0.11 \\
           & 8.29$\pm$0.04 & 8.30$\pm$0.05 & 8.77$\pm$0.06 & 7.99$\pm$0.04 & 7.53$\pm$0.07
           & 6.33$\pm$0.05 & 7.61$\pm$0.07 & 5.30$\pm$0.05 & 7.12$\pm$0.09 & 7.68 $\pm$0.11\\[1mm]
\object{HD 259135}\,A & 8.18$\pm$0.07 & 7.69$\pm$0.12 & 8.74$\pm$0.10 & ... & 7.52$\pm$0.10
             & ... & 7.50$\pm$0.06 & ... & ... & ... \\
             & 8.26$\pm$0.11 & 7.65$\pm$0.14 & 8.68$\pm$0.09 & 8.13$\pm$0.06
             & 7.42 & ... & 7.46$\pm$0.05 & ... & ... & ... \\
\object{HD 259135}\,B & 8.21$\pm$0.11 & 7.72$\pm$0.09 & 8.76$\pm$0.11 & ... & 7.50$\pm$0.10
             & ... & 7.44$\pm$0.06 & ... & ... & ... \\
             & 8.38$\pm$0.08 & 7.82$\pm$0.14 & 8.79$\pm$0.10 & ... & 7.60
             & ... & 7.48$\pm$0.05 & ... & 7.16 & 7.58\\[1mm]
\multicolumn{4}{l}{new applications:}\\[1mm]
HD\,149757 & 8.21$\pm$0.09 & 8.31$\pm$0.06 & 8.72$\pm$0.08 & 8.09 & 7.72 & ... & 7.76$\pm$0.08 & ... & ... & ... \\[1mm]
HD\,93827 & 8.20$\pm$0.10 & 8.25$\pm$0.09 & 8.62$\pm$0.09 & ...
             & 7.52 & 6.33$\pm$0.02 & $7.63\pm$0.08 & ... & 7.18 & 7.51$\pm$0.14 \\[1mm]
HD\,112092 & 8.37$\pm$0.04 & 7.83$\pm$0.07 & 8.78$\pm$0.08 & 8.10$\pm$0.07 & 7.55$\pm$0.02
             & 6.31$\pm$0.05 & 7.48$\pm$0.05 & 5.41$\pm$0.06 & 7.29$\pm$0.09 & 7.52$\pm$0.12 \\[1mm]
HD\,87015 & 8.49$\pm$0.10 & 8.07$\pm$0.13 & 8.77$\pm$0.08 & 7.98$\pm$0.09 & 7.40$\pm$0.06
             & 6.49$\pm$0.05 & 7.60$\pm$0.07 & ... & 7.14$\pm$0.08 & 7.51$\pm$0.14 \\[1mm]
HD\,17081 & 8.48$\pm$0.04 & 8.07$\pm$0.07 & 8.69$\pm$0.06 & 8.10$\pm$0.05 & 7.53$\pm$0.04
             & 6.31$\pm$0.01 & 7.67$\pm$0.10 & 5.70$\pm$0.03 & 7.31$\pm$0.08 & 7.70$\pm$0.08 \\[1mm]
HD\,77464\,A & 8.43$\pm$0.05 & 8.12$\pm$0.05 & 8.86$\pm$0.06 & 8.08$\pm$0.04 & 7.54$\pm$0.04
             & 6.35$\pm$0.06 & 7.61$\pm$0.05 & ... & 7.17$\pm$0.07 & 7.54$\pm$0.07 \\
HD\,77464\,B & 8.44$\pm$0.03 & 8.20$\pm$0.07 & 8.86$\pm$0.08 & 8.05$\pm$0.06 & 7.52$\pm$0.07
             & 6.38$\pm$0.08 & 7.56$\pm$0.09 & ... & 7.16$\pm$0.08 & 7.55$\pm$0.08 \\
\hline\\[-3mm]
CAS\tablefootmark{b,c,d}  & 8.35$\pm$0.04  & 7.79$\pm$0.04  & 8.76$\pm$0.05 & 8.09$\pm$0.05 & 7.56$\pm$0.05
                          & 6.30$\pm$0.07 & 7.50$\pm$0.06 & 5.36$\pm$0.14 & 7.14$\pm$0.06 & 7.52$\pm$0.03 \\
\hline
\end{tabular}
}
\tablefoot{For the benchmark stars the first row lists the literature values (references in Table \ref{table:results}), the second row the values derived in this work. Uncertainties are 1$\sigma$ values from the line-to-line scatter. If only one line was fitted, no uncertainty is provided.
\tablefoottext{a}{Phosphorus literature abundance from \citet{Aschenbrenneretal25}}
\tablefoottext{b}{\citet{NiPr12}}
\tablefoottext{c}{\citet{Przybillaetal13}}
\tablefoottext{d}{\citet{Aschenbrenneretal25}}
}
\end{table*}

\subsection{Limitations}\label{sect:limitations}
Although we cover a wide range of stellar parameters of OB stars with the present implementation of {\sc Saturn}, there are still limitations to the analysis method within the covered area in Fig.~\ref{fig:Kiel_diagram}. A large object class not suitable for analysis is the chemically peculiar stars. In our B$_1$ region lie the magnetic He-strong stars, which show helium number fractions of typically 0.2 to 0.3 \citep[e.g.][]{Zboriletal97}, while a few more extreme examples exist, with $y$\,=\,0.60 or even 0.95 \citep{Gonzalezetal19,Przybillaetal21}, which lead to markedly different atmospheric structures not covered here. At cooler temperatures towards and within the B$_2$ region, the He-weak, HgMn, and Bp stars \citep[e.g.][]{Smith96} are present. At lower rotational velocities ($\varv_\mathrm{rot}$\,<120\,km\,s$^{-1}$) atomic diffusion \citep{Michaud70} occurs in the then very stable atmospheres and leads to abundances outside the covered intervals, and in particular to very high abundances of otherwise unobserved (heavy) chemical species at optical wavelengths. Diffusion may even lead to accumulation of the $^3$He isotope, resulting in unusual profiles for some of the helium lines \citep[e.g.][]{Mazaetal14}, which we do not consider here due to its rarity of occurrence. On the other hand, the OBN and OBC stars \citep{Walborn76} within the parameter boundaries that are a consequence of mixing with nuclear-processed material should be covered by the emulated spectra. There may be very few exceptions, such as the enveloped-stripped B-type subgiant \object{HD 40494} \citep[$\gamma$~Col,][]{Irrgangetal22}, which has extreme abundances of carbon and nitrogen beyond the values considered here. Other rare objects, such as runaway stars from a supernova (SN) explosion in a tight binary system that were polluted by SN debris \citep[e.g. \object{HIP 60350},][]{Irrgangetal10}, can be analysed.

A second wide object class that is beyond the capabilities of analysis with {\sc Saturn} is that of fast-rotating classical Be stars \citep[e.g.][]{Riviniusetal13}. The second light introduced by the Keplerian decretion disc surrounding the star also requires to be modelled appropriately \citep[e.g.][]{Sigutetal09}. Limitations may also occur for stars without discs that rotate fast enough to lead to significant deviations from spherical geometry and gravity darkening. According to the study of \citet{Frematetal05}, these effects should be insignificant for angular velocity values relative to the critical one $\Omega/\Omega_\mathrm{c}$\,$<$\,0.6, and then gradually become more and more significant. We remind the reader that critical rotational velocities are about 700, 600, and 500\,km\,s$^{-1}$ for 20, 10, and 5\,$M_\odot$ stars on the ZAMS and about 430, 400, and 350\,km\,s$^{-1}$ on the terminal-age main sequence (TAMS).

In multiple-star systems, the composite spectrum can be well modelled as long as the stars are not interacting. Weak binary interactions such as small tidal deformations or small mutual irradiation effects can be accounted for by adjusting the $w_i$ factors in Eq.~\eqref{eq:composite_spectrum}, but in the case of, for example, (over)contact systems, our model spectra computed for single stars are no longer valid.

\section{Results}\label{section:results}
To test and validate the usage of our trained neural networks for future analysis of stellar spectra, we analysed five apparently slowly rotating OB-type stars previously studied on the basis of {\sc Ads} models in the literature\footnote{Each of the benchmark stars has been analysed multiple times before. While there is usually general agreement of results obtained with modern codes \citep[see, e.g.][]{Wessmayeretal2022,Aschenbrenneretal23}, there exist many differences in assumptions, methods and data used for the calculations, such that we refrain from detailed comparisons with all available results in the literature.}, as well as the spectrum of a detached eclipsing binary (DEB) system. The targets cover all five modelling regions shown in Fig.~\ref{fig:Kiel_diagram}. In addition, new applications showcase the capabilities of {\sc Saturn}.

The results for the stars analysed are summarised in Tables~\ref{table:results} and
\ref{tab:result_abundance}. The former lists the stellar parameters; the name of the object,
spectral type, instrument used for observation, signal-to-noise ratio of the spectrum, effective temperature,
surface gravity, helium abundance by number, microturbulence velocity, projected rotational velocity,
macroturbulence velocity, and a reference for literature values.
The latter table lists derived element abundances for up to ten metals.

Overall, we find excellent agreement with the values in the literature for atmospheric parameters
as well as element abundances. The small deviations can result from the improved atomic data
used for our model calculations or from the usage of different spectra for the analysis.
In the following, we provide additional notes for each object analysed, sorted in decreasing order of $T_\mathrm{eff}$.

\subsection{Benchmark star test sample}
\paragraph{HD\,214680 (10~Lac).}
This late O-type standard star is constant in radial velocity, and a $\sim$10\,mag (in the $I$ band) fainter star at $\sim$3\farcs6 distance \citep{Turneretal08} is irrelevant for spectroscopy. Therefore, 10~Lac can be viewed as a single star; for a detailed discussion of the star, see \citet{Aschenbrenneretal23}.
The derived atmospheric parameters and element abundances for the Morgan-Keenan
standard star 10 Lac are in excellent agreement with the literature values.
As an indicator for $\teff$, we used the simultaneous ionisation equilibria of \ion{He}{i/ii},
\ion{O}{ii/iii}, and \ion{Si}{iii/iv}. The noticeable higher helium abundance
derived in this work can be explained by using a different version
of {\sc Detail} to calculate the helium occupation numbers.
\cite{Aschenbrenneretal23} used opacity averaging with opacity distribution functions
to avoid the \ion{He}{i} singlet problem \citep{Najarroetal06}, which results from an overlap of the \ion{He}{i} 584\,{\AA} resonance line with two \ion{Fe}{iv} transitions. In order to keep all of the model
calculations consistent and to employ opacity sampling for all calculations, we use the
workaround by \cite{Pulsetal20} and set the oscillator strengths of the involved \ion{Fe}{iv}
transitions to $\log (gf)$\,=\,$-$5.\\[-9mm]
\paragraph{HD\,35299.}
To the best of our knowledge, HD\,35299 is a single star with constant radial velocity \citep{MoLe91,Chinietal12}. It was identified to show both $\beta$~Cephei and slowly pulsating B-star (SPB) variability by \citet{BaOz20}.
As $\teff$ indicator we used the simultaneous ionisation equilibria of \ion{O}{i/ii}, \ion{Ne}{i/ii}, and \ion{Si}{iii/iv}. For this star among our benchmark objects, intermediate-resolution spectra were also available. Tests of {\sc Saturn} at resolutions relevant for upcoming observations with WEAVE and 4MOST were therefore feasible. Consistent results on the stellar parameters are also obtained in that case; see Appendix~\ref{appendix:E} for details.\\[-9mm]
\paragraph{HD\,91316 (\texorpdfstring{$\rho$}{rho}~Leo).}
The star is a rare supergiant at high galactic latitude, which has reached its position as a runaway star from the galactic disc. Occultations by the moon have indicated a binary nature, with the companion resolved later by speckle interferometry, being located at a distance of 46.1\,mas, and showing a magnitude difference in $V$ of 1.5\,mag \citep{Tokovininetal10}; for a more detailed discussion of the star, see \citet{Wessmayeretal2023}. In order to compare with the literature values, we treated the spectrum as that of a supposedly single star. For the analysis, we used the ionisation equilibria of \ion{N}{ii/iii}, \ion{O}{i/ii}, \ion{Ne}{i/ii}, and \ion{Si}{iii/iv}. We obtained a slightly lower effective temperature compared to the previous literature value, however, all the derived parameters are still compatible.\\[-9mm]
\paragraph{HD\,160762 (\texorpdfstring{$\iota$}{iota}~Her).}
The $\beta$~Cep pulsator $\iota$~Her is a single-lined spectroscopic binary star (SB1), with a secondary of unknown spectral type in an eccentric ($e$\,=\,0.54$\pm$0.03) 111.5$\pm$0.1\,d orbit \citep{Baetzetal26}. The companion does not contribute to the spectrum and -- because $\iota$~Her does not show a CNO mixing signature -- is likely of significantly later spectral type than the primary, and not a white dwarf (which in view of the small minimum semi-major axis of 12.68$\pm$0.74\,$R_\odot$ would otherwise imply a previous mass exchange).  For deriving $\teff$, we used simultaneously the ionisation equilibria of
\ion{O}{i/ii}, \ion{Al}{ii/iii}, \ion{Si}{ii/iii}, and \ion{Fe}{ii/iii}. All the derived parameters and abundances agree excellently with the literature values except for the carbon abundance, where we derive a
$\sim$0.1\,dex higher value. The model atom for \ion{C}{ii} saw some extension with regard to the inclusion of additional energy levels in the quartet spin system, enabling the use of additional spectral lines for analysis but also leading to slightly different level populations overall. We also use a higher-quality spectrum for the analysis; both together likely result in the small shift in abundance. \\[-9mm]
\paragraph{HD\,164353 (67~Oph).}
This supergiant is a visual multiple star, with the closest companion being about 7\arcsec\ away and about 10\,mag fainter \citep{Masonetal01}. Therefore, it is effectively single for the spectroscopic analysis. We used ionisation equilibria of \ion{O}{i/ii}, \ion{Al}{ii/iii}, and \ion{Fe}{ii/iii} for the determination of $T_\mathrm{eff}$.\\[-9mm]
\paragraph{HD\,259135 (V578~Mon).}
This double-lined eclipsing binary is located in the open cluster \object{NGC 2244}, near the centre of the Rosette Nebula. The spectrum used for analysis was observed after the secondary eclipse, where the stars have a radial velocity separation of $\sim$\SI{200}{\kilo\meter\per\second}. We provide two solutions for each of the individual stars, using ionisation equilibria of \ion{He}{i/ii} and \ion{Si}{iii/iv} for the primary and \ion{O}{i/ii} for the secondary to determine $T_\mathrm{eff}$. The first solution is derived on the basis of a single epoch spectrum and we reproduce the same atmospheric parameters as previous works. The much higher uncertainties for the parameters of the secondary star are a consequence of its smaller contribution to the total flux.
For the second solution, we use the known surface gravities and the effective
temperature ratio of $T_2/T_1$\,=\,0.858$\pm$0.002 \citep{Garciaetal14} as fit constraints,
to derive the effective temperatures with lower uncertainties. We adopted a weight factor $w$\,=\,0.54$\pm$0.04 for the analysis (Eq.~\eqref{eq:composite_spectrum}). This is less than the expected value from pure geometric considerations based on the relative surface areas implied by the stellar radii ($w_\mathrm{geo}$\,=\,0.629$\pm$0.017).
However, deviations are expected due to mutual irradiation effects in this close binary system.
The resulting model fit to the observed spectrum is shown in Appendix~\ref{appendix:D}, Fig.~\ref{fig:V578fit}.

As a further test, we also calculated the spectroscopic distance $d_{\text{spec}}$\,=\,1468$\pm$83\,pc to V758~Mon using Eq.~(4) of \cite{Aschenbrenneretal24}. We corrected the measured apparent magnitude of $m_V$\,=\,8.541$\pm$0.007 \citep{Mermilliod97} for interstellar extinction by a value of $A_V$\,$\approx$\,1.4\,mag, found by fitting the SED, see Appendix~\ref{appendix:C}. For the radii, we adopted $R_1$\,=\,5.41$\pm$0.04\,$R_{\odot}$ and $R_2$\,=\,4.29$\pm$0.05\,$R_{\odot}$ \citep{Pavlovskietal18}. The spectroscopic distance is in excellent agreement with the cluster distance $d_{\text{NGC\,2244}}$\,=\,1464$^{+94}_{-84}$\,pc as derived by \citet{Aschenbrenneretal23} from considering {\it Gaia} Early Data Release 3 \citep[EDR3,][]{Gaia2016,Gaia2020} parallaxes for its massive star content. For comparison, the {\it Gaia} parallax based photogeometric distance to V578~Mon is $d$\,=\,1509$^{+102}_{-103}$\,pc \citep{Bailer-Jones_etal_2021} -- which is also in good agreement -- with a renormalised unit weight error (RUWE) of 1.3, hinting at the observed binarity.

\subsection{Applications}\label{sect:applications}
\paragraph{HD~149757 (\texorpdfstring{$\zeta$}{zet}~Oph).}
This runaway star is the closest O-type star
and shows a prominent bow-shock in the IR. It is a very fast-rotating single \citep{Hutteretal2021} star with broad spectral lines that show strong variability due to pulsations \citep[e.g.][]{Walkeretal2005,Kalitaetal2025}. Since the line broadening is dominated by the rotation of the star, we set $\zeta$\,=\,0\,km\,s$^{-1}$ for the analysis. In contrast to the other stars in the analysed sample, we determined $\varv \sin i$ by fitting multiple \ion{He}{i} lines and taking the mean and standard deviation of the individual best fitting velocities. To fit the atmospheric parameters, we used the ionisation equilibria of \ion{He}{i/ii} and \ion{O}{ii/iii} as temperature indicators and the wings of the Balmer lines to determine surface gravity. We note that the derived $\log(g)$ value in Table~\ref{table:results} is not corrected for centrifugal acceleration, the 'true' (polar) value is significantly larger at such a high rotation rate (see the footnote in Table~\ref{table:results}). We want to state the atmospheric parameters determined recently by \citet{deBurgosetal24}, $T_\mathrm{eff}$\,=\,30700$\pm$600\,K, $\log (g)$\,=\,3.30$\pm$0.09, $y$\,=\,0.130$\pm$0.030, and $\varv \sin i$\,=\,410\,km\,s$^{-1}$, which show reasonable agreement with our values; there are many additional studies available in the literature \citep[see e.g.][their Table~2]{Gordonetal18}. The star shows a CNO-mixing signature and enhanced helium, as expected for a fast rotator, but see Sect.~\ref{sect:fundamental}. The strong blending effects due to high rotation preclude the determination of the abundances of several elements, but magnesium and silicon are higher than CAS values. Certainly, the star shows pronounced deformation from spherical geometry and gravity darkening over its surface, such that our approach of an 'average' atmosphere can be expected to introduce an additional systematic error, see the discussion in Sect.~\ref{sect:limitations}. However, compared to typical uncertainties of abundances in massive stars to the extent of $\sim$0.3\,dex as discussed in the literature, these are still reasonable results if one assumes that they should be close to CAS values like for all OB stars in the solar neighbourhood investigated by us. The model fit to the observed spectrum is shown in Fig.~\ref{fig:HD149757fit}.\\[-9mm]
\paragraph{HD~93827.} This star of luminosity class between bright giant and supergiant is apparently single, located in the Carina spiral feature; it is the faintest and most distant of the stars investigated here. It has been studied in some aspects by \citet{Fraser_etal_10}, they adopted $T_\mathrm{eff}$\,=\,18500\,K from a spectral type calibration and $\log(g)$\,=\,2.70 from fitting H$\gamma$ and H$\delta$, and the very high $\varv \sin i$ for such an evolved object attracted our interest. We performed here the first comprehensive analysis of the stellar properties. We confirm the high rotational velocity, but its position in the Kiel diagram (Fig.~\ref{fig:Kiel_diagram}) locates it close to the TAMS, putting it near the end of core hydrogen burning, still not being subject to the rapid envelope extension (and spin-down) thereafter. The star also shows mixing of the surface layers with CNO-cycled material from the core, as expected at fast rotation, but otherwise abundances of the heavier elements are close to CAS values. The model fit to the observation is shown in Fig.~\ref{fig:HD93827fit}.\\[-9mm]
\paragraph{HD~112092 (\texorpdfstring{$\mu^1$}{mu1}~Cru).}
This subgiant was recently identified to show SPB pulsations \citep{BaOz20}, explaining the presence of some bumps in the observed spectral lines. It is constant in radial velocity \citep{Chinietal12}, apparently single, and its visual companion $\mu^2$~Cru is too distant to affect spectroscopy. We provide the first non-LTE analysis of this star since \citet{Kilian94}. All abundances are compatible with the CAS values, with sulphur showing a tendency to higher values. \\[-9mm]
\paragraph{HD~87015 (EO Leo).} The star is a fast rotator and apparently a single one. We present here the first comprehensive quantitative spectroscopic investigation; before, atmospheric parameters were determined by \citet[$T_\mathrm{eff}$\,=\,17900$\pm$300\,K, $\log(g)$\,=\,3.86$\pm$0.09]{Lyubimkovetal02} and \citet[$T_\mathrm{eff}$\,=\,16435$\pm$150\,K, $\log(g)$\,=\,3.694$\pm$0.020]{Huangetal10}, which are in agreement with our values in case of the former, and in disagreement in the case of the latter work (where both $T_\mathrm{eff}$ and $\log (g)$ were derived from fitting only one spectral line, H$\gamma$). Abundances roughly align with CAS values, the nitrogen abundance is elevated as expected for a fast rotator, but at high carbon abundance. Magnesium is slightly underabundant, and aluminium is slightly overabundant, which may be a selection effect due to the inaccessibility of some lines at high rotational velocity. In general, the star is nearly a clone of $\iota$~Her, but at high $\varv \sin i$.  A visual inspection of light curves observed with the Transiting Exoplanet Survey Satellite implies $\beta$~Cep pulsations, another property shared with $\iota$~Her.\\[-9mm]
\paragraph{HD~17081 (\texorpdfstring{$\pi$}{pi}~Cet).}
This is an SB1 star with a nearly circular orbit and a period of 7.45\,yr, the secondary probably being a white dwarf \citep{Lacyetal97}; for analysis purposes, it can be treated as single. It has been intensely studied in LTE \citep[e.g.][]{Adelmanetal02,Fossatietal2009} because it is a rare case of an apparently slowly rotating late B-type star that is not notably chemically peculiar. It has also been subject to a hybrid non-LTE abundance analysis \citep{Mashonkinaetal20}, based on the atmospheric parameters of \citet{Fossatietal2009}. As \citet{Mashonkinaetal20} used a different model atmosphere code, a different variant of the {\sc Detail} code, their own model atom implementations, and a different line-formation code with their own selection of line-formation parameters, we do not adopt this solution as a benchmark case. However, we note that their abundances (O, Ne, Na, Mg, Ca, Ti, Sr, and preliminary values of Fe) are close to the solar ones, with slightly enhanced carbon and silicon. For the latter two, we find the same and also enhanced abundances of N, P, S, and Fe, whereas the other elements are compatible with the CAS values. Our abundances, together with the enhanced helium, imply the possible occurrence of some mass transfer during the past asymptotic giant branch phase of the companion, and a small amount of atomic diffusion having taken place in $\pi$~Ceti's atmosphere since then. Finally, we want to note that the $T_\mathrm{eff}$ and $\log (g)$ determination shows some degeneracy with respect to the helium abundance, with some ionisation equilibria pointing towards a $\sim$300\,K higher effective temperature that would lead to a very similar quality of the overall spectral fit. Reassuringly, the present solution reproduces several photospheric non-LTE emission lines, such as the \ion{C}{i} 8335 and 9406\,{\AA} singlet lines.\\[-9mm]
\paragraph{HD~77464 (CV~Vel).}
This DEB system hosts two very similar B-type stars. Both are slowly rotating and show strong metal lines, separated by $\sim$135\,\si{\kilo\meter\per\second} in the spectrum used (see Fig. \ref{fig:HD77464fit}). Only the broad hydrogen and helium lines overlap and must be fitted simultaneously.
From our best global fit we obtained a weight of $w$\,=\,1.00$\pm$0.05, compatible with the ratio of surface areas calculated from the radii determined by \citet{Albrechtetal2014}, $w$\,=\,0.93$\pm$0.02. Similar to the case of V578~Mon, the solution based on the single-epoch spectrum alone leads to increased uncertainties, in particular for $\log(g)$, while the additional constraints from the earlier radial velocity and light curve solution \citep{Albrechtetal2014} result in very tight uncertainties. We note that our consideration of both, rotational and macroturbulent velocities, for the line broadening leads to significantly different $\varv \sin i$ values than those of \citet{Albrechtetal2014}. Another factor in this context may also be precession of the rotational axes due to their misalignment as reported by the same authors.
The chemical composition of both stars is identical and agrees with CAS values for most elements. However, we find slight elevated abundances of carbon and oxygen, while nitrogen is enriched by $>$0.3\,dex in both stars. A previous abundance analysis of the system's components by \citet[assuming $T_\mathrm{eff}$\,=\,18000\,K for both]{Yakutetal07} yielded overall systematically lower abundances, though still compatible with our values within their typical uncertainties of 0.2 to 0.6\,dex.

We also characterised the extinction towards HD\,77464, by fitting the SED using the reddening law of \citet{Fitzpatrick99}, where we set the ratio of total-to-selective extinction, $R_V$\,=\,$A_V/E(B-V)$\,=\,3.1 and obtained a colour excess of $E(B-V)$\,=\,0.034$\pm$0.010 from a least-square fit. The fitted SED is shown in Fig.~\ref{fig:SED_fits}. After correcting the measured magnitude of $m_V$\,=\,6.706$\pm$0.008 \citep{Mermilliod97} for extinction and by adopting the radii $R_1$\,=\,4.08$\pm$0.03\,$R_{\odot}$ and $R_2$\,=\,3.94$\pm$0.03\,$R_{\odot}$ \citep{Albrechtetal2014}, we calculated $d_{\text{spec}}$\,=\,598$\pm$9\,pc to the system. The \textit{Gaia} photogeometric distance is $d$\,=\,565$^{+10}_{-13}$\,pc \citep{Bailer-Jones_etal_2021}, with a RUWE of 0.84.

As a test of our derived parameters, we made sure that the position of our sample objects in the Kiel diagram ($\log(g)$ vs. $\log T_{\rm eff}$), the Hertzsprung-Russell diagram (HRD, $\log L/L_{\odot}$ vs. $\log T_{\rm eff}$) and the spectroscopic HRD \citep[sHRD, $\log(\mathscr{L}/\mathscr{L}_{\odot})$ vs. $\log T_\mathrm{eff}$, introduced by][with $\mathscr{L}$\,=\,$T_\mathrm{eff}^4/g$]{LaKu14} is similar relative to evolutionary tracks in all three diagrams. Finally, we also verified that the positions of the stars in the N/C versus N/O diagram \citep{Przybillaetal10,Maederetal14} are within the expected path for CNO mixing.

\begin{table*}
\caption{Fundamental stellar parameters.}
\label{tab:result_fundamental}
\small
\centering
\setlength{\tabcolsep}{1.5mm}
\begin{tabular}{lrrrrrrrrll}
\hline\hline
Object & $V$\tablefootmark{a} & $E(B-V)$\,(mag) & $R_V$ & $B.C.$\,(mag) & $M/M_\odot$ & $R/R_\odot$ & $\log L/L_\odot$ & $\log \tau_\mathrm{evol}$\,(yr) & $d_\mathrm{spec}$\,(pc) & $d_{Gaia}$\,(pc)\tablefootmark{b} \\ \hline
\\[-3mm]
HD\,149757 & 2.567$\pm$0.005 & 0.33$\pm$0.02 & 3.0$\pm$0.1 & $-$2.91$\pm$0.06 & 18.7$\pm$0.7 & 7.6$\pm$0.6 & 4.69$\pm$0.08 & $>$6.80$^{+0.04}_{-0.04}$ & 135$^{+11}_{-11}$ & 135$^{+12}_{-9}$ \\[0.5mm]
HD\,93827  & 9.320$\pm$0.010 & 0.32$\pm$0.02 & 3.1$\pm$0.1 & $-$2.24$\pm$0.10 & 12.9$\pm$0.8 & 11.2$\pm$0.8 & 4.54$\pm$0.09 & 7.20$^{+0.06}_{-0.05}$ & 3461$^{+266}_{-264}$ & 3484$^{+231}_{-194}$\\[0.5mm]
HD\,112092 & 4.032$\pm$0.014 & 0.04$\pm$0.01 & 3.1$\pm$0.1 & $-$2.00$\pm$0.03 & 7.0$\pm$0.1 & 3.3$\pm$0.2 & 3.24$\pm$0.04 & 6.98$^{+0.20}_{-0.35}$ & 113$^{+7}_{-6}$ & 124$^{+5}_{-3}$\\[0.5mm]
HD\,87015  & 5.655$\pm$0.041 & 0.01$\pm$0.01 & 3.1$\pm$0.1 & $-$1.62$\pm$0.05 & 6.3$\pm$0.2 & 4.6$\pm$0.3 & 3.28$\pm$0.05 & 7.68$^{+0.03}_{-0.04}$ & 310$^{+20}_{-20}$ & 308$^{+8}_{-8}$\\[0.5mm]
HD\,17081  & 4.241$\pm$0.006 & 0.00$\pm$0.01 & 3.1$\pm$0.1 & $-$0.80$\pm$0.03 & 4.0$\pm$0.1 & 4.0$\pm$0.2 & 2.60$\pm$0.03 & 8.19$^{+0.01}_{-0.01}$ & 109$^{+5}_{-4}$ & 119$^{+2}_{-3}$\\[0.5mm]
\hline
\end{tabular}
\tablefoot{Uncertainties are 1$\sigma$ values.
\tablefoottext{a}{\citet{Mermilliod97}}
\tablefoottext{b}{\citet{Bailer-Jones_etal_2021}}
}
\end{table*}

\subsection{Fundamental stellar parameters}\label{sect:fundamental}
In order to complete the analysis of the new targets, their fundamental stellar parameters are summarised in Table~\ref{tab:result_fundamental}. The entries are the Johnson $V$ magnitude, the colour excess $E(B-V)$, the ratio of total-to-selective extinction $R_V$, bolometric correction $B.C.$, mass $M$, radius $R$, luminosity $L$, evolutionary age $\tau_\mathrm{evol}$, spectroscopic distance $d_\mathrm{spec}$, and Gaia EDR3 distance $d_{Gaia}$.
We determined $E(B-V)$ and $R_V$ by fitting the SEDs of the individual objects, see Appendix \ref{appendix:C} for details.
In the case of HD\,112092, HD\,87015, and HD\,17081 we determined the mass by comparing the position of the stars in the sHRD with evolutionary tracks by \cite{Ekstroemetal12} and then calculated the additional parameters. In the case of the fast-rotating objects HD\,149757 and HD\,93827 we first calculated their radius from the measured parallax distance $d_{Gaia}$ and the $V$ magnitude corrected for extinction
\begin{equation}\label{eq:radius}
    \log(R/R_{\odot}) \approx 7.398-0.2\cdot (V-R_V\cdot E(B-V) - m_{\rm syn}) + \log(d/\textrm{pc}),
\end{equation}
where $m_{\rm syn}=-2.5\int F(\lambda)\cdot T(\lambda)\,{\rm d}\lambda\,+\,zp$ is a synthetic magnitude calculated from the theoretical astrophysical flux $F(\lambda)$, the filter response function $T(\lambda)$, and the zero-point $zp$ of the filter.
Then we corrected our derived $\log(g)$ values for centrifugal acceleration using the equation by \cite{RePuHe04}
\begin{equation}
    g_{\rm true} \approx g + \frac{\left(\varv \sin i\right)^2}{R}
\end{equation}
and determined the mass in the sHRD with the corrected surface gravity. For HD\,149757 we obtain $\log(g_{\rm true})$\,=\,3.91$\pm$0.07 and for HD\,93827 $\log(g_{\rm true})$\,=\,3.55$\pm$0.07 (the centrifugal acceleration is negligible for the other sample stars). The exceptional case of HD\,149757 is discussed in more detail further below. We note that the derived spectroscopic distance for both objects is not an independent distance measurement since we used the radius calculated from the parallax distance.

As our analysis results for the benchmark stars agree with those derived earlier in the literature, we refer for the discussion of their fundamental stellar parameters to \citet{NiPr14}, \citet{Wessmayeretal2022,Wessmayeretal2023} and \citet{Aschenbrenneretal23}. The combined radial-velocity and light curve analysis feasible for DEBs allows highly accurate and precise fundamental parameters to be derived; we therefore refer to the studies of \citet{Garciaetal14} for the data on V578~Mon and \citet{Albrechtetal2014} for CV~Vel.\\[-9mm]
\paragraph{HD\,149757~($\zeta$~Oph).}
The very high rotational speed of at least $\sim$90\% of the critical velocity makes $\zeta$~Oph one of the fastest rotating OB stars in the Milky Way known at present. Together with its runaway nature, it appears to be the product of binary interaction \citep{ReGo21}, where the mass overflow led to the spin-up and the supernova explosion of its companion made it speed away from its birth place. A little extra discussion is therefore in place when deriving its fundamental parameters.

The distance of $\zeta$~Oph is the crucial factor.
Three reasonable parallax measurements give values of 140$^{+16}_{-12}$\,pc \citep[\textit{Hipparcos},][]{ESA97}, 112$^{+3}_{-2}$\,pc \citep[new reduction of the \textit{Hipparcos} data,][]{vanLeeuwen07a}, and 135$^{+12}_{-9}$\,pc \citep[\textit{Gaia} EDR3,][]{Bailer-Jones_etal_2021}. The first two measurements lead to an association of $\zeta$~Oph with two different pulsars as the products of a binary supernova explosion in the Upper Scorpius subgroup \citep{Hoogerwerfetal01} or the Upper Centaurus Lupus subgroup \citep{Neuhaeuseretal20} of the Sco~OB2 association. The short \textit{Hipparcos} distance would result in a $\log L/L_\odot$\,$\approx$\,4.5, comparable to the least luminous O9.7\,V star in the sample of \citet{Aschenbrenneretal23}; the luminosity at the \textit{Gaia} (and the long \textit{Hipparcos}) distance, on the other hand, is comparable to those of the O9.2\,IV stars in that work. We therefore adopted the \textit{Gaia} distance, despite a poor RUWE value of 4.489 \citep[the star is brighter than the original bright star limit of \textit{Gaia}, and we speculate that variations of the star's photocentre due to the strong non-radial pulsations may also be an issue, with the parallax error of 0.6596 \,mas being similar in size than the measured angular diameter of the star, 0.462 to 0.540\,mas,][]{Gordonetal18}. The radius follows from Eq.~\eqref{eq:radius}, which is in good agreement with the radius value derived from the angular diameter. Our mass determination has to be viewed with more care, as we employed tracks for single star evolution. Ideally, a tailored binary evolution model similar to that of \citet{ReGo21} would need to be computed, meeting our observational constraints. But the differences between single and binary star evolution at the position of $\zeta$~Oph in the HRD (their Fig.~1) are not large (they will become important during further evolution), such that systematic errors can be expected to remain small. However, the value for the evolutionary age from our approach is only a lower limit. Finally, we note that the observed helium and CNO abundances are a consequence of the accretion of nuclear-processed material from the former companion star and subsequent thermohaline mixing because of the inverted mean molecular weight gradient near the surface.

\section{Summary and outlook}\label{section:summary}
In this paper, we present the first results from using {\sc Saturn}, a programme we developed to fit spectra of single and multiple OB-type stars. We used neural networks to emulate model spectra for a wide range of atmospheric parameters and account for lines of hydrogen, helium, and ten metals. Broadening with instrumental, rotational, and radial-tangential macroturbulence profiles is handled separately.
The neural networks were trained on a dataset we calculated with a hybrid non-LTE approach based on the {\sc Atlas}, \mbox{\sc Detail}, and {\sc Surface} codes, and which cover a wavelength range from 3000\,{\AA} to 10500\,{\AA}.

We first tested {\sc Saturn} by analysing five, apparently slowly rotating, single stars. Previous literature data on atmospheric parameters and element abundances were reproduced. We also analysed a single epoch spectrum of the DEB HD\,259135 (V578~Mon), and reproduced values from the literature derived from independent measurements of the radial-velocity and light curves. Of particular importance is the ability to match the observed spectra very closely at very high S/N, both in terms of completeness of the involved line list and in the match of the strengths and shapes of the spectral features.

As further applications, we used {\sc Saturn} to analyse an additional DEB and five single stars, among others the rapidly rotating star HD\,149757 ($\zeta$~Oph), and determined their atmospheric parameters and their chemical composition with high precision. This demonstrates the strength of {\sc Saturn} to also analyse fast-rotating stars with relative ease, where blended spectral lines dominate the spectra, putting the aforementioned completeness of the line lists to a test.

For the analysis of the objects we used available high-quality, high-resolution spectra, at $R$\,$\ge$\,40\,000. This opens up the possibility of analysing similar data for OB-type stars, be it new observations taken with state-of-the-art instrumentation, or by taking advantage of the vast existing telescope archives. A huge incentive for the development of {\sc Saturn} was to have a tool at hand to efficiently analyse the coming flood of data from spectroscopic facilities such as WEAVE and 4MOST, which will provide spectra for tens of thousands of OB-type stars with resolutions from $R$\,$\approx$\,4000 to 20\,000. By analysing low-resolution XSHOOTER spectra of the benchmark star HD\,35299, we could also show the capability of {\sc Saturn} to perform at lower spectral resolution, reproducing atmospheric parameters and chemical abundances at similar quality. However, the number of accessible chemical species is reduced because of increased instrumental broadening and can be expected to be further reduced at higher rotation velocities and for decreasing S/N, which will also increase the uncertainties of the atmospheric parameters.

The analysis framework realised in {\sc Saturn} can, of course, be extended to other types of stars. One direction could be subluminous O and B stars \citep[e.g.][]{Heber16}, ideal objects for hydrostatic analyses with {\sc Ads} even to higher effective temperatures than covered here because of their much higher gravities, $\log (g)$\,$\approx$\,5 to 6, and their negligible winds. Some first steps in this direction have already been taken \citep[e.g.][]{Przybillaetal06b,Schaffenrothetal21}. Atomic diffusion in their very stable atmospheres is an issue, though, which is noticeable through widely varying helium and metal abundances.

Extensions towards the modelling of cooler A-type stars are also aspired, to cover the full scope of existing investigations such as by \citet{Xiangetal2022} or \citet{Sun25}. Currently, we use a mixture of non-LTE and LTE spectrum synthesis with {\sc Ads} to cope with the much wider range of chemical species present in the spectra \citep[e.g.][]{Przybillaetal06,Przybillaetal17}. Clearly, a substantial extension of the non-LTE model atom database would be desirable. However, a word of caution is in order. The A-type stars with surface gravities as typical for the main sequence comprise to a considerable percentage objects of non-standard abundances, such as Am- and Ap-type stars, $\lambda$ Boo stars, metal-poor blue horizontal branch stars and blue stragglers. Slowly rotating A-type and late B-type stars pose in their entirety the probably most difficult star classes to be analysed because of their diversity, bringing pre-calculated models sooner or later to their limit.

In any case, {\sc Saturn} will provide a highly useful means for an efficient analysis of stellar spectra beyond the immediate applications presented here. We plan a public release of {\sc Saturn} in the future, once the ongoing model development for {\sc Ads} is completed.

\section*{Data availability}
Most optical spectra used in this paper are publicly available and can be downloaded from the ESO Science Portal and the CFHT Science Archive. Observed IUE spectrophotometry and Gaia XP spectra are available from the respective archives. Photometric measurements used can be accessed using the VizieR Catalogue Service.
The model SEDs, model spectra, and normalised spectra shown in the plots are available on Zenodo and can be accessed via the following link \mbox{DOI: {\href{https://doi.org/10.5281/zenodo.19387709}{{\normalfont\texttt{10.5281/zenodo.19387709}}}}}.
Missing data, model fits, or code outputs are available upon reasonable request.

\begin{acknowledgements}
The authors thank M.A. Urbaneja for valuable discussions and A. Ebenbichler for the provision of his DIB linelist. P.A.~acknowledges support of this work by grant of a Ph.D.~stipend from the Vice Rectorate for Research of the University of Innsbruck. The present work uses observations collected at the Centro Astron\'omico Hispano Alem\'an at Calar Alto (CAHA), operated jointly by the Max-Planck Institut für Astronomie and the Instituto de Astrof\'isica de Andalucía (CSIC), proposal H2005-2.2-016, and observations collected at the European Southern Observatory under ESO programme 091.C-0713(A). Based on data obtained from the ESO Science Archive Facility with DOIs: https://doi.org/10.18727/archive/24 and https://doi.org/10.18727/archive/33. It also employs observations made with the Nordic Optical Telescope, owned in collaboration by the University of Turku and Aarhus University, and operated jointly by Aarhus University, the University of Turku, and the University of Oslo, representing Denmark, Finland and Norway, the University of Iceland, and Stockholm University at the Observatorio del Roque de los Muchachos, La Palma, Spain, of the Instituto de Astrofisica de Canarias. The NOT data were obtained under programme ID P41-027.  This research used the facilities of the Canadian Astronomy Data Centre operated by the National Research Council of Canada with the support of the Canadian Space Agency.
\end{acknowledgements}

\bibliographystyle{aa}
\bibliography{biblio}

@ARTICLE{Przybillaetal00,
   author = {{Przybilla}, N. and {Butler}, K. and {Becker}, S.~R. and {Kudritzki}, R.~P. and 
	{Venn}, K.~A.},
    title = "{Non-LTE line formation for neutral oxygen. Model atom and first results on A-type stars}",
  journal = {\aap},
     year = 2000,
    month = jul,
   volume = 359,
    pages = {1085-1106},
   adsurl = {http://adsabs.harvard.edu/abs/2000A%26A...359.1085P}
}

@ARTICLE{NiPr08,
   author = {{Nieva}, M.~F. and {Przybilla}, N.},
    title = "{Carbon abundances of early B-type stars in the solar vicinity. Non-LTE line-formation for C II/III/IV and self-consistent atmospheric parameters}",
  journal = {\aap},
archivePrefix = "arXiv",
   eprint = {0711.3783},
     year = 2008,
    month = apr,
   volume = 481,
    pages = {199-216},
      doi = {10.1051/0004-6361:20078203},
   adsurl = {http://adsabs.harvard.edu/abs/2008A%26A...481..199N}
}

@ARTICLE{Przybillaetal01a,
   author = {{Przybilla}, N. and {Butler}, K. and {Becker}, S.~R. and {Kudritzki}, R.~P.
	},
    title = "{Non-LTE line formation for $\backslash$ion$\{$Mg$\}$$\{$I/II$\}$: Abundances and stellar parameters. Model atom and first results on A-type stars}",
  journal = {\aap},
     year = 2001,
    month = apr,
   volume = 369,
    pages = {1009-1026},
      doi = {10.1051/0004-6361:20010164},
   adsurl = {http://adsabs.harvard.edu/abs/2001A%26A...369.1009P}
}

@ARTICLE{Przybilla05,
   author = {{Przybilla}, N.},
    title = "{Non-LTE modelling of the He I 10830 {\AA} line in early-type main sequence stars}",
  journal = {\aap},
   eprint = {astro-ph/0508068},
     year = 2005,
    month = nov,
   volume = 443,
    pages = {293-296},
      doi = {10.1051/0004-6361:20053412},
   adsurl = {http://adsabs.harvard.edu/abs/2005A%26A...443..293P}
}

@ARTICLE{Moreletal06,
   author = {{Morel}, T. and {Butler}, K. and {Aerts}, C. and {Neiner}, C. and 
	{Briquet}, M.},
    title = "{Abundance analysis of prime B-type targets for asteroseismology. I. Nitrogen excess in slowly-rotating {$\beta$} Cephei stars}",
  journal = {\aap},
   eprint = {astro-ph/0607264},
     year = 2006,
    month = oct,
   volume = 457,
    pages = {651-663},
      doi = {10.1051/0004-6361:20065171},
   adsurl = {http://adsabs.harvard.edu/abs/2006A%26A...457..651M}
}

@ARTICLE{Becker98,
   author = {{Becker}, S.~R.},
    title = "{Non-LTE Line Formation for Iron-Group Elements in A Supergiants}",
  journal = {ASP Conf.~Ser.},
     year = 1998,
   volume = 131,
    pages = {137}
}

@ARTICLE{PrBu01,
   author = {{Przybilla}, N. and {Butler}, K.},
    title = "{Non-LTE line formation for N: Abundances and stellar parameters. Model atom and first results on BA-type stars}",
  journal = {\aap},
     year = 2001,
    month = dec,
   volume = 379,
    pages = {955-975},
      doi = {10.1051/0004-6361:20011393},
   adsurl = {http://adsabs.harvard.edu/abs/2001A%26A...379..955P}
}

@ARTICLE{PrBu04,
   author = {{Przybilla}, N. and {Butler}, K.},
    title = "{Non-LTE Line Formation for Hydrogen Revisited}",
  journal = {\apj},
   eprint = {astro-ph/0406458},
     year = 2004,
    month = jul,
   volume = 609,
    pages = {1181-1191},
      doi = {10.1086/421316},
   adsurl = {http://adsabs.harvard.edu/abs/2004ApJ...609.1181P}
}

@ARTICLE{NiPr06,
   author = {{Nieva}, M.~F. and {Przybilla}, N.},
    title = "{C II Abundances in Early-Type Stars: Solution to a Notorious Non-LTE Problem}",
  journal = {ApJ},
   eprint = {astro-ph/0602342},
     year = 2006,
    month = mar,
   volume = 639,
    pages = {L39-L42},
      doi = {10.1086/501124},
   adsurl = {http://adsabs.harvard.edu/abs/2006ApJ...639L..39N}
}

@ARTICLE{MoBu08,
   author = {{Morel}, T. and {Butler}, K.},
    title = "{The neon content of nearby B-type stars and its implications for the solar model problem}",
  journal = {\aap},
archivePrefix = "arXiv",
   eprint = {0806.0491},
     year = 2008,
    month = aug,
   volume = 487,
    pages = {307-315},
      doi = {10.1051/0004-6361:200809924},
   adsurl = {http://adsabs.harvard.edu/abs/2008A%26A...487..307M}
}

@ARTICLE{Vranckenetal96,
   author = {{Vrancken}, M. and {Butler}, K. and {Becker}, S.~R.},
    title = "{Non-LTE line formation for SII and SIII. I. Model atoms and first results.}",
  journal = {\aap},
     year = 1996,
    month = jul,
   volume = 311,
    pages = {661-668},
   adsurl = {http://adsabs.harvard.edu/abs/1996A%26A...311..661V}
}

@ARTICLE{Zboriletal97,
   author = {{Zboril}, M. and {North}, P. and {Glagolevskij}, Y.~V. and {Betrix}, F.
	},
    title = "{Properties of He-rich stars. I. Their evolutionary state and helium abundance.}",
  journal = {\aap},
     year = 1997,
    month = aug,
   volume = 324,
    pages = {949-958},
   adsurl = {http://adsabs.harvard.edu/abs/1997A%26A...324..949Z}
}

@ARTICLE{Cutrietal03,
   author = {{Cutri}, R.~M. and {Skrutskie}, M.~F. and {van Dyk}, S. and 
	{Beichman}, C.~A. and {Carpenter}, J.~M. and {Chester}, T. and 
	{Cambresy}, L. and {Evans}, T. and {Fowler}, J. and {Gizis}, J. and 
	{Howard}, E. and {Huchra}, J. and {Jarrett}, T. and {Kopan}, E.~L. and 
	{Kirkpatrick}, J.~D. and {Light}, R.~M. and {Marsh}, K.~A. and 
	{McCallon}, H. and {Schneider}, S. and {Stiening}, R. and {Sykes}, M. and 
	{Weinberg}, M. and {Wheaton}, W.~A. and {Wheelock}, S. and {Zacarias}, N.
	},
    title = "{VizieR Online Data Catalog: 2MASS All-Sky Catalog of Point Sources (Cutri+ 2003)}",
  journal = {VizieR Online Data Catalog},
     year = 2003,
    month = jun,
   volume = 2246,
   adsurl = {http://adsabs.harvard.edu/abs/2003yCat.2246....0C}
}

@ARTICLE{Mermilliod97,
   author = {{Mermilliod}, J.~C.},
    title = "{VizieR Online Data Catalog: Homogeneous Means in the UBV System (Mermilliod 1991)}",
  journal = {VizieR Online Data Catalog},
     year = 1997,
    month = sep,
   volume = 2168,
   adsurl = {http://adsabs.harvard.edu/abs/1997yCat.2168....0M}
}

@ARTICLE{Ekstroemetal12,
   author = {{Ekstr{\"o}m}, S. and {Georgy}, C. and {Eggenberger}, P. and 
	{Meynet}, G. and {Mowlavi}, N. and {Wyttenbach}, A. and {Granada}, A. and 
	{Decressin}, T. and {Hirschi}, R. and {Frischknecht}, U. and 
	{Charbonnel}, C. and {Maeder}, A.},
    title = "{Grids of stellar models with rotation. I. Models from 0.8 to 120 M$_{⊙}$ at solar metallicity (Z = 0.014)}",
  journal = {\aap},
archivePrefix = "arXiv",
   eprint = {1110.5049},
 primaryClass = "astro-ph.SR",
     year = 2012,
    month = jan,
   volume = 537,
      eid = {A146},
    pages = {A146},
      doi = {10.1051/0004-6361/201117751},
   adsurl = {http://adsabs.harvard.edu/abs/2012A%26A...537A.146E}
}

@ARTICLE{NiPr14,
   author = {{Nieva}, M.~F. and {Przybilla}, N.},
    title = "{Fundamental properties of nearby single early B-type stars}",
  journal = {\aap},
archivePrefix = "arXiv",
   eprint = {1412.1418},
 primaryClass = "astro-ph.SR",
     year = 2014,
    month = jun,
   volume = 566,
      eid = {A7},
    pages = {A7},
      doi = {10.1051/0004-6361/201423373},
   adsurl = {http://adsabs.harvard.edu/abs/2014A%26A...566A...7N}
}

@ARTICLE{Michaud70,
   author = {{Michaud}, G.},
    title = "{Diffusion Processes in Peculiar a Stars}",
  journal = {\apj},
     year = 1970,
    month = may,
   volume = 160,
    pages = {641},
      doi = {10.1086/150459},
   adsurl = {http://adsabs.harvard.edu/abs/1970ApJ...160..641M}
}

@ARTICLE{Smith96,
   author = {{Smith}, K.~C.},
    title = "{Chemically Peculiar Hot Stars}",
  journal = {\apss},
     year = 1996,
    month = mar,
   volume = 237,
    pages = {77},
      doi = {10.1007/BF02424427},
   adsurl = {http://adsabs.harvard.edu/abs/1996Ap%26SS.237...77S}
}

@ARTICLE{NiPr12,
   author = {{Nieva}, M.~F. and {Przybilla}, N.},
    title = "{Present-day cosmic abundances. A comprehensive study of nearby early B-type stars and implications for stellar and Galactic evolution and interstellar dust models}",
  journal = {\aap},
archivePrefix = "arXiv",
   eprint = {1203.5787},
 primaryClass = "astro-ph.SR",
     year = 2012,
    month = mar,
   volume = 539,
      eid = {A143},
    pages = {A143},
      doi = {10.1051/0004-6361/201118158},
   adsurl = {http://adsabs.harvard.edu/abs/2012A%26A...539A.143N}
}

@ARTICLE{Kurucz93,
   author = {{Kurucz}, R.},
    title = "{ATLAS9 Stellar Atmosphere Programs and 2 km/s grid.}",
  journal = {CD-ROM No.~13~(Cambridge, Mass.: SAO)},
     year = 1993,
   adsurl = {http://adsabs.harvard.edu/abs/1993KurCD..13.....K}
}

@ARTICLE{RyHu91,
   author = {{Rybicki}, G.~B. and {Hummer}, D.~G.},
    title = "{An accelerated lambda iteration method for multilevel radiative transfer. I - Non-overlapping lines with background continuum}",
  journal = {\aap},
     year = 1991,
    month = may,
   volume = 245,
    pages = {171-181},
   adsurl = {http://adsabs.harvard.edu/abs/1991A%26A...245..171R}
}

@ARTICLE{Przybillaetal13,
   author = {{Przybilla}, N. and {Nieva}, M.~F. and {Irrgang}, A. and {Butler}, K.
	},
    title = "{Hot stars and cosmic abundances}",
     year = 2013,
  journal = {EAS Publ.~Ser.},
   volume = 63,
    month = dec,
    pages = {13},
      doi = {10.1051/eas/1363002},
   adsurl = {http://adsabs.harvard.edu/abs/2013EAS....63...13P}
}

@ARTICLE{Kurucz96,
   author = {{Kurucz}, R.~L.},
    title = "{Status of the ATLAS 12 Opacity Sampling Program and of New Programs for Rosseland and for Distribution Function Opacity}",
     year = 1996,
  journal = {ASP Conf.~Ser.},
   volume = 108,
    pages = {160},
   adsurl = {http://adsabs.harvard.edu/abs/1996ASPC..108..160K}
}

@ARTICLE{Mazaetal14,
   author = {{Maza}, N.~L. and {Nieva}, M.~F. and {Przybilla}, N.},
    title = "{A non-LTE spectral analysis of the $^{3}$He and $^{4}$He isotopes in the HgMn star {$\kappa$} Cancri}",
  journal = {\aap},
archivePrefix = "arXiv",
   eprint = {1412.2052},
 primaryClass = "astro-ph.SR",
     year = 2014,
    month = dec,
   volume = 572,
      eid = {A112},
    pages = {A112},
      doi = {10.1051/0004-6361/201425037},
   adsurl = {http://adsabs.harvard.edu/abs/2014A%26A...572A.112M}
}

@ARTICLE{BuGi85,
   author = {{Butler}, K. and {Giddings}, J.~R.},
    title = "{xx}",
  journal = {Newsletter of Analysis of Astronomical Spectra, 9 (Univ. London)},
     year = 1985,
    month = aug,
    pages = {},
   editor = {Univ. London},
  adsurl = {},
}

@PHDTHESIS{Giddings81,
   author = {{Giddings}, J.~R.},
   school = {(Univ. London)},
     year = 1981,
    month = jun,
   adsurl = {http://adsabs.harvard.edu/abs/1981PhDT.......113G}
}

@ARTICLE{NiPr07,
   author = {{Nieva}, M.~F. and {Przybilla}, N.},
    title = "{Hydrogen and helium line formation in OB dwarfs and giants. A hybrid non-LTE approach}",
  journal = {\aap},
   eprint = {astro-ph/0608117},
     year = 2007,
    month = may,
   volume = 467,
    pages = {295-309},
      doi = {10.1051/0004-6361:20065757},
   adsurl = {http://adsabs.harvard.edu/abs/2007A%26A...467..295N}
}

@ARTICLE{vanLeeuwen07a,
   author = {{van Leeuwen}, F.},
    title = "{Validation of the new Hipparcos reduction}",
  journal = {\aap},
archivePrefix = "arXiv",
   eprint = {0708.1752},
     year = 2007,
    month = nov,
   volume = 474,
    pages = {653-664},
      doi = {10.1051/0004-6361:20078357},
   adsurl = {http://cdsads.u-strasbg.fr/abs/2007A%26A...474..653V}
}

@ARTICLE{Maederetal14,
       author = {{Maeder}, Andr{\'e} and {Przybilla}, Norbert and {Nieva}, Mar{\'\i}a~Fernanda and {Georgy}, Cyril and {Meynet}, Georges and {Ekstr{\"o}m}, Sylvia and {Eggenberger}, Patrick},
        title = "{Evolution of surface CNO abundances in massive stars}",
      journal = {\aap},
         year = 2014,
        month = may,
       volume = {565},
          eid = {A39},
        pages = {A39},
          doi = {10.1051/0004-6361/201220602},
archivePrefix = {arXiv},
       eprint = {1404.1020},
 primaryClass = {astro-ph.SR},
       adsurl = {https://ui.adsabs.harvard.edu/abs/2014A&A...565A..39M}
}

@ARTICLE{Przybillaetal10,
       author = {{Przybilla}, N. and {Firnstein}, M. and {Nieva}, M.~F. and {Meynet}, G. and  {Maeder}, A.},
        title = "{Mixing of CNO-cycled matter in massive stars}",
      journal = {\aap},
archivePrefix = "arXiv",
       eprint = {1005.2278},
 primaryClass = "astro-ph.SR",
         year = 2010,
        month = jul,
       volume = 517,
          eid = {A38},
        pages = {A38},
          doi = {10.1051/0004-6361/201014164},
       adsurl = {http://esoads.eso.org/abs/2010A%26A...517A..38P}
}

@PHDTHESIS{Irrgang14,
   author = {{Irrgang}, A.},
   school = {(Univ. Erlangen-Nuremberg)},
     year = 2014,
    month = aug
}

@ARTICLE{Irrgangetal14,
       author = {{Irrgang}, A. and {Przybilla}, N. and {Heber}, U. and {B{\"o}ck}, M. and {Hanke}, M. and {Nieva}, M.~F. and {Butler}, K.},
        title = "{A new method for an objective, {\ensuremath{\chi}}$^{2}$-based spectroscopic analysis of early-type stars. First results from its application to single and binary B- and late O-type stars}",
      journal = {\aap},
         year = 2014,
        month = may,
       volume = {565},
          eid = {A63},
        pages = {A63},
          doi = {10.1051/0004-6361/201323167},
archivePrefix = {arXiv},
       eprint = {1403.1122},
 primaryClass = {astro-ph.SR},
       adsurl = {https://ui.adsabs.harvard.edu/abs/2014A&A...565A..63I}
}

@ARTICLE{Przybillaetal06,
       author = {{Przybilla}, N. and {Butler}, K. and {Becker}, S.~R. and {Kudritzki}, R.~P.},
        title = "{Quantitative spectroscopy of BA-type supergiants}",
      journal = {\aap},
         year = 2006,
        month = jan,
       volume = {445},
       number = {3},
        pages = {1099-1126},
          doi = {10.1051/0004-6361:20053832},
archivePrefix = {arXiv},
       eprint = {astro-ph/0509669},
 primaryClass = {astro-ph},
       adsurl = {https://ui.adsabs.harvard.edu/abs/2006A&A...445.1099P}
}

@ARTICLE{Gaia2016,
       author = {{Gaia Collaboration} and {Prusti}, T. and {de Bruijne}, J.~H.~J. and {Brown}, A.~G.~A. and {Vallenari}, A. and {Babusiaux}, C. and {Bailer-Jones}, C.~A.~L. and {Bastian}, U. and {Biermann}, M. and {Evans}, D.~W. and {Eyer}, L. and {Jansen}, F. and {Jordi}, C. and {Klioner}, S.~A. and {Lammers}, U. and {Lindegren}, L. and {Luri}, X. and {Mignard}, F. and {Milligan}, D.~J. and {Panem}, C. and {Poinsignon}, V. and {Pourbaix}, D. and {Randich}, S. and {Sarri}, G. and {Sartoretti}, P. and {Siddiqui}, H.~I. and {Soubiran}, C. and {Valette}, V. and {van Leeuwen}, F. and {Walton}, N.~A. and {Aerts}, C. and {Arenou}, F. and {Cropper}, M. and {Drimmel}, R. and {H{\o}g}, E. and {Katz}, D. and {Lattanzi}, M.~G. and {O'Mullane}, W. and {Grebel}, E.~K. and {Holland}, A.~D. and {Huc}, C. and {Passot}, X. and {Bramante}, L. and {Cacciari}, C. and {Casta{\~n}eda}, J. and {Chaoul}, L. and {Cheek}, N. and {De Angeli}, F. and {Fabricius}, C. and {Guerra}, R. and {Hern{\'a}ndez}, J. and {Jean-Antoine-Piccolo}, A. and {Masana}, E. and {Messineo}, R. and {Mowlavi}, N. and {Nienartowicz}, K. and {Ord{\'o}{\~n}ez-Blanco}, D. and {Panuzzo}, P. and {Portell}, J. and {Richards}, P.~J. and {Riello}, M. and {Seabroke}, G.~M. and {Tanga}, P. and {Th{\'e}venin}, F. and {Torra}, J. and {Els}, S.~G. and {Gracia-Abril}, G. and {Comoretto}, G. and {Garcia-Reinaldos}, M. and {Lock}, T. and {Mercier}, E. and {Altmann}, M. and {Andrae}, R. and {Astraatmadja}, T.~L. and {Bellas-Velidis}, I. and {Benson}, K. and {Berthier}, J. and {Blomme}, R. and {Busso}, G. and {Carry}, B. and {Cellino}, A. and {Clementini}, G. and {Cowell}, S. and {Creevey}, O. and {Cuypers}, J. and {Davidson}, M. and {De Ridder}, J. and {de Torres}, A. and {Delchambre}, L. and {Dell'Oro}, A. and {Ducourant}, C. and {Fr{\'e}mat}, Y. and {Garc{\'\i}a-Torres}, M. and {Gosset}, E. and {Halbwachs}, J. -L. and {Hambly}, N.~C. and {Harrison}, D.~L. and {Hauser}, M. and {Hestroffer}, D. and {Hodgkin}, S.~T. and {Huckle}, H.~E. and {Hutton}, A. and {Jasniewicz}, G. and {Jordan}, S. and {Kontizas}, M. and {Korn}, A.~J. and {Lanzafame}, A.~C. and {Manteiga}, M. and {Moitinho}, A. and {Muinonen}, K. and {Osinde}, J. and {Pancino}, E. and {Pauwels}, T. and {Petit}, J. -M. and {Recio-Blanco}, A. and {Robin}, A.~C. and {Sarro}, L.~M. and {Siopis}, C. and {Smith}, M. and {Smith}, K.~W. and {Sozzetti}, A. and {Thuillot}, W. and {van Reeven}, W. and {Viala}, Y. and {Abbas}, U. and {Abreu Aramburu}, A. and {Accart}, S. and {Aguado}, J.~J. and {Allan}, P.~M. and {Allasia}, W. and {Altavilla}, G. and {{\'A}lvarez}, M.~A. and {Alves}, J. and {Anderson}, R.~I. and {Andrei}, A.~H. and {Anglada Varela}, E. and {Antiche}, E. and {Antoja}, T. and {Ant{\'o}n}, S. and {Arcay}, B. and {Atzei}, A. and {Ayache}, L. and {Bach}, N. and {Baker}, S.~G. and {Balaguer-N{\'u}{\~n}ez}, L. and {Barache}, C. and {Barata}, C. and {Barbier}, A. and {Barblan}, F. and {Baroni}, M. and {Barrado y Navascu{\'e}s}, D. and {Barros}, M. and {Barstow}, M.~A. and {Becciani}, U. and {Bellazzini}, M. and {Bellei}, G. and {Bello Garc{\'\i}a}, A. and {Belokurov}, V. and {Bendjoya}, P. and {Berihuete}, A. and {Bianchi}, L. and {Bienaym{\'e}}, O. and {Billebaud}, F. and {Blagorodnova}, N. and {Blanco-Cuaresma}, S. and {Boch}, T. and {Bombrun}, A. and {Borrachero}, R. and {Bouquillon}, S. and {Bourda}, G. and {Bouy}, H. and {Bragaglia}, A. and {Breddels}, M.~A. and {Brouillet}, N. and {Br{\"u}semeister}, T. and {Bucciarelli}, B. and {Budnik}, F. and {Burgess}, P. and {Burgon}, R. and {Burlacu}, A. and {Busonero}, D. and {Buzzi}, R. and {Caffau}, E. and {Cambras}, J. and {Campbell}, H. and {Cancelliere}, R. and {Cantat-Gaudin}, T. and {Carlucci}, T. and {Carrasco}, J.~M. and {Castellani}, M. and {Charlot}, P. and {Charnas}, J. and {Charvet}, P. and {Chassat}, F. and {Chiavassa}, A. and {Clotet}, M. and {Cocozza}, G. and {Collins}, R.~S. and {Collins}, P. and {Costigan}, G. and {Crifo}, F. and {Cross}, N.~J.~G. and {Crosta}, M. and {Crowley}, C. and {Dafonte}, C. and {Damerdji}, Y. and {Dapergolas}, A. and {David}, P. and {David}, M. and {De Cat}, P. and {de Felice}, F. and {de Laverny}, P. and {De Luise}, F. and {De March}, R. and {de Martino}, D. and {de Souza}, R. and {Debosscher}, J. and {del Pozo}, E. and {Delbo}, M. and {Delgado}, A. and {Delgado}, H.~E. and {di Marco}, F. and {Di Matteo}, P. and {Diakite}, S. and {Distefano}, E. and {Dolding}, C. and {Dos Anjos}, S. and {Drazinos}, P. and {Dur{\'a}n}, J. and {Dzigan}, Y. and {Ecale}, E. and {Edvardsson}, B. and {Enke}, H. and {Erdmann}, M. and {Escolar}, D. and {Espina}, M. and {Evans}, N.~W. and {Eynard Bontemps}, G. and {Fabre}, C. and {Fabrizio}, M. and {Faigler}, S. and {Falc{\~a}o}, A.~J. and {Farr{\`a}s Casas}, M. and {Faye}, F. and {Federici}, L. and {Fedorets}, G. and {Fern{\'a}ndez-Hern{\'a}ndez}, J. and {Fernique}, P. and {Fienga}, A. and {Figueras}, F. and {Filippi}, F. and {Findeisen}, K. and {Fonti}, A. and {Fouesneau}, M. and {Fraile}, E. and {Fraser}, M. and {Fuchs}, J. and {Furnell}, R. and {Gai}, M. and {Galleti}, S. and {Galluccio}, L. and {Garabato}, D. and {Garc{\'\i}a-Sedano}, F. and {Gar{\'e}}, P. and {Garofalo}, A. and {Garralda}, N. and {Gavras}, P. and {Gerssen}, J. and {Geyer}, R. and {Gilmore}, G. and {Girona}, S. and {Giuffrida}, G. and {Gomes}, M. and {Gonz{\'a}lez-Marcos}, A. and {Gonz{\'a}lez-N{\'u}{\~n}ez}, J. and {Gonz{\'a}lez-Vidal}, J.~J. and {Granvik}, M. and {Guerrier}, A. and {Guillout}, P. and {Guiraud}, J. and {G{\'u}rpide}, A. and {Guti{\'e}rrez-S{\'a}nchez}, R. and {Guy}, L.~P. and {Haigron}, R. and {Hatzidimitriou}, D. and {Haywood}, M. and {Heiter}, U. and {Helmi}, A. and {Hobbs}, D. and {Hofmann}, W. and {Holl}, B. and {Holland}, G. and {Hunt}, J.~A.~S. and {Hypki}, A. and {Icardi}, V. and {Irwin}, M. and {Jevardat de Fombelle}, G. and {Jofr{\'e}}, P. and {Jonker}, P.~G. and {Jorissen}, A. and {Julbe}, F. and {Karampelas}, A. and {Kochoska}, A. and {Kohley}, R. and {Kolenberg}, K. and {Kontizas}, E. and {Koposov}, S.~E. and {Kordopatis}, G. and {Koubsky}, P. and {Kowalczyk}, A. and {Krone-Martins}, A. and {Kudryashova}, M. and {Kull}, I. and {Bachchan}, R.~K. and {Lacoste-Seris}, F. and {Lanza}, A.~F. and {Lavigne}, J. -B. and {Le Poncin-Lafitte}, C. and {Lebreton}, Y. and {Lebzelter}, T. and {Leccia}, S. and {Leclerc}, N. and {Lecoeur-Taibi}, I. and {Lemaitre}, V. and {Lenhardt}, H. and {Leroux}, F. and {Liao}, S. and {Licata}, E. and {Lindstr{\o}m}, H.~E.~P. and {Lister}, T.~A. and {Livanou}, E. and {Lobel}, A. and {L{\"o}ffler}, W. and {L{\'o}pez}, M. and {Lopez-Lozano}, A. and {Lorenz}, D. and {Loureiro}, T. and {MacDonald}, I. and {Magalh{\~a}es Fernandes}, T. and {Managau}, S. and {Mann}, R.~G. and {Mantelet}, G. and {Marchal}, O. and {Marchant}, J.~M. and {Marconi}, M. and {Marie}, J. and {Marinoni}, S. and {Marrese}, P.~M. and {Marschalk{\'o}}, G. and {Marshall}, D.~J. and {Mart{\'\i}n-Fleitas}, J.~M. and {Martino}, M. and {Mary}, N. and {Matijevi{\v{c}}}, G. and {Mazeh}, T. and {McMillan}, P.~J. and {Messina}, S. and {Mestre}, A. and {Michalik}, D. and {Millar}, N.~R. and {Miranda}, B.~M.~H. and {Molina}, D. and {Molinaro}, R. and {Molinaro}, M. and {Moln{\'a}r}, L. and {Moniez}, M. and {Montegriffo}, P. and {Monteiro}, D. and {Mor}, R. and {Mora}, A. and {Morbidelli}, R. and {Morel}, T. and {Morgenthaler}, S. and {Morley}, T. and {Morris}, D. and {Mulone}, A.~F. and {Muraveva}, T. and {Musella}, I. and {Narbonne}, J. and {Nelemans}, G. and {Nicastro}, L. and {Noval}, L. and {Ord{\'e}novic}, C. and {Ordieres-Mer{\'e}}, J. and {Osborne}, P. and {Pagani}, C. and {Pagano}, I. and {Pailler}, F. and {Palacin}, H. and {Palaversa}, L. and {Parsons}, P. and {Paulsen}, T. and {Pecoraro}, M. and {Pedrosa}, R. and {Pentik{\"a}inen}, H. and {Pereira}, J. and {Pichon}, B. and {Piersimoni}, A.~M. and {Pineau}, F. -X. and {Plachy}, E. and {Plum}, G. and {Poujoulet}, E. and {Pr{\v{s}}a}, A. and {Pulone}, L. and {Ragaini}, S. and {Rago}, S. and {Rambaux}, N. and {Ramos-Lerate}, M. and {Ranalli}, P. and {Rauw}, G. and {Read}, A. and {Regibo}, S. and {Renk}, F. and {Reyl{\'e}}, C. and {Ribeiro}, R.~A. and {Rimoldini}, L. and {Ripepi}, V. and {Riva}, A. and {Rixon}, G. and {Roelens}, M. and {Romero-G{\'o}mez}, M. and {Rowell}, N. and {Royer}, F. and {Rudolph}, A. and {Ruiz-Dern}, L. and {Sadowski}, G. and {Sagrist{\`a} Sell{\'e}s}, T. and {Sahlmann}, J. and {Salgado}, J. and {Salguero}, E. and {Sarasso}, M. and {Savietto}, H. and {Schnorhk}, A. and {Schultheis}, M. and {Sciacca}, E. and {Segol}, M. and {Segovia}, J.~C. and {Segransan}, D. and {Serpell}, E. and {Shih}, I. -C. and {Smareglia}, R. and {Smart}, R.~L. and {Smith}, C. and {Solano}, E. and {Solitro}, F. and {Sordo}, R. and {Soria Nieto}, S. and {Souchay}, J. and {Spagna}, A. and {Spoto}, F. and {Stampa}, U. and {Steele}, I.~A. and {Steidelm{\"u}ller}, H. and {Stephenson}, C.~A. and {Stoev}, H. and {Suess}, F.~F. and {S{\"u}veges}, M. and {Surdej}, J. and {Szabados}, L. and {Szegedi-Elek}, E. and {Tapiador}, D. and {Taris}, F. and {Tauran}, G. and {Taylor}, M.~B. and {Teixeira}, R. and {Terrett}, D. and {Tingley}, B. and {Trager}, S.~C. and {Turon}, C. and {Ulla}, A. and {Utrilla}, E. and {Valentini}, G. and {van Elteren}, A. and {Van Hemelryck}, E. and {van Leeuwen}, M. and {Varadi}, M. and {Vecchiato}, A. and {Veljanoski}, J. and {Via}, T. and {Vicente}, D. and {Vogt}, S. and {Voss}, H. and {Votruba}, V. and {Voutsinas}, S. and {Walmsley}, G. and {Weiler}, M. and {Weingrill}, K. and {Werner}, D. and {Wevers}, T. and {Whitehead}, G. and {Wyrzykowski}, {\L}. and {Yoldas}, A. and {{\v{Z}}erjal}, M. and {Zucker}, S. and {Zurbach}, C. and {Zwitter}, T. and {Alecu}, A. and {Allen}, M. and {Allende Prieto}, C. and {Amorim}, A. and {Anglada-Escud{\'e}}, G. and {Arsenijevic}, V. and {Azaz}, S. and {Balm}, P. and {Beck}, M. and {Bernstein}, H. -H. and {Bigot}, L. and {Bijaoui}, A. and {Blasco}, C. and {Bonfigli}, M. and {Bono}, G. and {Boudreault}, S. and {Bressan}, A. and {Brown}, S. and {Brunet}, P. -M. and {Bunclark}, P. and {Buonanno}, R. and {Butkevich}, A.~G. and {Carret}, C. and {Carrion}, C. and {Chemin}, L. and {Ch{\'e}reau}, F. and {Corcione}, L. and {Darmigny}, E. and {de Boer}, K.~S. and {de Teodoro}, P. and {de Zeeuw}, P.~T. and {Delle Luche}, C. and {Domingues}, C.~D. and {Dubath}, P. and {Fodor}, F. and {Fr{\'e}zouls}, B. and {Fries}, A. and {Fustes}, D. and {Fyfe}, D. and {Gallardo}, E. and {Gallegos}, J. and {Gardiol}, D. and {Gebran}, M. and {Gomboc}, A. and {G{\'o}mez}, A. and {Grux}, E. and {Gueguen}, A. and {Heyrovsky}, A. and {Hoar}, J. and {Iannicola}, G. and {Isasi Parache}, Y. and {Janotto}, A. -M. and {Joliet}, E. and {Jonckheere}, A. and {Keil}, R. and {Kim}, D. -W. and {Klagyivik}, P. and {Klar}, J. and {Knude}, J. and {Kochukhov}, O. and {Kolka}, I. and {Kos}, J. and {Kutka}, A. and {Lainey}, V. and {LeBouquin}, D. and {Liu}, C. and {Loreggia}, D. and {Makarov}, V.~V. and {Marseille}, M.~G. and {Martayan}, C. and {Martinez-Rubi}, O. and {Massart}, B. and {Meynadier}, F. and {Mignot}, S. and {Munari}, U. and {Nguyen}, A. -T. and {Nordlander}, T. and {Ocvirk}, P. and {O'Flaherty}, K.~S. and {Olias Sanz}, A. and {Ortiz}, P. and {Osorio}, J. and {Oszkiewicz}, D. and {Ouzounis}, A. and {Palmer}, M. and {Park}, P. and {Pasquato}, E. and {Peltzer}, C. and {Peralta}, J. and {P{\'e}turaud}, F. and {Pieniluoma}, T. and {Pigozzi}, E. and {Poels}, J. and {Prat}, G. and {Prod'homme}, T. and {Raison}, F. and {Rebordao}, J.~M. and {Risquez}, D. and {Rocca-Volmerange}, B. and {Rosen}, S. and {Ruiz-Fuertes}, M.~I. and {Russo}, F. and {Sembay}, S. and {Serraller Vizcaino}, I. and {Short}, A. and {Siebert}, A. and {Silva}, H. and {Sinachopoulos}, D. and {Slezak}, E. and {Soffel}, M. and {Sosnowska}, D. and {Strai{\v{z}}ys}, V. and {ter Linden}, M. and {Terrell}, D. and {Theil}, S. and {Tiede}, C. and {Troisi}, L. and {Tsalmantza}, P. and {Tur}, D. and {Vaccari}, M. and {Vachier}, F. and {Valles}, P. and {Van Hamme}, W. and {Veltz}, L. and {Virtanen}, J. and {Wallut}, J. -M. and {Wichmann}, R. and {Wilkinson}, M.~I. and {Ziaeepour}, H. and {Zschocke}, S.},
        title = "{The Gaia mission}",
      journal = {\aap},
         year = 2016,
        month = nov,
       volume = {595},
          eid = {A1},
        pages = {A1},
          doi = {10.1051/0004-6361/201629272},
archivePrefix = {arXiv},
       eprint = {1609.04153},
 primaryClass = {astro-ph.IM},
       adsurl = {https://ui.adsabs.harvard.edu/abs/2016A&A...595A...1G}
}

@ARTICLE{Gaia2020,
       author = {{Gaia Collaboration} and {Brown}, A.~G.~A. and {Vallenari}, A. and {Prusti}, T. and {de Bruijne}, J.~H.~J. and {Babusiaux}, C. and {Biermann}, M. and {Creevey}, O.~L. and {Evans}, D.~W. and {Eyer}, L. and {Hutton}, A. and {Jansen}, F. and {Jordi}, C. and {Klioner}, S.~A. and {Lammers}, U. and {Lindegren}, L. and {Luri}, X. and {Mignard}, F. and {Panem}, C. and {Pourbaix}, D. and {Randich}, S. and {Sartoretti}, P. and {Soubiran}, C. and {Walton}, N.~A. and {Arenou}, F. and {Bailer-Jones}, C.~A.~L. and {Bastian}, U. and {Cropper}, M. and {Drimmel}, R. and {Katz}, D. and {Lattanzi}, M.~G. and {van Leeuwen}, F. and {Bakker}, J. and {Cacciari}, C. and {Casta{\~n}eda}, J. and {De Angeli}, F. and {Ducourant}, C. and {Fabricius}, C. and {Fouesneau}, M. and {Fr{\'e}mat}, Y. and {Guerra}, R. and {Guerrier}, A. and {Guiraud}, J. and {Jean-Antoine Piccolo}, A. and {Masana}, E. and {Messineo}, R. and {Mowlavi}, N. and {Nicolas}, C. and {Nienartowicz}, K. and {Pailler}, F. and {Panuzzo}, P. and {Riclet}, F. and {Roux}, W. and {Seabroke}, G.~M. and {Sordo}, R. and {Tanga}, P. and {Th{\'e}venin}, F. and {Gracia-Abril}, G. and {Portell}, J. and {Teyssier}, D. and {Altmann}, M. and {Andrae}, R. and {Bellas-Velidis}, I. and {Benson}, K. and {Berthier}, J. and {Blomme}, R. and {Brugaletta}, E. and {Burgess}, P.~W. and {Busso}, G. and {Carry}, B. and {Cellino}, A. and {Cheek}, N. and {Clementini}, G. and {Damerdji}, Y. and {Davidson}, M. and {Delchambre}, L. and {Dell'Oro}, A. and {Fern{\'a}ndez-Hern{\'a}ndez}, J. and {Galluccio}, L. and {Garc{\'\i}a-Lario}, P. and {Garcia-Reinaldos}, M. and {Gonz{\'a}lez-N{\'u}{\~n}ez}, J. and {Gosset}, E. and {Haigron}, R. and {Halbwachs}, J. -L. and {Hambly}, N.~C. and {Harrison}, D.~L. and {Hatzidimitriou}, D. and {Heiter}, U. and {Hern{\'a}ndez}, J. and {Hestroffer}, D. and {Hodgkin}, S.~T. and {Holl}, B. and {Jan{\ss}en}, K. and {Jevardat de Fombelle}, G. and {Jordan}, S. and {Krone-Martins}, A. and {Lanzafame}, A.~C. and {L{\"o}ffler}, W. and {Lorca}, A. and {Manteiga}, M. and {Marchal}, O. and {Marrese}, P.~M. and {Moitinho}, A. and {Mora}, A. and {Muinonen}, K. and {Osborne}, P. and {Pancino}, E. and {Pauwels}, T. and {Petit}, J. -M. and {Recio-Blanco}, A. and {Richards}, P.~J. and {Riello}, M. and {Rimoldini}, L. and {Robin}, A.~C. and {Roegiers}, T. and {Rybizki}, J. and {Sarro}, L.~M. and {Siopis}, C. and {Smith}, M. and {Sozzetti}, A. and {Ulla}, A. and {Utrilla}, E. and {van Leeuwen}, M. and {van Reeven}, W. and {Abbas}, U. and {Abreu Aramburu}, A. and {Accart}, S. and {Aerts}, C. and {Aguado}, J.~J. and {Ajaj}, M. and {Altavilla}, G. and {{\'A}lvarez}, M.~A. and {{\'A}lvarez Cid-Fuentes}, J. and {Alves}, J. and {Anderson}, R.~I. and {Anglada Varela}, E. and {Antoja}, T. and {Audard}, M. and {Baines}, D. and {Baker}, S.~G. and {Balaguer-N{\'u}{\~n}ez}, L. and {Balbinot}, E. and {Balog}, Z. and {Barache}, C. and {Barbato}, D. and {Barros}, M. and {Barstow}, M.~A. and {Bartolom{\'e}}, S. and {Bassilana}, J. -L. and {Bauchet}, N. and {Baudesson-Stella}, A. and {Becciani}, U. and {Bellazzini}, M. and {Bernet}, M. and {Bertone}, S. and {Bianchi}, L. and {Blanco-Cuaresma}, S. and {Boch}, T. and {Bombrun}, A. and {Bossini}, D. and {Bouquillon}, S. and {Bragaglia}, A. and {Bramante}, L. and {Breedt}, E. and {Bressan}, A. and {Brouillet}, N. and {Bucciarelli}, B. and {Burlacu}, A. and {Busonero}, D. and {Butkevich}, A.~G. and {Buzzi}, R. and {Caffau}, E. and {Cancelliere}, R. and {C{\'a}novas}, H. and {Cantat-Gaudin}, T. and {Carballo}, R. and {Carlucci}, T. and {Carnerero}, M.~I. and {Carrasco}, J.~M. and {Casamiquela}, L. and {Castellani}, M. and {Castro-Ginard}, A. and {Castro Sampol}, P. and {Chaoul}, L. and {Charlot}, P. and {Chemin}, L. and {Chiavassa}, A. and {Cioni}, M. -R.~L. and {Comoretto}, G. and {Cooper}, W.~J. and {Cornez}, T. and {Cowell}, S. and {Crifo}, F. and {Crosta}, M. and {Crowley}, C. and {Dafonte}, C. and {Dapergolas}, A. and {David}, M. and {David}, P. and {de Laverny}, P. and {De Luise}, F. and {De March}, R. and {De Ridder}, J. and {de Souza}, R. and {de Teodoro}, P. and {de Torres}, A. and {del Peloso}, E.~F. and {del Pozo}, E. and {Delbo}, M. and {Delgado}, A. and {Delgado}, H.~E. and {Delisle}, J. -B. and {Di Matteo}, P. and {Diakite}, S. and {Diener}, C. and {Distefano}, E. and {Dolding}, C. and {Eappachen}, D. and {Edvardsson}, B. and {Enke}, H. and {Esquej}, P. and {Fabre}, C. and {Fabrizio}, M. and {Faigler}, S. and {Fedorets}, G. and {Fernique}, P. and {Fienga}, A. and {Figueras}, F. and {Fouron}, C. and {Fragkoudi}, F. and {Fraile}, E. and {Franke}, F. and {Gai}, M. and {Garabato}, D. and {Garcia-Gutierrez}, A. and {Garc{\'\i}a-Torres}, M. and {Garofalo}, A. and {Gavras}, P. and {Gerlach}, E. and {Geyer}, R. and {Giacobbe}, P. and {Gilmore}, G. and {Girona}, S. and {Giuffrida}, G. and {Gomel}, R. and {Gomez}, A. and {Gonzalez-Santamaria}, I. and {Gonz{\'a}lez-Vidal}, J.~J. and {Granvik}, M. and {Guti{\'e}rrez-S{\'a}nchez}, R. and {Guy}, L.~P. and {Hauser}, M. and {Haywood}, M. and {Helmi}, A. and {Hidalgo}, S.~L. and {Hilger}, T. and {H{\l}adczuk}, N. and {Hobbs}, D. and {Holland}, G. and {Huckle}, H.~E. and {Jasniewicz}, G. and {Jonker}, P.~G. and {Juaristi Campillo}, J. and {Julbe}, F. and {Karbevska}, L. and {Kervella}, P. and {Khanna}, S. and {Kochoska}, A. and {Kontizas}, M. and {Kordopatis}, G. and {Korn}, A.~J. and {Kostrzewa-Rutkowska}, Z. and {Kruszy{\'n}ska}, K. and {Lambert}, S. and {Lanza}, A.~F. and {Lasne}, Y. and {Le Campion}, J. -F. and {Le Fustec}, Y. and {Lebreton}, Y. and {Lebzelter}, T. and {Leccia}, S. and {Leclerc}, N. and {Lecoeur-Taibi}, I. and {Liao}, S. and {Licata}, E. and {Lindstr{\o}m}, E.~P. and {Lister}, T.~A. and {Livanou}, E. and {Lobel}, A. and {Madrero Pardo}, P. and {Managau}, S. and {Mann}, R.~G. and {Marchant}, J.~M. and {Marconi}, M. and {Marcos Santos}, M.~M.~S. and {Marinoni}, S. and {Marocco}, F. and {Marshall}, D.~J. and {Martin Polo}, L. and {Mart{\'\i}n-Fleitas}, J.~M. and {Masip}, A. and {Massari}, D. and {Mastrobuono-Battisti}, A. and {Mazeh}, T. and {McMillan}, P.~J. and {Messina}, S. and {Michalik}, D. and {Millar}, N.~R. and {Mints}, A. and {Molina}, D. and {Molinaro}, R. and {Moln{\'a}r}, L. and {Montegriffo}, P. and {Mor}, R. and {Morbidelli}, R. and {Morel}, T. and {Morris}, D. and {Mulone}, A.~F. and {Munoz}, D. and {Muraveva}, T. and {Murphy}, C.~P. and {Musella}, I. and {Noval}, L. and {Ord{\'e}novic}, C. and {Orr{\`u}}, G. and {Osinde}, J. and {Pagani}, C. and {Pagano}, I. and {Palaversa}, L. and {Palicio}, P.~A. and {Panahi}, A. and {Pawlak}, M. and {Pe{\~n}alosa Esteller}, X. and {Penttil{\"a}}, A. and {Piersimoni}, A.~M. and {Pineau}, F. -X. and {Plachy}, E. and {Plum}, G. and {Poggio}, E. and {Poretti}, E. and {Poujoulet}, E. and {Pr{\v{s}}a}, A. and {Pulone}, L. and {Racero}, E. and {Ragaini}, S. and {Rainer}, M. and {Raiteri}, C.~M. and {Rambaux}, N. and {Ramos}, P. and {Ramos-Lerate}, M. and {Re Fiorentin}, P. and {Regibo}, S. and {Reyl{\'e}}, C. and {Ripepi}, V. and {Riva}, A. and {Rixon}, G. and {Robichon}, N. and {Robin}, C. and {Roelens}, M. and {Rohrbasser}, L. and {Romero-G{\'o}mez}, M. and {Rowell}, N. and {Royer}, F. and {Rybicki}, K.~A. and {Sadowski}, G. and {Sagrist{\`a} Sell{\'e}s}, A. and {Sahlmann}, J. and {Salgado}, J. and {Salguero}, E. and {Samaras}, N. and {Sanchez Gimenez}, V. and {Sanna}, N. and {Santove{\~n}a}, R. and {Sarasso}, M. and {Schultheis}, M. and {Sciacca}, E. and {Segol}, M. and {Segovia}, J.~C. and {S{\'e}gransan}, D. and {Semeux}, D. and {Shahaf}, S. and {Siddiqui}, H.~I. and {Siebert}, A. and {Siltala}, L. and {Slezak}, E. and {Smart}, R.~L. and {Solano}, E. and {Solitro}, F. and {Souami}, D. and {Souchay}, J. and {Spagna}, A. and {Spoto}, F. and {Steele}, I.~A. and {Steidelm{\"u}ller}, H. and {Stephenson}, C.~A. and {S{\"u}veges}, M. and {Szabados}, L. and {Szegedi-Elek}, E. and {Taris}, F. and {Tauran}, G. and {Taylor}, M.~B. and {Teixeira}, R. and {Thuillot}, W. and {Tonello}, N. and {Torra}, F. and {Torra}, J. and {Turon}, C. and {Unger}, N. and {Vaillant}, M. and {van Dillen}, E. and {Vanel}, O. and {Vecchiato}, A. and {Viala}, Y. and {Vicente}, D. and {Voutsinas}, S. and {Weiler}, M. and {Wevers}, T. and {Wyrzykowski}, {\L}. and {Yoldas}, A. and {Yvard}, P. and {Zhao}, H. and {Zorec}, J. and {Zucker}, S. and {Zurbach}, C. and {Zwitter}, T.},
        title = "{Gaia Early Data Release 3. Summary of the contents and survey properties}",
      journal = {\aap},
         year = 2021,
        month = may,
       volume = {649},
          eid = {A1},
        pages = {A1},
          doi = {10.1051/0004-6361/202039657},
archivePrefix = {arXiv},
       eprint = {2012.01533},
 primaryClass = {astro-ph.GA},
       adsurl = {https://ui.adsabs.harvard.edu/abs/2021A&A...649A...1G}
}

@ARTICLE{Bailer-Jones_etal_2021,
       author = {{Bailer-Jones}, C.~A.~L. and {Rybizki}, J. and {Fouesneau}, M. and {Demleitner}, M. and {Andrae}, R.},
        title = "{Estimating Distances from Parallaxes. V. Geometric and Photogeometric Distances to 1.47 Billion Stars in Gaia Early Data Release 3}",
      journal = {\aj},
         year = 2021,
        month = mar,
       volume = {161},
       number = {3},
          eid = {147},
        pages = {147},
          doi = {10.3847/1538-3881/abd806},
archivePrefix = {arXiv},
       eprint = {2012.05220},
 primaryClass = {astro-ph.SR},
       adsurl = {https://ui.adsabs.harvard.edu/abs/2021AJ....161..147B}
}

@ARTICLE{Fraser_etal_10,
       author = {{Fraser}, M. and {Dufton}, P.~L. and {Hunter}, I. and {Ryans}, R.~S.~I.},
        title = "{Atmospheric parameters and rotational velocities for a sample of Galactic B-type supergiants}",
      journal = {\mnras},
         year = 2010,
        month = may,
       volume = {404},
       number = {3},
        pages = {1306-1320},
          doi = {10.1111/j.1365-2966.2010.16392.x},
archivePrefix = {arXiv},
       eprint = {1001.3337},
 primaryClass = {astro-ph.SR},
       adsurl = {https://ui.adsabs.harvard.edu/abs/2010MNRAS.404.1306F}
}

@ARTICLE{hubeny_88,
       author = {{Hubeny}, I.},
        title = "{A computer program for calculating non-LTE model stellar atmospheres}",
      journal = {Comput. Phys. Commun.},
         year = 1988,
        month = dec,
       volume = {52},
       number = {1},
        pages = {103-132},
          doi = {10.1016/0010-4655(88)90177-4},
       adsurl = {https://ui.adsabs.harvard.edu/abs/1988CoPhC..52..103H}
}

@ARTICLE{HiMi98,
       author = {{Hillier}, D. John and {Miller}, D.~L.},
        title = "{The Treatment of Non-LTE Line Blanketing in Spherically Expanding Outflows}",
      journal = {\apj},
         year = 1998,
        month = mar,
       volume = {496},
       number = {1},
        pages = {407-427},
          doi = {10.1086/305350},
       adsurl = {https://ui.adsabs.harvard.edu/abs/1998ApJ...496..407H}
}

@ARTICLE{RePuHe04,
       author = {{Repolust}, T. and {Puls}, J. and {Herrero}, A.},
        title = "{Stellar and wind parameters of Galactic O-stars. The influence of line-blocking/blanketing}",
      journal = {\aap},
         year = 2004,
        month = feb,
       volume = {415},
        pages = {349-376},
          doi = {10.1051/0004-6361:20034594},
       adsurl = {https://ui.adsabs.harvard.edu/abs/2004A&A...415..349R}
}

@ARTICLE{Urbanejaetal08,
       author = {{Urbaneja}, Miguel A. and {Kudritzki}, Rolf~Peter and {Bresolin}, Fabio and {Przybilla}, Norbert and {Gieren}, Wolfgang and {Pietrzy{\'n}ski}, Grzegorz},
        title = "{The Araucaria Project: The Local Group Galaxy WLM{\textemdash}Distance and Metallicity from Quantitative Spectroscopy of Blue Supergiants}",
      journal = {\apj},
         year = 2008,
        month = sep,
       volume = {684},
       number = {1},
        pages = {118-135},
          doi = {10.1086/590334},
archivePrefix = {arXiv},
       eprint = {0805.3555},
 primaryClass = {astro-ph},
       adsurl = {https://ui.adsabs.harvard.edu/abs/2008ApJ...684..118U}
}

@ARTICLE{Kudritzkietal12,
       author = {{Kudritzki}, Rolf~Peter and {Urbaneja}, Miguel A. and {Gazak}, Zachary and {Bresolin}, Fabio and {Przybilla}, Norbert and {Gieren}, Wolfgang and {Pietrzy{\'n}ski}, Grzegorz},
        title = "{Quantitative Spectroscopy of Blue Supergiant Stars in the Disk of M81: Metallicity, Metallicity Gradient, and Distance}",
      journal = {\apj},
         year = 2012,
        month = mar,
       volume = {747},
       number = {1},
          eid = {15},
        pages = {15},
          doi = {10.1088/0004-637X/747/1/15},
archivePrefix = {arXiv},
       eprint = {1112.3643},
 primaryClass = {astro-ph.CO},
       adsurl = {https://ui.adsabs.harvard.edu/abs/2012ApJ...747...15K}
}

@ARTICLE{Urbanejaetal17,
       author = {{Urbaneja}, M.~A. and {Kudritzki}, R.~P. and {Gieren}, W. and {Pietrzy{\'n}ski}, G. and {Bresolin}, F. and {Przybilla}, N.},
        title = "{LMC Blue Supergiant Stars and the Calibration of the Flux-weighted Gravity-Luminosity Relationship}",
      journal = {\aj},
         year = 2017,
        month = sep,
       volume = {154},
       number = {3},
          eid = {102},
        pages = {102},
          doi = {10.3847/1538-3881/aa79a8},
archivePrefix = {arXiv},
       eprint = {1706.03967},
 primaryClass = {astro-ph.SR},
       adsurl = {https://ui.adsabs.harvard.edu/abs/2017AJ....154..102U}
}

@ARTICLE{Sanaetal12,
       author = {{Sana}, H. and {de Mink}, S.~E. and {de Koter}, A. and {Langer}, N. and {Evans}, C.~J. and {Gieles}, M. and {Gosset}, E. and {Izzard}, R.~G. and {Le Bouquin}, J. -B. and {Schneider}, F.~R.~N.},
        title = "{Binary Interaction Dominates the Evolution of Massive Stars}",
      journal = {Science},
         year = 2012,
        month = jul,
       volume = {337},
       number = {6093},
        pages = {444},
          doi = {10.1126/science.1223344},
archivePrefix = {arXiv},
       eprint = {1207.6397},
 primaryClass = {astro-ph.SR},
       adsurl = {https://ui.adsabs.harvard.edu/abs/2012Sci...337..444S}
}

@ARTICLE{Pulsetal05,
       author = {{Puls}, J. and {Urbaneja}, M.~A. and {Venero}, R. and {Repolust}, T. and {Springmann}, U. and {Jokuthy}, A. and {Mokiem}, M.~R.},
        title = "{Atmospheric NLTE-models for the spectroscopic analysis of blue stars with winds. II. Line-blanketed models}",
      journal = {\aap},
         year = 2005,
        month = may,
       volume = {435},
       number = {2},
        pages = {669-698},
          doi = {10.1051/0004-6361:20042365},
archivePrefix = {arXiv},
       eprint = {astro-ph/0411398},
 primaryClass = {astro-ph},
       adsurl = {https://ui.adsabs.harvard.edu/abs/2005A&A...435..669P}
}

@ARTICLE{Pauldrachetal01,
       author = {{Pauldrach}, A.~W.~A. and {Hoffmann}, T.~L. and {Lennon}, M.},
        title = "{Radiation-driven winds of hot luminous stars. XIII. A description of NLTE line blocking and blanketing towards realistic models for expanding atmospheres}",
      journal = {\aap},
         year = 2001,
        month = aug,
       volume = {375},
        pages = {161-195},
          doi = {10.1051/0004-6361:20010805},
       adsurl = {https://ui.adsabs.harvard.edu/abs/2001A&A...375..161P}
}

@ARTICLE{Graefeneretal02,
       author = {{Gr{\"a}fener}, G. and {Koesterke}, L. and {Hamann}, W. -R.},
        title = "{Line-blanketed model atmospheres for WR stars}",
      journal = {\aap},
         year = 2002,
        month = may,
       volume = {387},
        pages = {244-257},
          doi = {10.1051/0004-6361:20020269},
       adsurl = {https://ui.adsabs.harvard.edu/abs/2002A&A...387..244G}
}

@ARTICLE{Kudritzkietal14,
       author = {{Kudritzki}, Rolf~Peter and {Urbaneja}, Miguel A. and {Bresolin}, Fabio and {Hosek}, M. W. and {Przybilla}, Norbert},
        title = "{Stellar Metallicity of the Extended Disk and Distance of the Spiral Galaxy NGC 3621}",
      journal = {\apj},
         year = 2014,
        month = jun,
       volume = {788},
       number = {1},
          eid = {56},
        pages = {56},
          doi = {10.1088/0004-637X/788/1/56},
archivePrefix = {arXiv},
       eprint = {1404.7244},
 primaryClass = {astro-ph.GA},
       adsurl = {https://ui.adsabs.harvard.edu/abs/2014ApJ...788...56K}
}

@ARTICLE{Pfeifferetal98,
       author = {{Pfeiffer}, M.~J. and {Frank}, C. and {Baumueller}, D. and {Fuhrmann}, K. and {Gehren}, T.},
        title = "{FOCES - a fibre optics Cassegrain Echelle spectrograph}",
      journal = {\aaps},
         year = 1998,
        month = jun,
       volume = {130},
        pages = {381-393},
          doi = {10.1051/aas:1998231},
       adsurl = {https://ui.adsabs.harvard.edu/abs/1998A&AS..130..381P}
}

@ARTICLE{Kauferetal99,
       author = {{Kaufer}, A. and {Stahl}, O. and {Tubbesing}, S. and {N{\o}rregaard}, P. and {Avila}, G. and {Francois}, P. and {Pasquini}, L. and {Pizzella}, A.},
        title = "{Commissioning FEROS, the new high-resolution spectrograph at La-Silla.}",
      journal = {The Messenger},
         year = 1999,
        month = mar,
       volume = {95},
        pages = {8-12},
       adsurl = {https://ui.adsabs.harvard.edu/abs/1999Msngr..95....8K}
}

@ARTICLE{Przybillaetal06b,
       author = {{Przybilla}, N. and {Nieva}, M.~F. and {Edelmann}, H.},
        title = "{NLTE Analyses of Sdb Stars: Progress and Prospects}",
      journal = {Baltic Astronomy},
         year = 2006,
        month = jan,
       volume = {15},
        pages = {107-114},
archivePrefix = {arXiv},
       eprint = {astro-ph/0512151},
 primaryClass = {astro-ph},
       adsurl = {https://ui.adsabs.harvard.edu/abs/2006BaltA..15..107P}
}

@ARTICLE{Schaffenrothetal21,
       author = {{Schaffenroth}, V. and {Casewell}, S.~L. and {Schneider}, D. and {Kilkenny}, D. and {Geier}, S. and {Heber}, U. and {Irrgang}, A. and {Przybilla}, N. and {Marsh}, T.~R. and {Littlefair}, S.~P. and {Dhillon}, V.~S.},
        title = "{A quantitative in-depth analysis of the prototype sdB+BD system SDSS J08205+0008 revisited in the Gaia era}",
      journal = {\mnras},
         year = 2021,
        month = mar,
       volume = {501},
       number = {3},
        pages = {3847-3870},
          doi = {10.1093/mnras/staa3661},
archivePrefix = {arXiv},
       eprint = {2011.10013},
 primaryClass = {astro-ph.SR},
       adsurl = {https://ui.adsabs.harvard.edu/abs/2021MNRAS.501.3847S}
}

@ARTICLE{Przybillaetal21,
       author = {{Przybilla}, N. and {Fossati}, L. and {Jeffery}, C.~S.},
        title = "{HD 144941: the most extreme helium-strong star}",
      journal = {\aap},
         year = 2021,
        month = oct,
       volume = {654},
          eid = {A119},
        pages = {A119},
          doi = {10.1051/0004-6361/202141625},
archivePrefix = {arXiv},
       eprint = {2110.11267},
 primaryClass = {astro-ph.SR},
       adsurl = {https://ui.adsabs.harvard.edu/abs/2021A&A...654A.119P}
}

@ARTICLE{TrBe09,
       author = {{Tremblay}, P. -E. and {Bergeron}, P.},
        title = "{Spectroscopic Analysis of DA White Dwarfs: Stark Broadening of Hydrogen Lines Including Nonideal Effects}",
      journal = {\apj},
         year = 2009,
        month = may,
       volume = {696},
       number = {2},
        pages = {1755-1770},
          doi = {10.1088/0004-637X/696/2/1755},
archivePrefix = {arXiv},
       eprint = {0902.4182},
 primaryClass = {astro-ph.SR},
       adsurl = {https://ui.adsabs.harvard.edu/abs/2009ApJ...696.1755T}
}

@ARTICLE{Beauchampsetal97,
       author = {{Beauchamp}, A. and {Wesemael}, F. and {Bergeron}, P.},
        title = "{Spectroscopic Studies of DB White Dwarfs: Improved Stark Profiles for Optical Transitions of Neutral Helium}",
      journal = {\apjs},
         year = 1997,
        month = feb,
       volume = {108},
       number = {2},
        pages = {559-573},
          doi = {10.1086/312961},
       adsurl = {https://ui.adsabs.harvard.edu/abs/1997ApJS..108..559B}
}

@ARTICLE{Hubenyetal94,
       author = {{Hubeny}, I. and {Hummer}, D.~G. and {Lanz}, T.},
        title = "{NLTE model stellar atmospheres with line blanketing near the series limits.}",
      journal = {\aap},
         year = 1994,
        month = feb,
       volume = {282},
        pages = {151-167},
       adsurl = {https://ui.adsabs.harvard.edu/abs/1994A&A...282..151H}
}

@ARTICLE{HuMi88,
       author = {{Hummer}, D.~G. and {Mihalas}, Dimitri},
        title = "{The Equation of State for Stellar Envelopes. I. an Occupation Probability Formalism for the Truncation of Internal Partition Functions}",
      journal = {\apj},
         year = 1988,
        month = aug,
       volume = {331},
        pages = {794},
          doi = {10.1086/166600},
       adsurl = {https://ui.adsabs.harvard.edu/abs/1988ApJ...331..794H}
}

@ARTICLE{LaKu14,
       author = {{Langer}, N. and {Kudritzki}, R.~P.},
        title = "{The spectroscopic Hertzsprung-Russell diagram}",
      journal = {\aap},
         year = 2014,
        month = apr,
       volume = {564},
          eid = {A52},
        pages = {A52},
          doi = {10.1051/0004-6361/201423374},
archivePrefix = {arXiv},
       eprint = {1403.2212},
 primaryClass = {astro-ph.SR},
       adsurl = {https://ui.adsabs.harvard.edu/abs/2014A&A...564A..52L}
}

@ARTICLE{Przybillaetal01b,
       author = {{Przybilla}, N. and {Butler}, K. and {Kudritzki}, R.~P.},
        title = "{Non-LTE line-formation for neutral and singly-ionized carbon. Model atom and first results on BA-type stars}",
      journal = {\aap},
         year = 2001,
        month = dec,
       volume = {379},
        pages = {936-954},
          doi = {10.1051/0004-6361:20011384},
       adsurl = {https://ui.adsabs.harvard.edu/abs/2001A&A...379..936P}
}

@ARTICLE{Gonzalez-Toraetal22,
       author = {{Gonz{\'a}lez-Tor{\`a}}, G. and {Urbaneja}, M.~A. and {Przybilla}, N. and {Dreizler}, S. and {Roth}, M.~M. and {Kamann}, S. and {Castro}, N.},
        title = "{MUSE crowded field 3D spectroscopy in NGC 300. II. Quantitative spectroscopy of BA-type supergiants}",
      journal = {\aap},
         year = 2022,
        month = feb,
       volume = {658},
          eid = {A117},
        pages = {A117},
          doi = {10.1051/0004-6361/202142372},
archivePrefix = {arXiv},
       eprint = {2201.01311},
 primaryClass = {astro-ph.GA},
       adsurl = {https://ui.adsabs.harvard.edu/abs/2022A&A...658A.117G}
}

@ARTICLE{FFT04,
       author = {{Froese Fischer}, Charlotte and {Tachiev}, Georgio},
        title = "{Breit-Pauli energy levels, lifetimes, and transition probabilities for the beryllium-like to neon-like sequences}",
      journal = {At. Data and Nucl. Data Tables},
         year = 2004,
        month = may,
       volume = {87},
       number = {1},
        pages = {1-184},
          doi = {10.1016/j.adt.2004.02.001},
       adsurl = {https://ui.adsabs.harvard.edu/abs/2004ADNDT..87....1F}
}

@ARTICLE{FFTI06,
       author = {{Froese Fischer}, Charlotte and {Tachiev}, Georgio and {Irimia}, Andrei},
        title = "{Relativistic energy levels, lifetimes, and transition probabilities for the sodium-like to argon-like sequences}",
      journal = {At. Data and Nucl. Data Tables},
         year = 2006,
        month = sep,
       volume = {92},
       number = {5},
        pages = {607-812},
          doi = {10.1016/j.adt.2006.03.001},
       adsurl = {https://ui.adsabs.harvard.edu/abs/2006ADNDT..92..607F}
}

@ARTICLE{fitzpatrick99,
       author = {{Fitzpatrick}, Edward L.},
        title = "{Correcting for the Effects of Interstellar Extinction}",
      journal = {\pasp},
         year = 1999,
        month = jan,
       volume = {111},
       number = {755},
        pages = {63-75},
          doi = {10.1086/316293},
archivePrefix = {arXiv},
       eprint = {astro-ph/9809387},
 primaryClass = {astro-ph},
       adsurl = {https://ui.adsabs.harvard.edu/abs/1999PASP..111...63F}
}

@ARTICLE{Aschenbrenneretal23,
       author = {{Aschenbrenner}, P. and {Przybilla}, N. and {Butler}, K.},
        title = "{Quantitative spectroscopy of late O-type main-sequence stars with a hybrid non-LTE method}",
      journal = {\aap},
         year = 2023,
        month = mar,
       volume = {671},
          eid = {A36},
        pages = {A36},
          doi = {10.1051/0004-6361/202244906},
archivePrefix = {arXiv},
       eprint = {2301.09462},
 primaryClass = {astro-ph.SR},
       adsurl = {https://ui.adsabs.harvard.edu/abs/2023A&A...671A..36A}
}

@ARTICLE{Wessmayeretal2023,
       author = {{We{\ss}mayer}, D. and {Przybilla}, N. and {Ebenbichler}, A. and {Aschenbrenner}, P. and {Butler}, K.},
        title = "{The blue supergiant Sher 25 revisited in the Gaia era}",
      journal = {\aap},
         year = 2023,
        month = sep,
       volume = {677},
          eid = {A175},
        pages = {A175},
          doi = {10.1051/0004-6361/202347253},
archivePrefix = {arXiv},
       eprint = {2308.06164},
 primaryClass = {astro-ph.SR},
       adsurl = {https://ui.adsabs.harvard.edu/abs/2023A&A...677A.175W}
}

@ARTICLE{Pavlovskietal18,
       author = {{Pavlovski}, K. and {Southworth}, J. and {Tamajo}, E.},
        title = "{Physical properties and CNO abundances for high-mass stars in four main-sequence detached eclipsing binaries: V478 Cyg, AH Cep, V453 Cyg, and V578 Mon}",
      journal = {\mnras},
         year = 2018,
        month = dec,
       volume = {481},
       number = {3},
        pages = {3129-3147},
          doi = {10.1093/mnras/sty2516},
archivePrefix = {arXiv},
       eprint = {1809.04061},
 primaryClass = {astro-ph.SR},
       adsurl = {https://ui.adsabs.harvard.edu/abs/2018MNRAS.481.3129P}
}

@ARTICLE{Wessmayeretal2022,
       author = {{We{\ss}mayer}, D. and {Przybilla}, N. and {Butler}, K.},
        title = "{Quantitative spectroscopy of B-type supergiants}",
      journal = {\aap},
         year = 2022,
        month = dec,
       volume = {668},
          eid = {A92},
        pages = {A92},
          doi = {10.1051/0004-6361/202243973},
archivePrefix = {arXiv},
       eprint = {2208.02692},
 primaryClass = {astro-ph.SR},
       adsurl = {https://ui.adsabs.harvard.edu/abs/2022A&A...668A..92W}
}

@misc{paszkeetal2019,
      title={PyTorch: An Imperative Style, High-Performance Deep Learning Library}, 
      author={Adam Paszke and Sam Gross and Francisco Massa and Adam Lerer and James Bradbury and Gregory Chanan and Trevor Killeen and Zeming Lin and Natalia Gimelshein and Luca Antiga and Alban Desmaison and Andreas Köpf and Edward Yang and Zach DeVito and Martin Raison and Alykhan Tejani and Sasank Chilamkurthy and Benoit Steiner and Lu Fang and Junjie Bai and Soumith Chintala},
      year={2019},
      eprint={1912.01703},
      archivePrefix={arXiv},
      primaryClass={cs.LG},
      url={https://arxiv.org/abs/1912.01703}, 
}

@ARTICLE{Pulsetal20,
       author = {{Puls}, J. and {Najarro}, F. and {Sundqvist}, J.~O. and {Sen}, K.},
        title = "{Atmospheric NLTE models for the spectroscopic analysis of blue stars with winds. V. Complete comoving frame transfer, and updated modeling of X-ray emission}",
      journal = {\aap},
         year = 2020,
        month = oct,
       volume = {642},
          eid = {A172},
        pages = {A172},
          doi = {10.1051/0004-6361/202038464},
archivePrefix = {arXiv},
       eprint = {2011.02310},
 primaryClass = {astro-ph.SR},
       adsurl = {https://ui.adsabs.harvard.edu/abs/2020A&A...642A.172P}
}

@ARTICLE{Najarroetal06,
       author = {{Najarro}, F. and {Hillier}, D.~J. and {Puls}, J. and {Lanz}, T. and {Martins}, F.},
        title = "{On the sensitivity of He I singlet lines to the Fe IV model atom in O stars}",
      journal = {\aap},
         year = 2006,
        month = sep,
       volume = {456},
       number = {2},
        pages = {659-664},
          doi = {10.1051/0004-6361:20054489},
archivePrefix = {arXiv},
       eprint = {astro-ph/0605211},
 primaryClass = {astro-ph},
       adsurl = {https://ui.adsabs.harvard.edu/abs/2006A&A...456..659N}
}

@ARTICLE{Aschenbrenneretal24,
       author = {{Aschenbrenner}, P. and {Przybilla}, N.},
        title = "{Quantitative spectroscopy of multiple OB stars: I. The quadruple system HD 37061 at the centre of Messier 43}",
      journal = {\aap},
         year = 2024,
        month = nov,
       volume = {691},
          eid = {A361},
        pages = {A361},
          doi = {10.1051/0004-6361/202451878},
archivePrefix = {arXiv},
       eprint = {2410.23229},
 primaryClass = {astro-ph.SR},
       adsurl = {https://ui.adsabs.harvard.edu/abs/2024A&A...691A.361A}
}

@book{Gray05,
       author = {{Gray}, D.~F.},
        title = "{The Observation and Analysis of Stellar Photospheres}",
     edition  = "3rd",
         year = 2005,
        month = sep,
    publisher = "Cambridge: Cambridge University Press",
       adsurl = {https://ui.adsabs.harvard.edu/abs/2005oasp.book.....G}
}

@article{Horniketal89,
title = {Multilayer feedforward networks are universal approximators},
journal = {Neural Netw},
volume = {2},
number = {5},
pages = {359-366},
year = {1989},
issn = {0893-6080},
doi = {https://doi.org/10.1016/0893-6080(89)90020-8},
url = {https://www.sciencedirect.com/science/article/pii/0893608089900208},
author = {Kurt Hornik and Maxwell Stinchcombe and Halbert White}
}

@ARTICLE{Metropolisetal53,
       author = {{Metropolis}, Nicholas and {Rosenbluth}, Arianna W. and {Rosenbluth}, Marshall N. and {Teller}, Augusta H. and {Teller}, Edward},
        title = "{Equation of State Calculations by Fast Computing Machines}",
      journal = {\jcp},
         year = 1953,
        month = jun,
       volume = {21},
       number = {6},
        pages = {1087-1092},
          doi = {10.1063/1.1699114},
       adsurl = {https://ui.adsabs.harvard.edu/abs/1953JChPh..21.1087M}
}

@ARTICLE{Irrgangetal22,
       author = {{Irrgang}, Andreas and {Przybilla}, Norbert and {Meynet}, Georges},
        title = "{{\ensuremath{\gamma}} Columbae as a recently stripped pulsating core of a massive star}",
      journal = {Nature Astronomy},
         year = 2022,
        month = dec,
       volume = {6},
        pages = {1414-1420},
          doi = {10.1038/s41550-022-01809-6},
archivePrefix = {arXiv},
       eprint = {2211.00358},
 primaryClass = {astro-ph.SR},
       adsurl = {https://ui.adsabs.harvard.edu/abs/2022NatAs...6.1414I}
}

@ARTICLE{Garciaetal14,
       author = {{Garcia}, E.~V. and {Stassun}, Keivan G. and {Pavlovski}, K. and {Hensberge}, H. and {G{\'o}mez Maqueo Chew}, Y. and {Claret}, A.},
        title = "{A Strict Test of Stellar Evolution Models: The Absolute Dimensions of the Massive Benchmark Eclipsing Binary V578 Mon}",
      journal = {\aj},
         year = 2014,
        month = sep,
       volume = {148},
       number = {3},
          eid = {39},
        pages = {39},
          doi = {10.1088/0004-6256/148/3/39},
archivePrefix = {arXiv},
       eprint = {1405.0739},
 primaryClass = {astro-ph.SR},
       adsurl = {https://ui.adsabs.harvard.edu/abs/2014AJ....148...39G}
}

@misc{kingma2017,
      title={Adam: A Method for Stochastic Optimization}, 
      author={Diederik P. Kingma and Jimmy Ba},
      year={2017},
      eprint={1412.6980},
      archivePrefix={arXiv},
      primaryClass={cs.LG},
      url={https://arxiv.org/abs/1412.6980}, 
}

@ARTICLE{Cutrietal21,
       author = {{Cutri}, R.~M. and {Wright}, E.~L. and {Conrow}, T. and {Fowler}, J.~W. and {Eisenhardt}, P.~R.~M. and {Grillmair}, C. and {Kirkpatrick}, J.~D. and {Masci}, F. and {McCallon}, H.~L. and {Wheelock}, S.~L. and {Fajardo-Acosta}, S. and {Yan}, L. and {Benford}, D. and {Harbut}, M. and {Jarrett}, T. and {Lake}, S. and {Leisawitz}, D. and {Ressler}, M.~E. and {Stanford}, S.~A. and {Tsai}, C. -W. and {Liu}, F. and {Helou}, G. and {Mainzer}, A. and {Gettngs}, D. and {Gonzalez}, A. and {Hoffman}, D. and {Marsh}, K.~A. and {Padgett}, D. and {Skrutskie}, M.~F. and {Beck}, R. and {Papin}, M. and {Wittman}, M.},
        title = "{VizieR Online Data Catalog: AllWISE Data Release (Cutri+ 2013)}",
      journal = {VizieR Online Data Catalog},
         year = 2021,
        month = feb,
          eid = {II/328},
        pages = {II/328},
       adsurl = {https://ui.adsabs.harvard.edu/abs/2014yCat.2328....0C}
}

@ARTICLE{Gaia2023,
       author = {{Gaia Collaboration} and {Vallenari}, A. and {Brown}, A.~G.~A. and {Prusti}, T. and {de Bruijne}, J.~H.~J. and {Arenou}, F. and {Babusiaux}, C. and {Biermann}, M. and {Creevey}, O.~L. and {Ducourant}, C. and {Evans}, D.~W. and {Eyer}, L. and {Guerra}, R. and {Hutton}, A. and {Jordi}, C. and {Klioner}, S.~A. and {Lammers}, U.~L. and {Lindegren}, L. and {Luri}, X. and {Mignard}, F. and {Panem}, C. and {Pourbaix}, D. and {Randich}, S. and {Sartoretti}, P. and {Soubiran}, C. and {Tanga}, P. and {Walton}, N.~A. and {Bailer-Jones}, C.~A.~L. and {Bastian}, U. and {Drimmel}, R. and {Jansen}, F. and {Katz}, D. and {Lattanzi}, M.~G. and {van Leeuwen}, F. and {Bakker}, J. and {Cacciari}, C. and {Casta{\~n}eda}, J. and {De Angeli}, F. and {Fabricius}, C. and {Fouesneau}, M. and {Fr{\'e}mat}, Y. and {Galluccio}, L. and {Guerrier}, A. and {Heiter}, U. and {Masana}, E. and {Messineo}, R. and {Mowlavi}, N. and {Nicolas}, C. and {Nienartowicz}, K. and {Pailler}, F. and {Panuzzo}, P. and {Riclet}, F. and {Roux}, W. and {Seabroke}, G.~M. and {Sordo}, R. and {Th{\'e}venin}, F. and {Gracia-Abril}, G. and {Portell}, J. and {Teyssier}, D. and {Altmann}, M. and {Andrae}, R. and {Audard}, M. and {Bellas-Velidis}, I. and {Benson}, K. and {Berthier}, J. and {Blomme}, R. and {Burgess}, P.~W. and {Busonero}, D. and {Busso}, G. and {C{\'a}novas}, H. and {Carry}, B. and {Cellino}, A. and {Cheek}, N. and {Clementini}, G. and {Damerdji}, Y. and {Davidson}, M. and {de Teodoro}, P. and {Nu{\~n}ez Campos}, M. and {Delchambre}, L. and {Dell'Oro}, A. and {Esquej}, P. and {Fern{\'a}ndez-Hern{\'a}ndez}, J. and {Fraile}, E. and {Garabato}, D. and {Garc{\'\i}a-Lario}, P. and {Gosset}, E. and {Haigron}, R. and {Halbwachs}, J. -L. and {Hambly}, N.~C. and {Harrison}, D.~L. and {Hern{\'a}ndez}, J. and {Hestroffer}, D. and {Hodgkin}, S.~T. and {Holl}, B. and {Jan{\ss}en}, K. and {Jevardat de Fombelle}, G. and {Jordan}, S. and {Krone-Martins}, A. and {Lanzafame}, A.~C. and {L{\"o}ffler}, W. and {Marchal}, O. and {Marrese}, P.~M. and {Moitinho}, A. and {Muinonen}, K. and {Osborne}, P. and {Pancino}, E. and {Pauwels}, T. and {Recio-Blanco}, A. and {Reyl{\'e}}, C. and {Riello}, M. and {Rimoldini}, L. and {Roegiers}, T. and {Rybizki}, J. and {Sarro}, L.~M. and {Siopis}, C. and {Smith}, M. and {Sozzetti}, A. and {Utrilla}, E. and {van Leeuwen}, M. and {Abbas}, U. and {{\'A}brah{\'a}m}, P. and {Abreu Aramburu}, A. and {Aerts}, C. and {Aguado}, J.~J. and {Ajaj}, M. and {Aldea-Montero}, F. and {Altavilla}, G. and {{\'A}lvarez}, M.~A. and {Alves}, J. and {Anders}, F. and {Anderson}, R.~I. and {Anglada Varela}, E. and {Antoja}, T. and {Baines}, D. and {Baker}, S.~G. and {Balaguer-N{\'u}{\~n}ez}, L. and {Balbinot}, E. and {Balog}, Z. and {Barache}, C. and {Barbato}, D. and {Barros}, M. and {Barstow}, M.~A. and {Bartolom{\'e}}, S. and {Bassilana}, J. -L. and {Bauchet}, N. and {Becciani}, U. and {Bellazzini}, M. and {Berihuete}, A. and {Bernet}, M. and {Bertone}, S. and {Bianchi}, L. and {Binnenfeld}, A. and {Blanco-Cuaresma}, S. and {Blazere}, A. and {Boch}, T. and {Bombrun}, A. and {Bossini}, D. and {Bouquillon}, S. and {Bragaglia}, A. and {Bramante}, L. and {Breedt}, E. and {Bressan}, A. and {Brouillet}, N. and {Brugaletta}, E. and {Bucciarelli}, B. and {Burlacu}, A. and {Butkevich}, A.~G. and {Buzzi}, R. and {Caffau}, E. and {Cancelliere}, R. and {Cantat-Gaudin}, T. and {Carballo}, R. and {Carlucci}, T. and {Carnerero}, M.~I. and {Carrasco}, J.~M. and {Casamiquela}, L. and {Castellani}, M. and {Castro-Ginard}, A. and {Chaoul}, L. and {Charlot}, P. and {Chemin}, L. and {Chiaramida}, V. and {Chiavassa}, A. and {Chornay}, N. and {Comoretto}, G. and {Contursi}, G. and {Cooper}, W.~J. and {Cornez}, T. and {Cowell}, S. and {Crifo}, F. and {Cropper}, M. and {Crosta}, M. and {Crowley}, C. and {Dafonte}, C. and {Dapergolas}, A. and {David}, M. and {David}, P. and {de Laverny}, P. and {De Luise}, F. and {De March}, R. and {De Ridder}, J. and {de Souza}, R. and {de Torres}, A. and {del Peloso}, E.~F. and {del Pozo}, E. and {Delbo}, M. and {Delgado}, A. and {Delisle}, J. -B. and {Demouchy}, C. and {Dharmawardena}, T.~E. and {Di Matteo}, P. and {Diakite}, S. and {Diener}, C. and {Distefano}, E. and {Dolding}, C. and {Edvardsson}, B. and {Enke}, H. and {Fabre}, C. and {Fabrizio}, M. and {Faigler}, S. and {Fedorets}, G. and {Fernique}, P. and {Fienga}, A. and {Figueras}, F. and {Fournier}, Y. and {Fouron}, C. and {Fragkoudi}, F. and {Gai}, M. and {Garcia-Gutierrez}, A. and {Garcia-Reinaldos}, M. and {Garc{\'\i}a-Torres}, M. and {Garofalo}, A. and {Gavel}, A. and {Gavras}, P. and {Gerlach}, E. and {Geyer}, R. and {Giacobbe}, P. and {Gilmore}, G. and {Girona}, S. and {Giuffrida}, G. and {Gomel}, R. and {Gomez}, A. and {Gonz{\'a}lez-N{\'u}{\~n}ez}, J. and {Gonz{\'a}lez-Santamar{\'\i}a}, I. and {Gonz{\'a}lez-Vidal}, J.~J. and {Granvik}, M. and {Guillout}, P. and {Guiraud}, J. and {Guti{\'e}rrez-S{\'a}nchez}, R. and {Guy}, L.~P. and {Hatzidimitriou}, D. and {Hauser}, M. and {Haywood}, M. and {Helmer}, A. and {Helmi}, A. and {Sarmiento}, M.~H. and {Hidalgo}, S.~L. and {Hilger}, T. and {H{\l}adczuk}, N. and {Hobbs}, D. and {Holland}, G. and {Huckle}, H.~E. and {Jardine}, K. and {Jasniewicz}, G. and {Jean-Antoine Piccolo}, A. and {Jim{\'e}nez-Arranz}, {\'O}. and {Jorissen}, A. and {Juaristi Campillo}, J. and {Julbe}, F. and {Karbevska}, L. and {Kervella}, P. and {Khanna}, S. and {Kontizas}, M. and {Kordopatis}, G. and {Korn}, A.~J. and {K{\'o}sp{\'a}l}, {\'A}. and {Kostrzewa-Rutkowska}, Z. and {Kruszy{\'n}ska}, K. and {Kun}, M. and {Laizeau}, P. and {Lambert}, S. and {Lanza}, A.~F. and {Lasne}, Y. and {Le Campion}, J. -F. and {Lebreton}, Y. and {Lebzelter}, T. and {Leccia}, S. and {Leclerc}, N. and {Lecoeur-Taibi}, I. and {Liao}, S. and {Licata}, E.~L. and {Lindstr{\o}m}, H.~E.~P. and {Lister}, T.~A. and {Livanou}, E. and {Lobel}, A. and {Lorca}, A. and {Loup}, C. and {Madrero Pardo}, P. and {Magdaleno Romeo}, A. and {Managau}, S. and {Mann}, R.~G. and {Manteiga}, M. and {Marchant}, J.~M. and {Marconi}, M. and {Marcos}, J. and {Marcos Santos}, M.~M.~S. and {Mar{\'\i}n Pina}, D. and {Marinoni}, S. and {Marocco}, F. and {Marshall}, D.~J. and {Martin Polo}, L. and {Mart{\'\i}n-Fleitas}, J.~M. and {Marton}, G. and {Mary}, N. and {Masip}, A. and {Massari}, D. and {Mastrobuono-Battisti}, A. and {Mazeh}, T. and {McMillan}, P.~J. and {Messina}, S. and {Michalik}, D. and {Millar}, N.~R. and {Mints}, A. and {Molina}, D. and {Molinaro}, R. and {Moln{\'a}r}, L. and {Monari}, G. and {Mongui{\'o}}, M. and {Montegriffo}, P. and {Montero}, A. and {Mor}, R. and {Mora}, A. and {Morbidelli}, R. and {Morel}, T. and {Morris}, D. and {Muraveva}, T. and {Murphy}, C.~P. and {Musella}, I. and {Nagy}, Z. and {Noval}, L. and {Oca{\~n}a}, F. and {Ogden}, A. and {Ordenovic}, C. and {Osinde}, J.~O. and {Pagani}, C. and {Pagano}, I. and {Palaversa}, L. and {Palicio}, P.~A. and {Pallas-Quintela}, L. and {Panahi}, A. and {Payne-Wardenaar}, S. and {Pe{\~n}alosa Esteller}, X. and {Penttil{\"a}}, A. and {Pichon}, B. and {Piersimoni}, A.~M. and {Pineau}, F. -X. and {Plachy}, E. and {Plum}, G. and {Poggio}, E. and {Pr{\v{s}}a}, A. and {Pulone}, L. and {Racero}, E. and {Ragaini}, S. and {Rainer}, M. and {Raiteri}, C.~M. and {Rambaux}, N. and {Ramos}, P. and {Ramos-Lerate}, M. and {Re Fiorentin}, P. and {Regibo}, S. and {Richards}, P.~J. and {Rios Diaz}, C. and {Ripepi}, V. and {Riva}, A. and {Rix}, H. -W. and {Rixon}, G. and {Robichon}, N. and {Robin}, A.~C. and {Robin}, C. and {Roelens}, M. and {Rogues}, H.~R.~O. and {Rohrbasser}, L. and {Romero-G{\'o}mez}, M. and {Rowell}, N. and {Royer}, F. and {Ruz Mieres}, D. and {Rybicki}, K.~A. and {Sadowski}, G. and {S{\'a}ez N{\'u}{\~n}ez}, A. and {Sagrist{\`a} Sell{\'e}s}, A. and {Sahlmann}, J. and {Salguero}, E. and {Samaras}, N. and {Sanchez Gimenez}, V. and {Sanna}, N. and {Santove{\~n}a}, R. and {Sarasso}, M. and {Schultheis}, M. and {Sciacca}, E. and {Segol}, M. and {Segovia}, J.~C. and {S{\'e}gransan}, D. and {Semeux}, D. and {Shahaf}, S. and {Siddiqui}, H.~I. and {Siebert}, A. and {Siltala}, L. and {Silvelo}, A. and {Slezak}, E. and {Slezak}, I. and {Smart}, R.~L. and {Snaith}, O.~N. and {Solano}, E. and {Solitro}, F. and {Souami}, D. and {Souchay}, J. and {Spagna}, A. and {Spina}, L. and {Spoto}, F. and {Steele}, I.~A. and {Steidelm{\"u}ller}, H. and {Stephenson}, C.~A. and {S{\"u}veges}, M. and {Surdej}, J. and {Szabados}, L. and {Szegedi-Elek}, E. and {Taris}, F. and {Taylor}, M.~B. and {Teixeira}, R. and {Tolomei}, L. and {Tonello}, N. and {Torra}, F. and {Torra}, J. and {Torralba Elipe}, G. and {Trabucchi}, M. and {Tsounis}, A.~T. and {Turon}, C. and {Ulla}, A. and {Unger}, N. and {Vaillant}, M.~V. and {van Dillen}, E. and {van Reeven}, W. and {Vanel}, O. and {Vecchiato}, A. and {Viala}, Y. and {Vicente}, D. and {Voutsinas}, S. and {Weiler}, M. and {Wevers}, T. and {Wyrzykowski}, {\L}. and {Yoldas}, A. and {Yvard}, P. and {Zhao}, H. and {Zorec}, J. and {Zucker}, S. and {Zwitter}, T.},
        title = "{Gaia Data Release 3. Summary of the content and survey properties}",
      journal = {\aap},
         year = 2023,
        month = jun,
       volume = {674},
          eid = {A1},
        pages = {A1},
          doi = {10.1051/0004-6361/202243940},
archivePrefix = {arXiv},
       eprint = {2208.00211},
 primaryClass = {astro-ph.GA},
       adsurl = {https://ui.adsabs.harvard.edu/abs/2023A&A...674A...1G}
}

@ARTICLE{Tingetal2019,
       author = {{Ting}, Yuan-Sen and {Conroy}, Charlie and {Rix}, Hans-Walter and {Cargile}, Phillip},
        title = "{The Payne: Self-consistent ab initio Fitting of Stellar Spectra}",
      journal = {\apj},
         year = 2019,
        month = jul,
       volume = {879},
       number = {2},
          eid = {69},
        pages = {69},
          doi = {10.3847/1538-4357/ab2331},
archivePrefix = {arXiv},
       eprint = {1804.01530},
 primaryClass = {astro-ph.SR},
       adsurl = {https://ui.adsabs.harvard.edu/abs/2019ApJ...879...69T}
}

@ARTICLE{Xiangetal2022,
       author = {{Xiang}, Maosheng and {Rix}, Hans-Walter and {Ting}, Yuan-Sen and {Kudritzki}, Rolf-Peter and {Conroy}, Charlie and {Zari}, Eleonora and {Shi}, Jian-Rong and {Przybilla}, Norbert and {Ramirez-Tannus}, Maria and {Tkachenko}, Andrew and {Gebruers}, Sarah and {Liu}, Xiao-Wei},
        title = "{Stellar labels for hot stars from low-resolution spectra. I. The HotPayne method and results for 330 000 stars from LAMOST DR6}",
      journal = {\aap},
         year = 2022,
        month = jun,
       volume = {662},
          eid = {A66},
        pages = {A66},
          doi = {10.1051/0004-6361/202141570},
archivePrefix = {arXiv},
       eprint = {2108.02878},
 primaryClass = {astro-ph.SR},
       adsurl = {https://ui.adsabs.harvard.edu/abs/2022A&A...662A..66X}
}

@ARTICLE{Aschenbrenneretal25,
       author = {{Aschenbrenner}, P. and {Butler}, K. and {Przybilla}, N.},
        title = "{The present-day cosmic phosphorus abundance}",
      journal = {\aap},
         year = 2025,
        month = jun,
       volume = {698},
          eid = {A164},
        pages = {A164},
          doi = {10.1051/0004-6361/202554356},
archivePrefix = {arXiv},
       eprint = {2505.08428},
 primaryClass = {astro-ph.SR},
       adsurl = {https://ui.adsabs.harvard.edu/abs/2025A&A...698A.164A}
}

@PHDTHESIS{Payne-Gaposchkin25,
       author = {{Payne-Gaposchkin}, Cecilia Helena},
        title = "{Stellar atmospheres: A contribution to the observational study of high temperature in the reversing layers of stars}",
       school = {(Radcliffe College, Cambridge, MA)},
         year = 1925,
        month = jan,
       adsurl = {https://ui.adsabs.harvard.edu/abs/1925PhDT.........6P}
}

@ARTICLE{Gerasimovic29,
       author = {{Gerasimovi{\v{c}}}, B.~P.},
        title = "{Note on the deviation of stellar atmospheres from thermodynamic equilibrium}",
      journal = {\mnras},
         year = 1929,
        month = jan,
       volume = {89},
        pages = {272},
          doi = {10.1093/mnras/89.3.272},
       adsurl = {https://ui.adsabs.harvard.edu/abs/1929MNRAS..89..272G}
}

@BOOK{HuMi15,
       author = {{Hubeny}, Ivan and {Mihalas}, Dimitri},
        title = "{Theory of Stellar Atmospheres}",
    publisher = {Princeton University Press, Princeton},
         year = 2015,
       adsurl = {https://ui.adsabs.harvard.edu/abs/2015tsaa.book.....H}
}

@ARTICLE{HaGr03,
       author = {{Hamann}, W. -R. and {Gr{\"a}fener}, G.},
        title = "{A temperature correction method for expanding atmospheres}",
      journal = {\aap},
         year = 2003,
        month = nov,
       volume = {410},
        pages = {993-1000},
          doi = {10.1051/0004-6361:20031308},
       adsurl = {https://ui.adsabs.harvard.edu/abs/2003A&A...410..993H}
}

@ARTICLE{Sanderetal15,
       author = {{Sander}, A. and {Shenar}, T. and {Hainich}, R. and {G{\'\i}menez-Garc{\'\i}a}, A. and {Todt}, H. and {Hamann}, W. -R.},
        title = "{On the consistent treatment of the quasi-hydrostatic layers in hot star atmospheres}",
      journal = {\aap},
         year = 2015,
        month = may,
       volume = {577},
          eid = {A13},
        pages = {A13},
          doi = {10.1051/0004-6361/201425356},
archivePrefix = {arXiv},
       eprint = {1503.01338},
 primaryClass = {astro-ph.SR},
       adsurl = {https://ui.adsabs.harvard.edu/abs/2015A&A...577A..13S}
}

@ARTICLE{KrKu17,
       author = {{Krti{\v{c}}ka}, J. and {Kub{\'a}t}, J.},
        title = "{Comoving frame models of hot star winds. II. Reduction of O star wind mass-loss rates in global models}",
      journal = {\aap},
         year = 2017,
        month = oct,
       volume = {606},
          eid = {A31},
        pages = {A31},
          doi = {10.1051/0004-6361/201730723},
archivePrefix = {arXiv},
       eprint = {1706.06194},
 primaryClass = {astro-ph.SR},
       adsurl = {https://ui.adsabs.harvard.edu/abs/2017A&A...606A..31K}
}

@ARTICLE{Evansetal05,
       author = {{Evans}, C.~J. and {Smartt}, S.~J. and {Lee}, J. -K. and {Lennon}, D.~J. and {Kaufer}, A. and {Dufton}, P.~L. and {Trundle}, C. and {Herrero}, A. and {Sim{\'o}n-D{\'\i}az}, S. and {de Koter}, A. and {Hamann}, W. -R. and {Hendry}, M.~A. and {Hunter}, I. and {Irwin}, M.~J. and {Korn}, A.~J. and {Kudritzki}, R. -P. and {Langer}, N. and {Mokiem}, M.~R. and {Najarro}, F. and {Pauldrach}, A.~W.~A. and {Przybilla}, N. and {Puls}, J. and {Ryans}, R.~S.~I. and {Urbaneja}, M.~A. and {Venn}, K.~A. and {Villamariz}, M.~R.},
        title = "{The VLT-FLAMES survey of massive stars: Observations in the Galactic clusters NGC 3293, NGC 4755 and NGC 6611}",
      journal = {\aap},
         year = 2005,
        month = jul,
       volume = {437},
       number = {2},
        pages = {467-482},
          doi = {10.1051/0004-6361:20042446},
archivePrefix = {arXiv},
       eprint = {astro-ph/0503655},
 primaryClass = {astro-ph},
       adsurl = {https://ui.adsabs.harvard.edu/abs/2005A&A...437..467E}
}

@ARTICLE{Evansetal11,
       author = {{Evans}, C.~J. and {Taylor}, W.~D. and {H{\'e}nault-Brunet}, V. and {Sana}, H. and {de Koter}, A. and {Sim{\'o}n-D{\'\i}az}, S. and {Carraro}, G. and {Bagnoli}, T. and {Bastian}, N. and {Bestenlehner}, J.~M. and {Bonanos}, A.~Z. and {Bressert}, E. and {Brott}, I. and {Campbell}, M.~A. and {Cantiello}, M. and {Clark}, J.~S. and {Costa}, E. and {Crowther}, P.~A. and {de Mink}, S.~E. and {Doran}, E. and {Dufton}, P.~L. and {Dunstall}, P.~R. and {Friedrich}, K. and {Garcia}, M. and {Gieles}, M. and {Gr{\"a}fener}, G. and {Herrero}, A. and {Howarth}, I.~D. and {Izzard}, R.~G. and {Langer}, N. and {Lennon}, D.~J. and {Ma{\'\i}z Apell{\'a}niz}, J. and {Markova}, N. and {Najarro}, F. and {Puls}, J. and {Ramirez}, O.~H. and {Sab{\'\i}n-Sanjuli{\'a}n}, C. and {Smartt}, S.~J. and {Stroud}, V.~E. and {van Loon}, J. Th. and {Vink}, J.~S. and {Walborn}, N.~R.},
        title = "{The VLT-FLAMES Tarantula Survey. I. Introduction and observational overview}",
      journal = {\aap},
         year = 2011,
        month = jun,
       volume = {530},
          eid = {A108},
        pages = {A108},
          doi = {10.1051/0004-6361/201116782},
archivePrefix = {arXiv},
       eprint = {1103.5386},
 primaryClass = {astro-ph.SR},
       adsurl = {https://ui.adsabs.harvard.edu/abs/2011A&A...530A.108E}
}

@ARTICLE{Blommeetal22,
       author = {{Blomme}, R. and {Daflon}, S. and {Gebran}, M. and {Herrero}, A. and {Lobel}, A. and {Mahy}, L. and {Martins}, F. and {Morel}, T. and {Berlanas}, S.~R. and {Blaz{\`e}re}, A. and {Fr{\'e}mat}, Y. and {Gosset}, E. and {Ma{\'\i}z Apell{\'a}niz}, J. and {Santos}, W. and {Semaan}, T. and {Sim{\'o}n-D{\'\i}az}, S. and {Volpi}, D. and {Holgado}, G. and {Jim{\'e}nez-Esteban}, F. and {Nieva}, M.~F. and {Przybilla}, N. and {Gilmore}, G. and {Randich}, S. and {Negueruela}, I. and {Prusti}, T. and {Vallenari}, A. and {Alfaro}, E.~J. and {Bensby}, T. and {Bragaglia}, A. and {Flaccomio}, E. and {Francois}, P. and {Korn}, A.~J. and {Lanzafame}, A. and {Pancino}, E. and {Smiljanic}, R. and {Bergemann}, M. and {Carraro}, G. and {Franciosini}, E. and {Gonneau}, A. and {Heiter}, U. and {Hourihane}, A. and {Jofr{\'e}}, P. and {Magrini}, L. and {Morbidelli}, L. and {Sacco}, G.~G. and {Worley}, C.~C. and {Zaggia}, S.},
        title = "{The Gaia-ESO Survey: The analysis of the hot-star spectra}",
      journal = {\aap},
         year = 2022,
        month = may,
       volume = {661},
          eid = {A120},
        pages = {A120},
          doi = {10.1051/0004-6361/202142349},
archivePrefix = {arXiv},
       eprint = {2202.08662},
 primaryClass = {astro-ph.SR},
       adsurl = {https://ui.adsabs.harvard.edu/abs/2022A&A...661A.120B}
}

@ARTICLE{Moreletal22,
       author = {{Morel}, T. and {Blaz{\`e}re}, A. and {Semaan}, T. and {Gosset}, E. and {Zorec}, J. and {Fr{\'e}mat}, Y. and {Blomme}, R. and {Daflon}, S. and {Lobel}, A. and {Nieva}, M.~F. and {Przybilla}, N. and {Gebran}, M. and {Herrero}, A. and {Mahy}, L. and {Santos}, W. and {Tautvai{\v{s}}ien{\.{e}}}, G. and {Gilmore}, G. and {Randich}, S. and {Alfaro}, E.~J. and {Bergemann}, M. and {Carraro}, G. and {Damiani}, F. and {Franciosini}, E. and {Morbidelli}, L. and {Pancino}, E. and {Worley}, C.~C. and {Zaggia}, S.},
        title = "{The Gaia-ESO survey: A spectroscopic study of the young open cluster NGC 3293}",
      journal = {\aap},
         year = 2022,
        month = sep,
       volume = {665},
          eid = {A108},
        pages = {A108},
          doi = {10.1051/0004-6361/202244112},
archivePrefix = {arXiv},
       eprint = {2207.12792},
 primaryClass = {astro-ph.SR},
       adsurl = {https://ui.adsabs.harvard.edu/abs/2022A&A...665A.108M}
}

@ARTICLE{Castroetal18,
       author = {{Castro}, N. and {Crowther}, P.~A. and {Evans}, C.~J. and {Mackey}, J. and {Castro-Rodriguez}, N. and {Vink}, J.~S. and {Melnick}, J. and {Selman}, F.},
        title = "{Mapping the core of the Tarantula Nebula with VLT-MUSE. I. Spectral and nebular content around R136}",
      journal = {\aap},
         year = 2018,
        month = jun,
       volume = {614},
          eid = {A147},
        pages = {A147},
          doi = {10.1051/0004-6361/201732084},
archivePrefix = {arXiv},
       eprint = {1802.01597},
 primaryClass = {astro-ph.GA},
       adsurl = {https://ui.adsabs.harvard.edu/abs/2018A&A...614A.147C}
}

@ARTICLE{Evansetal19,
       author = {{Evans}, C.~J. and {Castro}, N. and {Gonzalez}, O.~A. and {Garcia}, M. and {Bastian}, N. and {Cioni}, M. -R.~L. and {Clark}, J.~S. and {Davies}, B. and {Ferguson}, A.~M.~N. and {Kamann}, S. and {Lennon}, D.~J. and {Patrick}, L.~R. and {Vink}, J.~S. and {Weisz}, D.~R.},
        title = "{First stellar spectroscopy in Leo P}",
      journal = {\aap},
         year = 2019,
        month = feb,
       volume = {622},
          eid = {A129},
        pages = {A129},
          doi = {10.1051/0004-6361/201834145},
archivePrefix = {arXiv},
       eprint = {1901.01295},
 primaryClass = {astro-ph.SR},
       adsurl = {https://ui.adsabs.harvard.edu/abs/2019A&A...622A.129E}
}

@ARTICLE{Jinetal24,
       author = {{Jin}, Shoko and {Trager}, Scott C. and {Dalton}, Gavin B. and {Aguerri}, J. Alfonso L. and {Drew}, J.~E. and {Falc{\'o}n-Barroso}, Jes{\'u}s and {G{\"a}nsicke}, Boris T. and {Hill}, Vanessa and {Iovino}, Angela and {Pieri}, Matthew M. and {Poggianti}, Bianca M. and {Smith}, D.~J.~B. and {Vallenari}, Antonella and {Abrams}, Don Carlos and {Aguado}, David S. and {Antoja}, Teresa and {Arag{\'o}n-Salamanca}, Alfonso and {Ascasibar}, Yago and {Babusiaux}, Carine and {Balcells}, Marc and {Barrena}, R. and {Battaglia}, Giuseppina and {Belokurov}, Vasily and {Bensby}, Thomas and {Bonifacio}, Piercarlo and {Bragaglia}, Angela and {Carrasco}, Esperanza and {Carrera}, Ricardo and {Cornwell}, Daniel J. and {Dom{\'\i}nguez-Palmero}, Lilian and {Duncan}, Kenneth J. and {Famaey}, Benoit and {Fari{\~n}a}, Cecilia and {Gonzalez}, Oscar A. and {Guest}, Steve and {Hatch}, Nina A. and {Hess}, Kelley M. and {Hoskin}, Matthew J. and {Irwin}, Mike and {Knapen}, Johan H. and {Koposov}, Sergey E. and {Kuchner}, Ulrike and {Laigle}, Clotilde and {Lewis}, Jim and {Longhetti}, Marcella and {Lucatello}, Sara and {M{\'e}ndez-Abreu}, Jairo and {Mercurio}, Amata and {Molaeinezhad}, Alireza and {Mongui{\'o}}, Maria and {Morrison}, Sean and {Murphy}, David N.~A. and {Peralta de Arriba}, Luis and {P{\'e}rez}, Isabel and {P{\'e}rez-R{\`a}fols}, Ignasi and {Pic{\'o}}, Sergio and {Raddi}, Roberto and {Romero-G{\'o}mez}, Merc{\`e} and {Royer}, Fr{\'e}d{\'e}ric and {Siebert}, Arnaud and {Seabroke}, George M. and {Som}, Debopam and {Terrett}, David and {Thomas}, Guillaume and {Wesson}, Roger and {Worley}, C. Clare and {Alfaro}, Emilio J. and {Allende Prieto}, Carlos and {Alonso-Santiago}, Javier and {Amos}, Nicholas J. and {Ashley}, Richard P. and {Balaguer-N{\'u}{\~n}ez}, Lola and {Balbinot}, Eduardo and {Bellazzini}, Michele and {Benn}, Chris R. and {Berlanas}, Sara R. and {Bernard}, Edouard J. and {Best}, Philip and {Bettoni}, Daniela and {Bianco}, Andrea and {Bishop}, Georgia and {Blomqvist}, Michael and {Boeche}, Corrado and {Bolzonella}, Micol and {Bonoli}, Silvia and {Bosma}, Albert and {Britavskiy}, Nikolay and {Busarello}, Gianni and {Caffau}, Elisabetta and {Cantat-Gaudin}, Tristan and {Castro-Ginard}, Alfred and {Couto}, Guilherme and {Carbajo-Hijarrubia}, Juan and {Carter}, David and {Casamiquela}, Laia and {Conrado}, Ana M. and {Corcho-Caballero}, Pablo and {Costantin}, Luca and {Deason}, Alis and {de Burgos}, Abel and {De Grandi}, Sabrina and {Di Matteo}, Paola and {Dom{\'\i}nguez-G{\'o}mez}, Jes{\'u}s and {Dorda}, Ricardo and {Drake}, Alyssa and {Dutta}, Rajeshwari and {Erkal}, Denis and {Feltzing}, Sofia and {Ferr{\'e}-Mateu}, Anna and {Feuillet}, Diane and {Figueras}, Francesca and {Fossati}, Matteo and {Franciosini}, Elena and {Frasca}, Antonio and {Fumagalli}, Michele and {Gallazzi}, Anna and {Garc{\'\i}a-Benito}, Rub{\'e}n and {Gentile Fusillo}, Nicola and {Gebran}, Marwan and {Gilbert}, James and {Gledhill}, T.~M. and {Gonz{\'a}lez Delgado}, Rosa M. and {Greimel}, Robert and {Guarcello}, Mario Giuseppe and {Guerra}, Jose and {Gullieuszik}, Marco and {Haines}, Christopher P. and {Hardcastle}, Martin J. and {Harris}, Amy and {Haywood}, Misha and {Helmi}, Amina and {Hernandez}, Nauzet and {Herrero}, Artemio and {Hughes}, Sarah and {Ir{\v{s}}i{\v{c}}}, Vid and {Jablonka}, Pascale and {Jarvis}, Matt J. and {Jordi}, Carme and {Kondapally}, Rohit and {Kordopatis}, Georges and {Krogager}, Jens-Kristian and {La Barbera}, Francesco and {Lam}, Man I. and {Larsen}, S{\o}ren S. and {Lemasle}, Bertrand and {Lewis}, Ian J. and {Lhom{\'e}}, Emilie and {Lind}, Karin and {Lodi}, Marcello and {Longobardi}, Alessia and {Lonoce}, Ilaria and {Magrini}, Laura and {Ma{\'\i}z Apell{\'a}niz}, Jes{\'u}s and {Marchal}, Olivier and {Marco}, Amparo and {Martin}, Nicolas F. and {Matsuno}, Tadafumi and {Maurogordato}, Sophie and {Merluzzi}, Paola and {Miralda-Escud{\'e}}, Jordi and {Molinari}, Emilio and {Monari}, Giacomo and {Morelli}, Lorenzo and {Mottram}, Christopher J. and {Naylor}, Tim and {Negueruela}, Ignacio and {O{\~n}orbe}, Jose and {Pancino}, Elena and {Peirani}, S{\'e}bastien and {Peletier}, Reynier F. and {Pozzetti}, Lucia and {Rainer}, Monica and {Ramos}, Pau and {Read}, Shaun C. and {Rossi}, Elena Maria and {R{\"o}ttgering}, Huub J.~A. and {Rubi{\~n}o-Mart{\'\i}n}, Jose Alberto and {Sabater}, Jose and {San Juan}, Jos{\'e} and {Sanna}, Nicoletta and {Schallig}, Ellen and {Schiavon}, Ricardo P. and {Schultheis}, Mathias and {Serra}, Paolo and {Shimwell}, Timothy W. and {Sim{\'o}n-D{\'\i}az}, Sergio and {Smith}, Russell J. and {Sordo}, Rosanna and {Sorini}, Daniele and {Soubiran}, Caroline and {Starkenburg}, Else and {Steele}, Iain A. and {Stott}, John and {Stuik}, Remko and {Tolstoy}, Eline and {Tortora}, Crescenzo and {Tsantaki}, Maria and {Van der Swaelmen}, Mathieu and {van Weeren}, Reinout J. and {Vergani}, Daniela},
        title = "{The wide-field, multiplexed, spectroscopic facility WEAVE: Survey design, overview, and simulated implementation}",
      journal = {\mnras},
         year = 2024,
        month = may,
       volume = {530},
       number = {3},
        pages = {2688-2730},
          doi = {10.1093/mnras/stad557},
archivePrefix = {arXiv},
       eprint = {2212.03981},
 primaryClass = {astro-ph.IM},
       adsurl = {https://ui.adsabs.harvard.edu/abs/2024MNRAS.530.2688J}
}

@ARTICLE{deJongetal19,
       author = {{de Jong}, R.~S. and {Agertz}, O. and {Berbel}, A.~A. and {Aird}, J. and {Alexander}, D.~A. and {Amarsi}, A. and {Anders}, F. and {Andrae}, R. and {Ansarinejad}, B. and {Ansorge}, W. and {Antilogus}, P. and {Anwand-Heerwart}, H. and {Arentsen}, A. and {Arnadottir}, A. and {Asplund}, M. and {Auger}, M. and {Azais}, N. and {Baade}, D. and {Baker}, G. and {Baker}, S. and {Balbinot}, E. and {Baldry}, I.~K. and {Banerji}, M. and {Barden}, S. and {Barklem}, P. and {Barth{\'e}l{\'e}my-Mazot}, E. and {Battistini}, C. and {Bauer}, S. and {Bell}, C.~P.~M. and {Bellido-Tirado}, O. and {Bellstedt}, S. and {Belokurov}, V. and {Bensby}, T. and {Bergemann}, M. and {Bestenlehner}, J.~M. and {Bielby}, R. and {Bilicki}, M. and {Blake}, C. and {Bland-Hawthorn}, J. and {Boeche}, C. and {Boland}, W. and {Boller}, T. and {Bongard}, S. and {Bongiorno}, A. and {Bonifacio}, P. and {Boudon}, D. and {Brooks}, D. and {Brown}, M.~J.~I. and {Brown}, R. and {Br{\"u}ggen}, M. and {Brynnel}, J. and {Brzeski}, J. and {Buchert}, T. and {Buschkamp}, P. and {Caffau}, E. and {Caillier}, P. and {Carrick}, J. and {Casagrande}, L. and {Case}, S. and {Casey}, A. and {Cesarini}, I. and {Cescutti}, G. and {Chapuis}, D. and {Chiappini}, C. and {Childress}, M. and {Christlieb}, N. and {Church}, R. and {Cioni}, M. -R.~L. and {Cluver}, M. and {Colless}, M. and {Collett}, T. and {Comparat}, J. and {Cooper}, A. and {Couch}, W. and {Courbin}, F. and {Croom}, S. and {Croton}, D. and {Daguis{\'e}}, E. and {Dalton}, G. and {Davies}, L.~J.~M. and {Davis}, T. and {de Laverny}, P. and {Deason}, A. and {Dionies}, F. and {Disseau}, K. and {Doel}, P. and {D{\"o}scher}, D. and {Driver}, S.~P. and {Dwelly}, T. and {Eckert}, D. and {Edge}, A. and {Edvardsson}, B. and {Youssoufi}, D.~E. and {Elhaddad}, A. and {Enke}, H. and {Erfanianfar}, G. and {Farrell}, T. and {Fechner}, T. and {Feiz}, C. and {Feltzing}, S. and {Ferreras}, I. and {Feuerstein}, D. and {Feuillet}, D. and {Finoguenov}, A. and {Ford}, D. and {Fotopoulou}, S. and {Fouesneau}, M. and {Frenk}, C. and {Frey}, S. and {Gaessler}, W. and {Geier}, S. and {Gentile Fusillo}, N. and {Gerhard}, O. and {Giannantonio}, T. and {Giannone}, D. and {Gibson}, B. and {Gillingham}, P. and {Gonz{\'a}lez-Fern{\'a}ndez}, C. and {Gonzalez-Solares}, E. and {Gottloeber}, S. and {Gould}, A. and {Grebel}, E.~K. and {Gueguen}, A. and {Guiglion}, G. and {Haehnelt}, M. and {Hahn}, T. and {Hansen}, C.~J. and {Hartman}, H. and {Hauptner}, K. and {Hawkins}, K. and {Haynes}, D. and {Haynes}, R. and {Heiter}, U. and {Helmi}, A. and {Aguayo}, C.~H. and {Hewett}, P. and {Hinton}, S. and {Hobbs}, D. and {Hoenig}, S. and {Hofman}, D. and {Hook}, I. and {Hopgood}, J. and {Hopkins}, A. and {Hourihane}, A. and {Howes}, L. and {Howlett}, C. and {Huet}, T. and {Irwin}, M. and {Iwert}, O. and {Jablonka}, P. and {Jahn}, T. and {Jahnke}, K. and {Jarno}, A. and {Jin}, S. and {Jofre}, P. and {Johl}, D. and {Jones}, D. and {J{\"o}nsson}, H. and {Jordan}, C. and {Karovicova}, I. and {Khalatyan}, A. and {Kelz}, A. and {Kennicutt}, R. and {King}, D. and {Kitaura}, F. and {Klar}, J. and {Klauser}, U. and {Kneib}, J. -P. and {Koch}, A. and {Koposov}, S. and {Kordopatis}, G. and {Korn}, A. and {Kosmalski}, J. and {Kotak}, R. and {Kovalev}, M. and {Kreckel}, K. and {Kripak}, Y. and {Krumpe}, M. and {Kuijken}, K. and {Kunder}, A. and {Kushniruk}, I. and {Lam}, M.~I. and {Lamer}, G. and {Laurent}, F. and {Lawrence}, J. and {Lehmitz}, M. and {Lemasle}, B. and {Lewis}, J. and {Li}, B. and {Lidman}, C. and {Lind}, K. and {Liske}, J. and {Lizon}, J. -L. and {Loveday}, J. and {Ludwig}, H. -G. and {McDermid}, R.~M. and {Maguire}, K. and {Mainieri}, V. and {Mali}, S. and {Mandel}, H.},
        title = "{4MOST: Project overview and information for the First Call for Proposals}",
      journal = {The Messenger},
         year = 2019,
        month = mar,
       volume = {175},
        pages = {3-11},
          doi = {10.18727/0722-6691/5117},
archivePrefix = {arXiv},
       eprint = {1903.02464},
 primaryClass = {astro-ph.IM},
       adsurl = {https://ui.adsabs.harvard.edu/abs/2019Msngr.175....3D}
}

@ARTICLE{deBurgosetal24,
       author = {{de Burgos}, A. and {Sim{\'o}n-D{\'\i}az}, S. and {Urbaneja}, M.~A. and {Puls}, J.},
        title = "{The IACOB project. X. Large-scale quantitative spectroscopic analysis of Galactic luminous blue stars}",
      journal = {\aap},
         year = 2024,
        month = jul,
       volume = {687},
          eid = {A228},
        pages = {A228},
          doi = {10.1051/0004-6361/202348808},
archivePrefix = {arXiv},
       eprint = {2312.00241},
 primaryClass = {astro-ph.SR},
       adsurl = {https://ui.adsabs.harvard.edu/abs/2024A&A...687A.228D}
}

@INPROCEEDINGS{SimonDiazetal20,
       author = {{Sim{\'o}n-D{\'\i}az}, S. and {P{\'e}rez Prieto}, J.~A. and {Holgado}, G. and {de Burgos}, A. and {Iacob Team}},
        title = "{The IACOB spectroscopic database. New interface and second data release}",
    booktitle = {XIV.0 Scientific Meeting (virtual) of the Spanish Astronomical Society},
         year = 2020,
        month = jul,
          eid = {187},
        pages = {187},
       adsurl = {https://ui.adsabs.harvard.edu/abs/2020sea..confE.187S}
}

@ARTICLE{Vinketal23,
       author = {{Vink}, Jorick S. and {Mehner}, A. and {Crowther}, P.~A. and {Fullerton}, A. and {Garcia}, M. and {Martins}, F. and {Morrell}, N. and {Oskinova}, L.~M. and {St-Louis}, N. and {ud-Doula}, A. and {Sander}, A.~A.~C. and {Sana}, H. and {Bouret}, J. -C. and {Kub{\'a}tov{\'a}}, B. and {Marchant}, P. and {Martins}, L.~P. and {Wofford}, A. and {van Loon}, J. Th. and {Grace Telford}, O. and {G{\"o}tberg}, Y. and {Bowman}, D.~M. and {Erba}, C. and {Kalari}, V.~M. and {Abdul-Masih}, M. and {Alkousa}, T. and {Backs}, F. and {Barbosa}, C.~L. and {Berlanas}, S.~R. and {Bernini-Peron}, M. and {Bestenlehner}, J.~M. and {Blomme}, R. and {Bodensteiner}, J. and {Brands}, S.~A. and {Evans}, C.~J. and {David-Uraz}, A. and {Driessen}, F.~A. and {Dsilva}, K. and {Geen}, S. and {G{\'o}mez-Gonz{\'a}lez}, V.~M.~A. and {Grassitelli}, L. and {Hamann}, W. -R. and {Hawcroft}, C. and {Herrero}, A. and {Higgins}, E.~R. and {John Hillier}, D. and {Ignace}, R. and {Istrate}, A.~G. and {Kaper}, L. and {Kee}, N.~D. and {Kehrig}, C. and {Keszthelyi}, Z. and {Klencki}, J. and {de Koter}, A. and {Kuiper}, R. and {Laplace}, E. and {Larkin}, C.~J.~K. and {Lefever}, R.~R. and {Leitherer}, C. and {Lennon}, D.~J. and {Mahy}, L. and {Ma{\'\i}z Apell{\'a}niz}, J. and {Maravelias}, G. and {Marcolino}, W. and {McLeod}, A.~F. and {de Mink}, S.~E. and {Najarro}, F. and {Oey}, M.~S. and {Parsons}, T.~N. and {Pauli}, D. and {Pedersen}, M.~G. and {Prinja}, R.~K. and {Ramachandran}, V. and {Ram{\'\i}rez-Tannus}, M.~C. and {Sabhahit}, G.~N. and {Schootemeijer}, A. and {Reyero Serantes}, S. and {Shenar}, T. and {Stringfellow}, G.~S. and {Sudnik}, N. and {Tramper}, F. and {Wang}, L.},
        title = "{X-Shooting ULLYSES: Massive stars at low metallicity. I. Project description}",
      journal = {\aap},
         year = 2023,
        month = jul,
       volume = {675},
          eid = {A154},
        pages = {A154},
          doi = {10.1051/0004-6361/202245650},
archivePrefix = {arXiv},
       eprint = {2305.06376},
 primaryClass = {astro-ph.SR},
       adsurl = {https://ui.adsabs.harvard.edu/abs/2023A&A...675A.154V}
}

@ARTICLE{Chinietal12,
       author = {{Chini}, R. and {Hoffmeister}, V.~H. and {Nasseri}, A. and {Stahl}, O. and {Zinnecker}, H.},
        title = "{A spectroscopic survey on the multiplicity of high-mass stars}",
      journal = {\mnras},
         year = 2012,
        month = aug,
       volume = {424},
       number = {3},
        pages = {1925-1929},
          doi = {10.1111/j.1365-2966.2012.21317.x},
archivePrefix = {arXiv},
       eprint = {1205.5238},
 primaryClass = {astro-ph.SR},
       adsurl = {https://ui.adsabs.harvard.edu/abs/2012MNRAS.424.1925C}
}

@article{Albrechtetal2014,
doi = {10.1088/0004-637X/785/2/83},
url = {https://doi.org/10.1088/0004-637X/785/2/83},
year = {2014},
month = {mar},
publisher = {The American Astronomical Society},
volume = {785},
number = {2},
pages = {83},
author = {Albrecht, Simon and Winn, Joshua N. and Torres, Guillermo and Fabrycky, Daniel C. and Setiawan, Johny and Gillon, Michaël and Jehin, Emmanuel and Triaud, Amaury and Queloz, Didier and Snellen, Ignas and Eggleton, Peter},
title = {THE BANANA PROJECT. V. MISALIGNED AND PRECESSING STELLAR ROTATION AXES IN CV VELORUM*},
journal = {\apj}
}

@ARTICLE{Fossatietal2009,
       author = {{Fossati}, L. and {Ryabchikova}, T. and {Bagnulo}, S. and {Alecian}, E. and {Grunhut}, J. and {Kochukhov}, O. and {Wade}, G.},
        title = "{The chemical abundance analysis of normal early A- and late B-type stars}",
      journal = {\aap},
         year = 2009,
        month = sep,
       volume = {503},
       number = {3},
        pages = {945-962},
          doi = {10.1051/0004-6361/200811561},
archivePrefix = {arXiv},
       eprint = {0906.5269},
 primaryClass = {astro-ph.SR},
       adsurl = {https://ui.adsabs.harvard.edu/abs/2009A&A...503..945F}
}

@ARTICLE{Floresetal23,
       author = {{Flores}, R.~M. and {Corral}, L.~J. and {Fierro-Santill{\'a}n}, C.~R. and {Navarro}, S.~G.},
        title = "{Stellar parameter estimation in O-type stars using artificial neural networks}",
      journal = {Astron. Comput.},
         year = 2023,
        month = oct,
       volume = {45},
          eid = {100760},
        pages = {100760},
          doi = {10.1016/j.ascom.2023.100760},
       adsurl = {https://ui.adsabs.harvard.edu/abs/2023A&C....4500760R}
}

@ARTICLE{Mokiemetal05,
       author = {{Mokiem}, M.~R. and {de Koter}, A. and {Puls}, J. and {Herrero}, A. and {Najarro}, F. and {Villamariz}, M.~R.},
        title = "{Spectral analysis of early-type stars using a genetic algorithm based fitting method}",
      journal = {\aap},
         year = 2005,
        month = oct,
       volume = {441},
       number = {2},
        pages = {711-733},
          doi = {10.1051/0004-6361:20053522},
archivePrefix = {arXiv},
       eprint = {astro-ph/0506751},
 primaryClass = {astro-ph},
       adsurl = {https://ui.adsabs.harvard.edu/abs/2005A&A...441..711M}
}

@ARTICLE{Tamajoetal11,
       author = {{Tamajo}, E. and {Pavlovski}, K. and {Southworth}, J.},
        title = "{Constrained fitting of disentangled binary star spectra: application to V615 Persei in the open cluster h Persei}",
      journal = {\aap},
         year = 2011,
        month = feb,
       volume = {526},
          eid = {A76},
        pages = {A76},
          doi = {10.1051/0004-6361/201015913},
archivePrefix = {arXiv},
       eprint = {1012.2244},
 primaryClass = {astro-ph.SR},
       adsurl = {https://ui.adsabs.harvard.edu/abs/2011A&A...526A..76T}
}

@ARTICLE{Pavlovskietal23,
       author = {{Pavlovski}, K. and {Southworth}, J. and {Tkachenko}, A. and {Van Reeth}, T. and {Tamajo}, E.},
        title = "{High-mass eclipsing binaries: A testbed for models of interior structure and evolution. Accurate fundamental properties and surface chemical composition for V1034 Sco, GL Car, V573 Car, and V346 Cen}",
      journal = {\aap},
         year = 2023,
        month = mar,
       volume = {671},
          eid = {A139},
        pages = {A139},
          doi = {10.1051/0004-6361/202244980},
archivePrefix = {arXiv},
       eprint = {2301.04215},
 primaryClass = {astro-ph.SR},
       adsurl = {https://ui.adsabs.harvard.edu/abs/2023A&A...671A.139P}
}

@ARTICLE{Arayaetal25,
       author = {{Araya}, I. and {Cur{\'e}}, M. and {Machuca}, N. and {Venero}, R.~O.~J. and {Cu{\'e}llar}, S. and {Arcos}, C. and {Cidale}, L.~S.},
        title = "{ISOSCELES project: A grid-based quantitative spectroscopic analysis of massive stars}",
      journal = {\aap},
         year = 2025,
        month = dec,
       volume = {704},
          eid = {A77},
        pages = {A77},
          doi = {10.1051/0004-6361/202555509},
archivePrefix = {arXiv},
       eprint = {2511.03184},
 primaryClass = {astro-ph.SR},
       adsurl = {https://ui.adsabs.harvard.edu/abs/2025A&A...704A..77A}
}

@ARTICLE{Rozanskietal25,
       author = {{R{\'o}{\.z}a{\'n}ski}, Tomasz and {Ting}, Yuan-Sen and {Jab{\l}o{\'n}ska}, Maja},
        title = "{TransformerPayne: Enhancing Spectral Emulation Accuracy and Data Efficiency by Capturing Long-range Correlations}",
      journal = {\apj},
         year = 2025,
        month = feb,
       volume = {980},
       number = {1},
          eid = {66},
        pages = {66},
          doi = {10.3847/1538-4357/ad9b99},
archivePrefix = {arXiv},
       eprint = {2407.05751},
 primaryClass = {astro-ph.IM},
       adsurl = {https://ui.adsabs.harvard.edu/abs/2025ApJ...980...66R}
}

@ARTICLE{Nessetal15,
       author = {{Ness}, M. and {Hogg}, David W. and {Rix}, H.-W. and {Ho}, Anna. Y.~Q. and {Zasowski}, G.},
        title = "{The Cannon: A data-driven approach to Stellar Label Determination}",
      journal = {\apj},
         year = 2015,
        month = jul,
       volume = {808},
       number = {1},
          eid = {16},
        pages = {16},
          doi = {10.1088/0004-637X/808/1/16},
archivePrefix = {arXiv},
       eprint = {1501.07604},
 primaryClass = {astro-ph.SR},
       adsurl = {https://ui.adsabs.harvard.edu/abs/2015ApJ...808...16N}
}

@ARTICLE{Mayoretal2003,
       author = {{Mayor}, M. and {Pepe}, F. and {Queloz}, D. and {Bouchy}, F. and {Rupprecht}, G. and {Lo Curto}, G. and {Avila}, G. and {Benz}, W. and {Bertaux}, J.-L. and {Bonfils}, X. and {Dall}, Th. and {Dekker}, H. and {Delabre}, B. and {Eckert}, W. and {Fleury}, M. and {Gilliotte}, A. and {Gojak}, D. and {Guzman}, J.~C. and {Kohler}, D. and {Lizon}, J.-L. and {Longinotti}, A. and {Lovis}, C. and {Megevand}, D. and {Pasquini}, L. and {Reyes}, J. and {Sivan}, J.-P. and {Sosnowska}, D. and {Soto}, R. and {Udry}, S. and {van Kesteren}, A. and {Weber}, L. and {Weilenmann}, U.},
        title = "{Setting New Standards with HARPS}",
      journal = {The Messenger},
         year = 2003,
        month = dec,
       volume = {114},
        pages = {20-24},
       adsurl = {https://ui.adsabs.harvard.edu/abs/2003Msngr.114...20M}
}

@ARTICLE{Telting2014,
       author = {{Telting}, J.~H. and {Avila}, G. and {Buchhave}, L. and {Frandsen}, S. and {Gandolfi}, D. and {Lindberg}, B. and {Stempels}, H.~C. and {Prins}, S. and {NOT staff}},
        title = "{FIES: The high-resolution Fiber-fed Echelle Spectrograph at the Nordic Optical Telescope}",
      journal = {Astron. Nachr.},
         year = 2014,
        month = jan,
       volume = {335},
       number = {1},
        pages = {41},
          doi = {10.1002/asna.201312007},
       adsurl = {https://ui.adsabs.harvard.edu/abs/2014AN....335...41T}
}

@ARTICLE{ManDon03,
       author = {{Manset}, N. and {Donati}, J.-F.},
        title = "{ESPaDOnS; an exhelle spectro-polarimetric device for the observation of stars}",
      journal = {Proc. SPIE},
         year = 2003,
        month = feb,
       volume = {4843},
        pages = {425-436},
          doi = {10.1117/12.458230},
       adsurl = {https://ui.adsabs.harvard.edu/abs/2003SPIE.4843..425M}
}

@ARTICLE{DiSa90,
       author = {{Dimitrijevic}, M.~S. and {Sahal-Brechot}, S.},
        title = "{Stark broadening of He I lines.}",
      journal = {\aaps},
         year = 1990,
        month = mar,
       volume = {82},
        pages = {519-529},
       adsurl = {https://ui.adsabs.harvard.edu/abs/1990A&AS...82..519D}
}

\begin{appendix}
\twocolumn
\setcounter{section}{1}
\begin{figure}[ht!]
    \centering
    \includegraphics[width=0.99\linewidth]{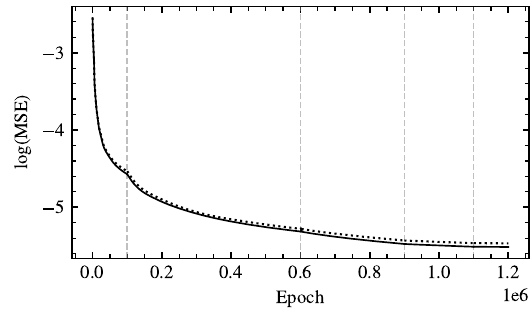}
    \caption{Mean squared error as a function of training epoch. The solid line is the
    MSE for the training data and the dotted line for the test data. The dashed vertical
    lines mark the epochs where the learning rate was reduced.}
    \label{fig:loss}
\end{figure}
\setcounter{section}{0}
\section{Neural network training}\label{appendix:A}
For the training of the neural networks, we used 90\% of the calculated models, the remaining 10\% were used for testing.
We trained the neural networks by minimising the mean squared error (MSE) over
all $n$ training models with $m$ wavelength points each
\begin{equation}
    \text{MSE}=\frac{1}{n}\sum_{i=1}^n \frac{1}{m}\sum_{j=1}^m \left(y_i(\lambda_j) - f_i(\lambda_j)\right)^2 \,,
\end{equation}
where $y_i(\lambda_j)$ is the model flux and $f_i(\lambda_j)$ the predicted flux
at wavelength point $\lambda_j$.
We used the Adam optimiser \citep{kingma2017} to minimise the MSE, with an initial learning rate of \num[output-exponent-marker = E]{1e-2}.
After a fixed number of training epochs (1E5, 5E5, 3E5, 2E5) we reduced the learning
rate to one tenth of the previous value. After 1.2 million epochs in total, we stopped
the training. The behaviour of the MSE over the training epochs is shown in Fig. \ref{fig:loss}.
The MSE of the test and training data follow exactly the same trend, indicating
that no over-fitting of the training data occurred.
The training of one neural network (i.e. a network as described in the main text with
14 input neurons and 7500 output neurons) with 4500 training models
took less than 45\,min on a consumer graphics processing unit.

\section{Microturbulence}\label{appendix:B}
The usual approach -- adjusting the microturbulence so that the
fitted abundance of individual lines of a single element $\varepsilon_p$ becomes
independent of the equivalent width $W_p$ -- assumes a linear relation
\begin{equation}
    \varepsilon_p=k(\xi)\cdot W_p+d(\xi)+\sigma_{\varepsilon_p}\,.
    \label{eq:linear}
\end{equation}
In Eq.~\eqref{eq:linear} $p$ is an index for the different spectral lines,
$k(\xi)$ the slope as a function of microturbulence,
$d(\xi)$ the intersection and
$\sigma_{\varepsilon_p} \sim \mathcal{N}(0,\,\sigma_\varepsilon^2)$ the assumed normal distributed uncertainties
of the individual line abundances.
Calculating the expected variance ${\rm E}[(\varepsilon_p-{\rm E}[\varepsilon_p])^2]$ of the
individual line abundances $\varepsilon_p$, one finds
\begin{equation}
    {\rm E}[(\varepsilon_p-{\rm E}[\varepsilon_p])^2]=\sigma_\varepsilon^2+k(\xi)^2\sigma_W^2\,,
    \label{eq:micro_var}
\end{equation}
where $\sigma_W^2$ is the variance of the
distribution of equivalent widths. For a finite and non-zero
$\sigma_W^2$, the minimum of Eq.~\eqref{eq:micro_var} is at $k(\xi$$=$$\xi_0)$\,=\,0, proving the
equivalence of minimising the variance of individual line
abundances to the approach of strength-independent
line abundances.\\

To determine the microturbulence of the stars analysed in this paper, we use Eq. \eqref{eq:likelihood} to fit multiple lines of one element
with $\xi$ and the abundance $\varepsilon$ as free parameters. To estimate the most likely parameters $\xi$ and $\varepsilon$, we assume the true parameters are $\xi_0$ and $\varepsilon_0$ and that each line is fitted by the true microturbulence value $\xi_0$ and the abundance $\varepsilon_p$ (linearisation of Eq. \eqref{eq:linear})
\begin{equation}
    \varepsilon_p \approx \varepsilon_0 + k\cdot (\xi-\xi_0) \cdot W_p + \sigma_{\varepsilon_p}\,.
\end{equation}
Also linearising the model flux for the $p$-th spectral line and assuming constant derivatives at all wavelength points one obtains
\begin{align}
\label{eq:linearise}
    f(\xi,\,\varepsilon)_{\lambda_i,\textrm{mod}} &\approx f(\xi_0,\,\varepsilon_p)_{\lambda_i,\textrm{mod}} \notag \\
       &~~~~~~~~~ + (\xi-\xi_0) \left.\frac{\partial f(\xi,\,\varepsilon)_{\lambda_i,\textrm{mod}}}{\partial \xi}\right|_{\xi=\xi_0} \notag \\
       &~~~~~~~~~ + (\varepsilon-\varepsilon_p) \left.\frac{\partial f(\xi,\,\varepsilon)_{\lambda_i,\textrm{mod}}}{\partial \varepsilon}\right|_{\varepsilon=\varepsilon_p} \notag \\
     &\approx \bar{f}_{\lambda_i,\textrm{mod}} + (\xi-\xi_0) \cdot \Delta_{\xi} + (\varepsilon-\varepsilon_p) \cdot \Delta_{\varepsilon}\,.
\end{align}
Inserting Eq. \eqref{eq:linearise} in Eq. \eqref{eq:likelihood} and assuming constant uncertainties ($\sigma^2_{\lambda_i,\mathrm{stat}}+\sigma^2_{\lambda_i,\mathrm{sys}}$\,$\approx$\,$\textrm{const.}$) the gradient of the negative log-likelihood can be calculated. Setting the gradient to zero we find
\begin{align*}
    \xi &= \xi_0 + \frac{\frac{1}{|p|}\cdot \sum_p W_p \sigma_{\varepsilon_p}}{k\cdot \left( \left(\frac{1}{|p|}\sum_p W_p\right)^2 - \frac{1}{|p|}\sum_p W_p^2 \right) } \\
    \varepsilon &= \varepsilon_0 + (\xi-\xi_0)\cdot \left( \frac{k}{|p|}\sum_p W_p - \frac{\Delta_\xi}{\Delta_\varepsilon} \right)\,,
\end{align*}
where $|p|=\sum_p1$ is the number of spectral lines used for fitting.
Finally, if the individual abundance uncertainties $\sigma_{\varepsilon_p}$ are independent of the equivalent widths $W_p$, we expect $\sum_iW_i\sigma_i$\,=\,0 since ${\rm E}[W_p\sigma_{\varepsilon_p}]$\,=\,${\rm E}[W_p]\cdot {\rm E}[\sigma_{\varepsilon_p}]$\,=\,0 and consequently the parameters maximising the likelihood function are the true parameters $\xi$\,=\,$\xi_0$ and $\varepsilon$\,=\,$\varepsilon_0$.

\section{SED fits
}\label{appendix:C}
To constrain the interstellar extinction to the binary systems, we calculated the
theoretical SED by using the sum of the {\sc Atlas9} model fluxes of the
two individual stars, scaled by the squared radii.
The model was fitted to observed Johnson $UBV$ \citep{Mermilliod97},
2MASS $JHK$ \citep{Cutrietal03}, and WISE $W1$ to $W4$ \citep{Cutrietal21} photometry.
In the case of HD\,77464 spectra taken with the International Ultraviolet Explorer
(IUE) are available from the Mikulski Archive for
Space Telescopes (MAST\footnote{\url{https://archive.stsci.edu/iue/}}).
To account for interstellar extinction, we used
the reddening law of \citet{Fitzpatrick99}, parametrised by the colour
excess $E(B-V)$ and the ratio of total-to-selective extinction, $R_V$\,=\,$A_V/E(B-V)$.
From a least-squares fit we obtained $R_V$\,=\,2.95$\pm$0.04 and $E(B-V)$\,=\,0.474$\pm$0.006 for HD\,259135.
For HD\,77464 we set $R_V$\,=\,3.1 and obtained $E(B-V)$\,=\,0.034$\pm$0.010.
The best fit for both systems is shown in Fig.~\ref{fig:SED_fits}.

\begin{figure}[ht]
    \centering
    \includegraphics[width=0.98\linewidth]{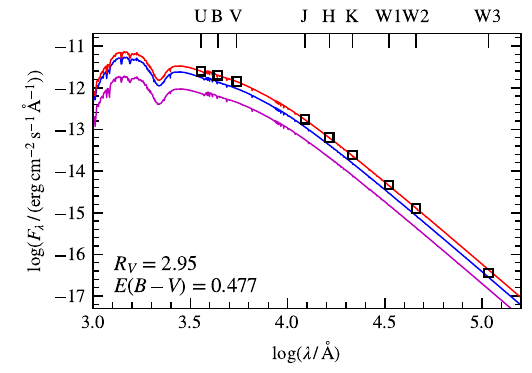}
    \includegraphics[width=0.98\linewidth]{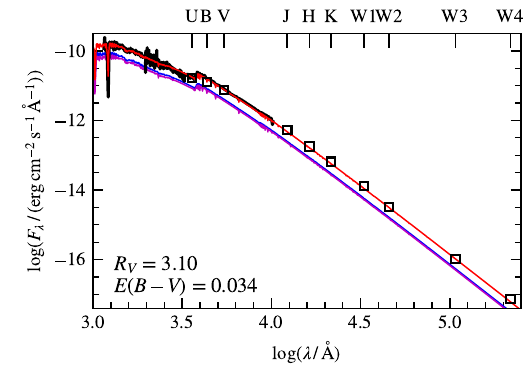}
    \caption{Spectral energy distribution fits of the reddened combined \mbox{{\sc Atlas9}} model fluxes (red)
    to photometric measurements (black squares) and observed IUE spectrophotometry and Gaia XP spectra \citep{Gaia2023} (black line), if available.
    The contributions of the primary and secondary are shown in blue and magenta, respectively.
    Top panel: SED fit for HD\,259135 (V578\,Mon); bottom panel: SED fit for HD\,77464 (CV\,Vel).}
    \label{fig:SED_fits}
\end{figure}

We also determined interstellar extinction towards the new single targets (Sect.~\ref{sect:applications}). Information on the IUE spectrophotometric data that was employed for the comparison of the observed SED with the model fluxes is summarised in Table~\ref{table:IUE}, supplemented by Johnson, 2MASS, and WISE photometry  as in the case of the binaries.

\section{Spectrum fits}\label{appendix:D}
We show a comparison between the observed spectrum from 3950\AA-4750\AA ~and the global best-fitting model, created with the neural networks, for HD\,259135 (V578 Mon) in Fig.~\ref{fig:V578fit},
for HD\,149757 ($\zeta$~Oph) in Fig.~\ref{fig:HD149757fit}, for HD\,93827 in Fig.~\ref{fig:HD93827fit},
and for HD\,77464 (CV Vel) in Fig.~\ref{fig:HD77464fit}.
Except for the latter, the objects show more or less prominent diffuse interstellar bands (DIBs). In addition, interstellar lines such as the \ion{Ca}{ii} H and K lines are also present. We note the high quality of the fits at the very high S/N of the observed spectra, which implies overall completeness of the line list, correctness of oscillator strengths, proper determination of atmospheric parameters, and accurate and precise determination of level populations.
The remaining discrepancies can best be seen in the residuals. The largest differences arise from the DIBs and the interstellar lines, which are not modelled. Inherent to the models are inaccuracies due to the oscillator strengths, which, to a good part, are responsible for the line-to-line abundance scatter. Uncertainties of less than 10-20\% should be typical for the oscillator strengths of many of the metal lines investigated here, but some may reach 50\%, as reflected by the quality markers assigned in the data collection of the National Institute of Standards and Technology\footnote{\url{https://www.nist.gov/pml/atomic-spectra-database}}.
In the case of HD\,149757, additional deviations stem from pulsations of the star, which are not part of the modelling.

\setcounter{section}{3}
\begin{table}
\caption{IUE spectrophotometry used in the present work.}
\label{table:IUE}
\centering
{\small
\begin{tabular}{lllll}
\hline\hline
Object & SW & Date & LW & Date\\
\hline
HD\,149757 &  P32894 & 1988-02-10 & P12637 & 1988-02-10\\
HD\,93827  &  P48400 & 1993-08-16 & P26160 & 1993-08-16\\
HD\,112092 &  P21955 & 1984-01-07 & P18239 & 1990-06-29\\
HD\,87015  &  P21988 & 1984-01-11 & P02597 & 1984-01-11\\
HD\,17081  &  P16255 & 1982-02-04 & R12501 & 1982-02-03\\
HD\,77464  &  P50160 & 1994-03-03 & P27555 & 1994-03-03\\
\hline
\end{tabular}
}
\end{table}
\setcounter{section}{4}

\begin{figure*}
    \centering
    \includegraphics[width=0.99\linewidth]{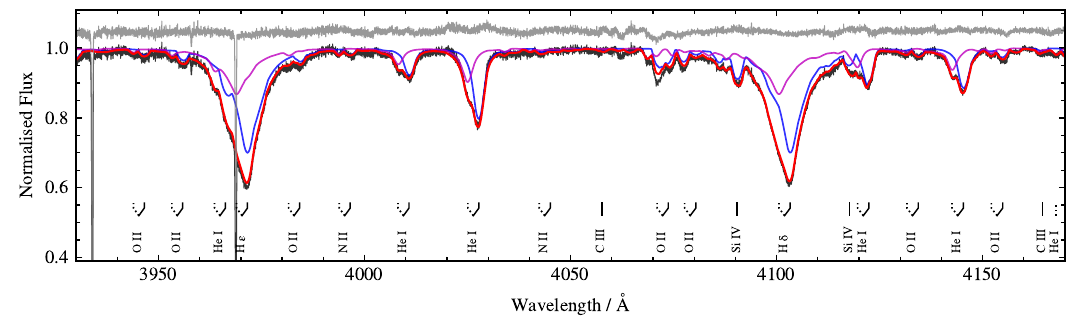}
    \includegraphics[width=0.99\linewidth]{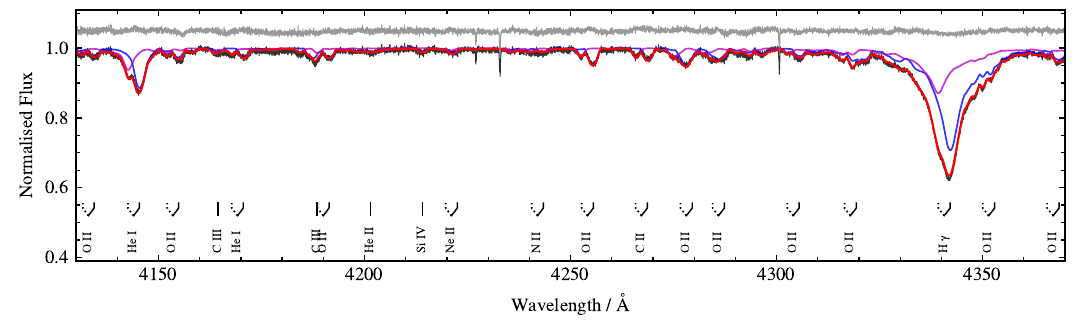}
    \includegraphics[width=0.99\linewidth]{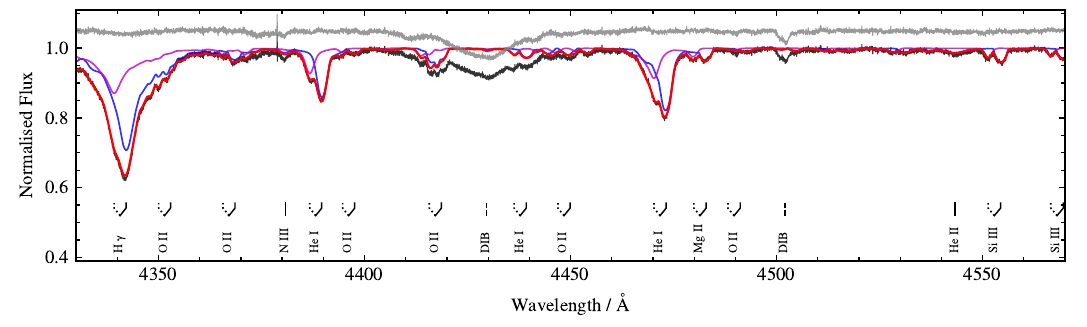}
    \includegraphics[width=0.99\linewidth]{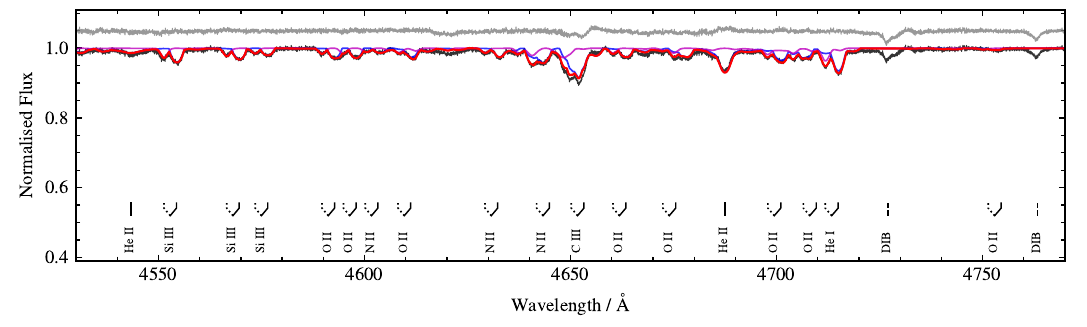}
    \caption{Comparison between the spectrum of the benchmark DEB HD\,259135 (V578\,Mon) in black and the global best-fitting model in red. The difference between observed and model flux is shown in grey at an offset of $1.05$.
    The flux contributions of the primary and secondary star are shown in blue and magenta, respectively.
             The strongest spectral lines for the primary and secondary star are
             marked via solid and dotted lines, respectively. The DIBs are marked by dashed lines.}
    \label{fig:V578fit}
\end{figure*}

\begin{figure*}
    \centering
    \includegraphics[width=0.99\linewidth]{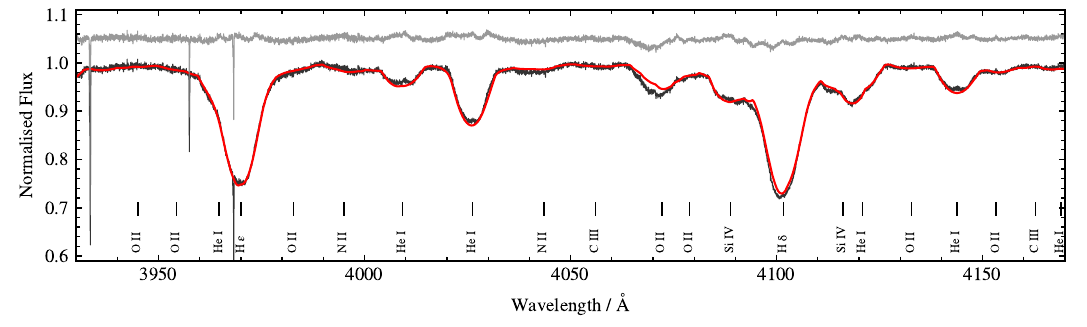}
    \includegraphics[width=0.99\linewidth]{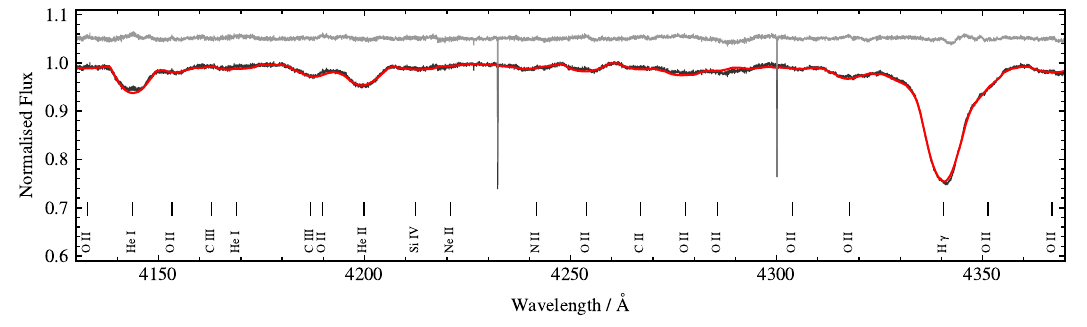}
    \includegraphics[width=0.99\linewidth]{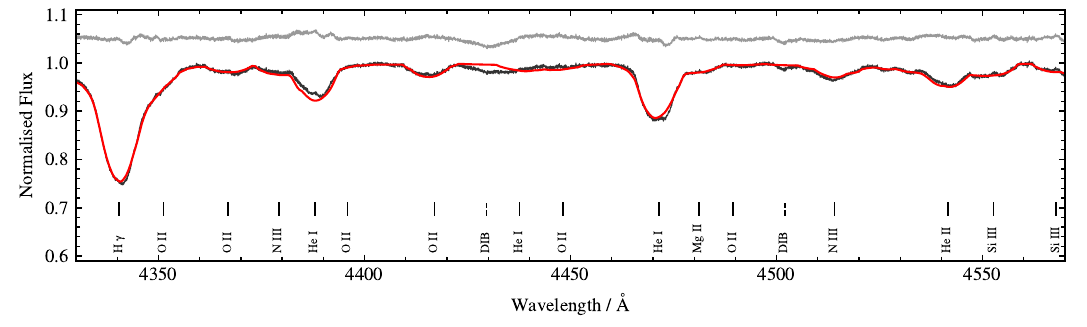}
    \includegraphics[width=0.99\linewidth]{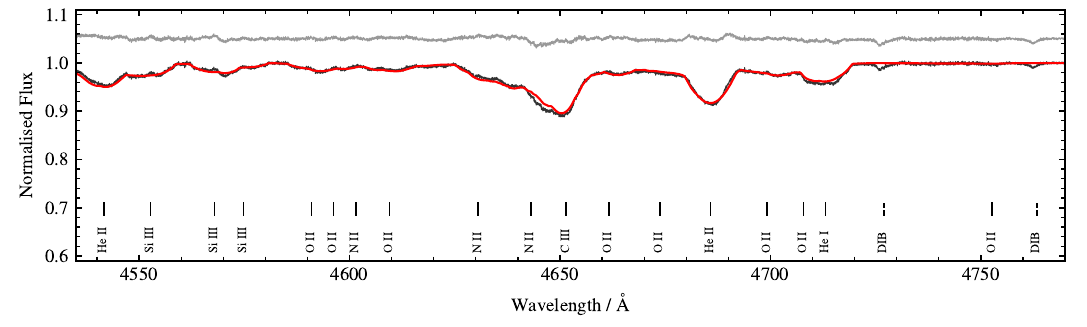}
    \caption{Comparison between the spectrum of the very fast rotator HD\,149757 ($\zeta$\,Oph) in black and the global best-fitting model in red. The difference between observed and model flux is shown in grey at an offset of $1.05$.
    The strongest spectral lines are marked by solid lines, DIBs are marked by dashed lines.}
    \label{fig:HD149757fit}
\end{figure*}

\begin{figure*}
    \centering
    \includegraphics[width=0.99\linewidth]{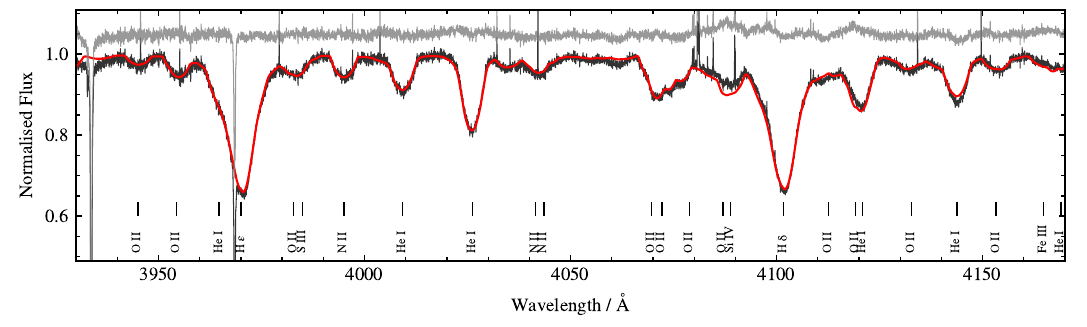}
    \includegraphics[width=0.99\linewidth]{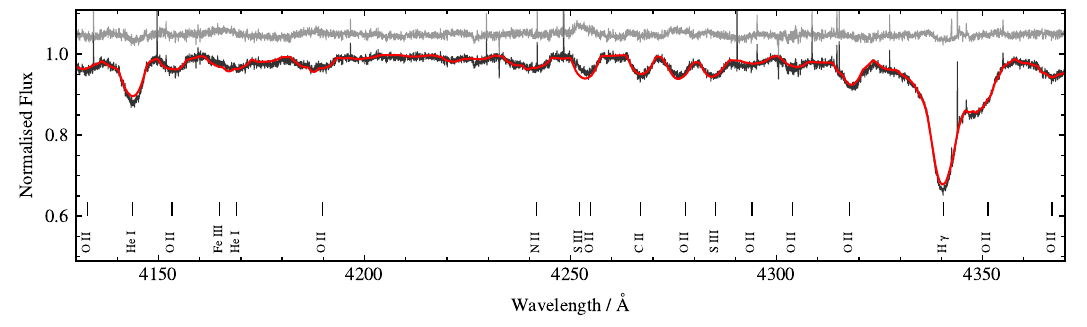}
    \includegraphics[width=0.99\linewidth]{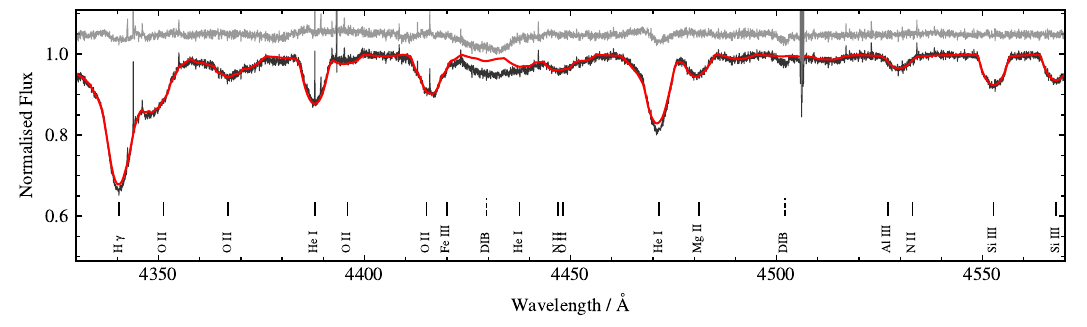}
    \includegraphics[width=0.99\linewidth]{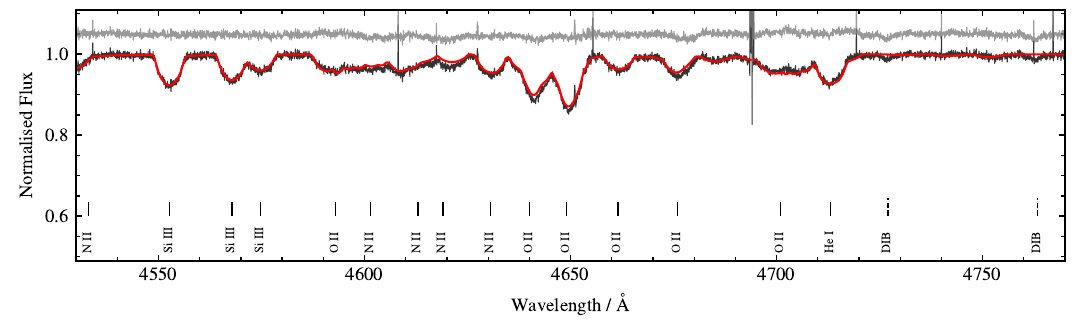}
    \caption{Comparison between the spectrum of the fast-rotating bright giant/supergiant HD\,93827 in black and the global best-fitting model in red. The difference between observed and model flux is shown in grey at an offset of $1.05$.
    The strongest spectral lines are marked via solid lines.
    The DIBs are indicated by dashed lines.}
    \label{fig:HD93827fit}
\end{figure*}

\begin{figure*}
    \centering
    \includegraphics[width=0.99\linewidth]{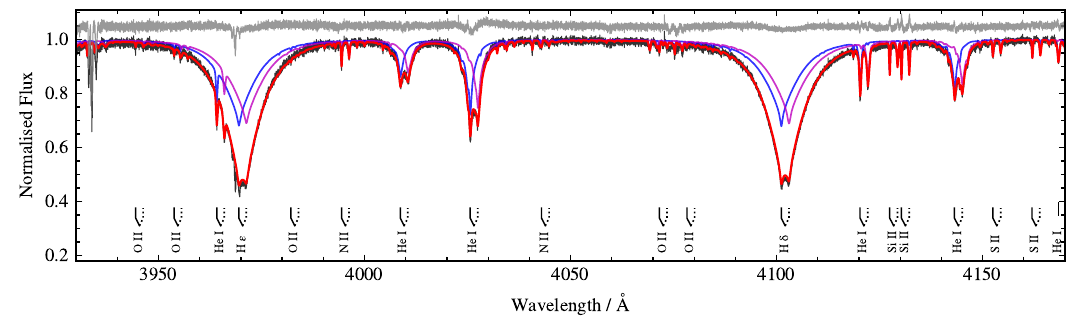}
    \includegraphics[width=0.99\linewidth]{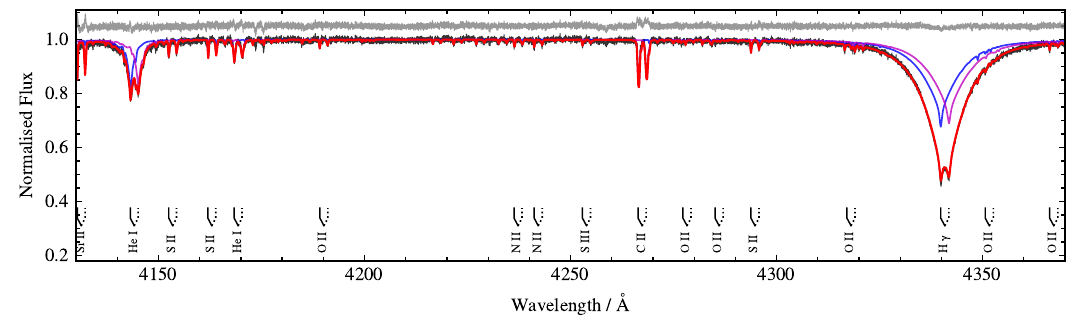}
    \includegraphics[width=0.99\linewidth]{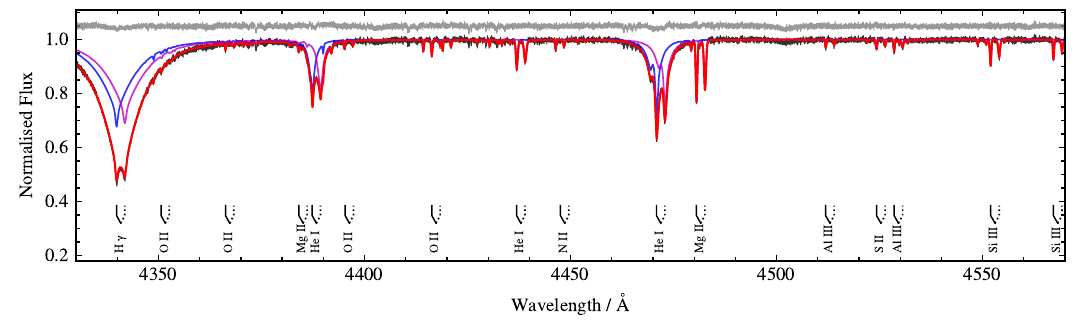}
    \includegraphics[width=0.99\linewidth]{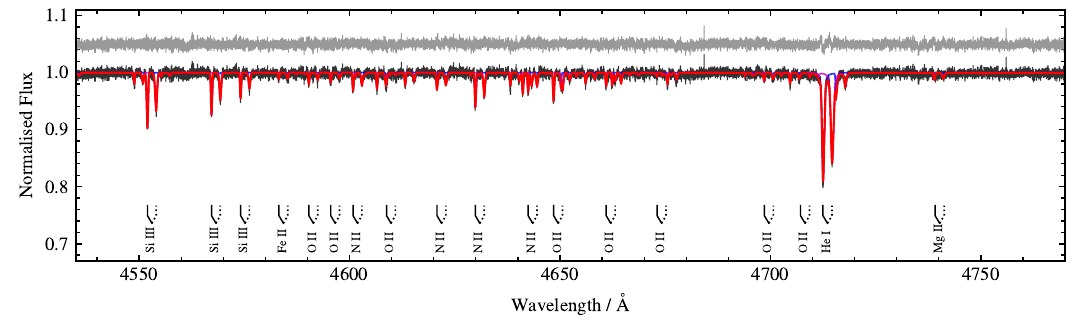}
    \caption{Comparison between the spectrum of the DEB HD\,77464 (CV Vel) in black and the global best-fitting model in red.
            The difference between observed and model flux is shown in grey at an offset of $1.05$.
    The flux contributions of the primary and secondary star are shown in blue and magenta, respectively. The strongest spectral lines for the primary and secondary star are
             marked via solid and dotted lines, respectively.}
    \label{fig:HD77464fit}
\end{figure*}

\setcounter{section}{4}
\section{Tests with lower-resolution spectra}\label{appendix:E}
The availability of XSHOOTER data for our benchmark star HD\,35299 facilitates testing of {\sc Saturn} at spectral resolutions that are relevant for intermediate-resolution spectroscopy with XSHOOTER itself, as widely employed for OB-star observations in the XShootU survey \citep{Vinketal23}, and for future large spectroscopic surveys with WEAVE and 4MOST.
WEAVE will provide a wavelength coverage of about 3660 to 9590\,{\AA} (with some small gaps) in low-resolution mode ($R$\,$\approx$\,5000) and three regions (4040-4650, 4730-5450 and 5950-6850\,{\AA}) at $R$\,$\approx$\,20\,000 in the higher-resolution mode. 4MOST has similar characteristics, a wavelength coverage of about 3700 to 9500\,{\AA} in low-resolution mode ($R$\,$\approx$\,4000 to 7700) and, again, three regions at about 3926–4355, 5160–5730 and 6100–6790\,{\AA} at $R$\,$\approx$\,18\,000 to 21\,000 in the higher-resolution mode. Many indicators for the stellar parameter and abundance determination will be accessible such that quantitative spectroscopy of OB-type stars will face only minor restrictions, mainly  losing some redundancy.

The results of the stellar parameter determination from XSHOOTER spectra are compared to the solution from the analysis of the high-resolution FOCES data in Table~\ref{table:mid_res}. The $R$\,$\approx$\,18\,000 data cover the visual spectral range (VIS) from 5595 to 10240\,{\AA}, and the lower-resolution data the UV and blue range (UVB) from 3000 to 5595\,{\AA}. Excellent agreement of $T_\mathrm{eff}$, $\log (g)$, helium abundance and microturbulence within the error bars is found, with a slight trend towards the recovery of lower temperatures and lower gravity values at lower spectral resolution. The increased uncertainty in $\log(g)$ from the VIS analysis results from H$\alpha$ being the only Balmer line available in that wavelength range, while the remainder of the Balmer series is covered in the UVB range.
The combined rotational and macroturbulent velocity determination shows some small discrepancies, but one has to stress that for such an apparently slow rotator as HD\,35299 the instrumental broadening dominates the line broadening at XSHOOTER resolution. One can expect that for faster rotators the agreement will be better.

The derived metal abundances for the different spectra are listed in Table~\ref{tab:result_abundance_HD5299},
again we find an excellent agreement. At lower spectral resolution, many of the weak spectral lines are no
longer distinguishable from the continuum; an example in the case of HD\,35299 are the phosphorus lines. For fast-rotating stars, even more spectral lines would be affected since they are already shallow due to rotational broadening. We further note that lines from some species are available in either the VIS or the UVB wavelength range, but not in both.

\clearpage

\setcounter{section}{5}
\begin{table*}[ht!]
\centering
\caption{Comparison of stellar parameters derived at different spectral resolutions.}
\label{table:mid_res}
\small
\setlength{\tabcolsep}{1.5mm}
\begin{tabular}{llrccccccc}
\hline\hline
Object & Instrument & $R$ & S/N & $T_\mathrm{eff}$ & $\log (g)$ & $y$
 & \multicolumn{1}{c}{$\xi$} & $\varv \sin i$ & \multicolumn{1}{c}{$\zeta$} \\
\cline{8-10}
 & & & & K & (cgs) & by number &\multicolumn{3}{c}{km\,s$^{-1}$} \\
\hline
HD\,35299  & FOCES & 40\,000 & 350 & 23391$\pm$521 & 4.19$\pm$0.06 & 0.092$\pm$0.002
           & 0.5$\pm$0.6 & 1.4$\pm$0.8 & 8.3$\pm$0.2 \\[1mm]
           & XSHOOTER & 18\,340 & 420 & 23230$\pm$453 & 4.19$\pm$0.09 & 0.090$\pm$0.003
           & 0.8$\pm$0.6 & 3.8$\pm$2.5 & 7.8$\pm$1.6 \\[1mm]
           & XSHOOTER & 9861 & 640 & 23142$\pm$342 & 4.16$\pm$0.04 & 0.092$\pm$0.007
           & 1.0$\pm$0.6 & 0.5$\pm$0.3 & 0.5$\pm$0.3 \\[1mm]
           & XSHOOTER & 5453 & 550 & 23292$\pm$479 & 4.17$\pm$0.06 & 0.099$\pm$0.007
           & 1.3$\pm$0.7 & 1.8$\pm$1.3 & 1.7$\pm$1.3 \\
\hline
 \end{tabular}
 \end{table*}
\begin{table*}[ht!]
\caption{Metal abundances $\varepsilon (X)$\,=\,$\log (X/\mathrm{H})+12$ (by number) for the star HD\,35299,
derived at different spectral resolutions.}
\label{tab:result_abundance_HD5299}
\centering
\small
\setlength{\tabcolsep}{1.5mm}
\begin{tabular}{l@{\hskip 4mm}lllllllllc}
\hline\hline
$R$ & C & N & O & Ne & Mg  & Al & Si & P\tablefootmark{a} & S & Fe  \\ \hline
\\[-3mm]
40\,000  & 8.35$\pm$0.12 & 7.81$\pm$0.07 & 8.85$\pm$0.09 & 8.10$\pm$0.07 & 7.52$\pm$0.05
         & 6.28$\pm$0.04 & 7.52$\pm$0.08 & 5.19$\pm$0.04 & 7.17$\pm$0.08 & 7.49$\pm$0.10 \\[1mm]
18\,340  & 8.33$\pm$0.10 & 7.84$\pm$0.07 & 8.88$\pm$0.08 & ... & 7.51
         & 6.29$\pm$0.06 & 7.51$\pm$0.11 & ... & 7.25$\pm$0.12 & 7.53$\pm$0.14 \\[1mm]
9861     & 8.31$\pm$0.06 & 7.77$\pm$0.06 & 8.83$\pm$0.06 & 8.06$\pm$0.05 & 7.51$\pm$0.06
         & 6.35 & 7.59 & ... & 7.19$\pm$0.12 & 7.52$\pm$0.11 \\[1mm]
5453     & 8.26$\pm$0.13 & 7.88$\pm$0.11 & 8.88$\pm$0.08 & ... & 7.59
         & 6.33$\pm$0.08 & 7.61$\pm$0.06 & ... & 7.10$\pm$0.13 & 7.50$\pm$0.14 \\
\hline
\end{tabular}
\end{table*}

\end{appendix}

\end{document}